\documentclass[%
 aip,
 amsmath,amssymb,
 preprint,%
]{revtex4-1}

\usepackage{graphicx}% Include figure files
\usepackage{dcolumn}% Align table columns on decimal point
\usepackage{bm}% bold math
\usepackage[utf8]{inputenc}
\usepackage[T1]{fontenc}
\usepackage{mathptmx}
\usepackage[version=4]{mhchem}
\usepackage{easyReview}

\begin{document}

%\preprint{AIP/123-QED}

\title[Heavy-Impact Vibrational Excitation and Dissociation Processes in \ce{CO2}]{Heavy-Impact Vibrational Excitation and Dissociation Processes in \ce{CO2}}
% Force line breaks with \\

\author{J. Vargas}
 \email{joao.f.vargas@tecnico.ulisboa.pt}
\affiliation{
Instituto de Plasmas e Fus\~{a}o Nuclear, Instituto Superior T\'{e}cnico, Universidade de Lisboa, Av. Rovisco Pais 1, Lisboa, 1049-001, Portugal%\\This line break forced with \textbackslash\textbackslash
}%
\author{B. Lopez}%
\affiliation{ 
Department of Aerospace Engineering, University of Illinois at Urbana-Champaign, 306 Talbot Lab, 104 S. Wright St. Urbana, 61801, IL, USA%\\This line break forced with \textbackslash\textbackslash
}%

\author{M. Lino da Silva}
\affiliation{%
Instituto de Plasmas e Fus\~{a}o Nuclear, Instituto Superior T\'{e}cnico, Universidade de Lisboa, Av. Rovisco Pais 1, Lisboa, 1049-001, Portugal%\\This line break forced% with \\
}%

\date{\today}

\begin{abstract}
A heavy-impact vibrational excitation and dissociation model for \ce{CO2} is presented. This state-to-state model is based on the Forced Harmonic Oscillator (FHO) theory which is more accurate than current state of the art kinetic models of \ce{CO2} based on First Order Perturbation Theory. The first excited triplet state $^{3}$B$_{2}$ of \ce{CO2}, including its vibrational structure, is considered in our model, and a more consistent approach to \ce{CO2} dissociation is also proposed. The model is benchmarked against a few academic 0D cases and compared to decomposition time measurements in a shock tube. Our model is shown to have reasonable predictive capabilities, and the \ce{CO2 + O <-> CO + O2} is found to have a key influence on the dissociation dynamics of \ce{CO2} shocked flows, warranting further theoretical studies. We conclude this study with a discussion on the theoretical improvements that are still required for a more consistent analysis of the vibrational dynamics of \ce{CO2}, discussing the concept of vibrational chaos and its possible application to \ce{CO2}. The necessity for further experimental works to calibrate such state-to-state models is also discussed, with a proposed roadmap for novel experiments in shocked flows.
\end{abstract}

\maketitle

\section{Introduction}

The modeling of \ce{CO2} nonequilibrium vibrational excitation and dissociation/recombination processes is a research topic that has been studied since the mid-20th Century. This was initially driven by applications such as \ce{CO2} lasers, combustion, and the design of planetary exploration Spacecraft for Venus and Mars, two planets whose atmosphere is mainly composed of \ce{CO2}. Since the beginning of the 21st Century, the modeling of \ce{CO2} nonequilibrium processes faces renewed interest \cite{Bongers2017,Bogaerts2019}, which is once again driven by applications. While the sizing of planetary exploration Spacecraft still remains a major research driver, in the scope of the planning for robotic and crewed exploration of planet Mars, new applications have emerged, such as for example the low-temperature plasma reforming of \ce{CO2} \cite{Guerra2017}. As a polyatomic molecule with three vibrational degrees of freedom (symmetric stretch $s$; bending $b$; assymetric stretch $a$), \ce{CO2} exhibits a very complex and at times puzzling array of diverse physical-chemical processes, many of which are only qualitatively known. This means that in parallel with the application-based research, a great deal of theoretical research is also lacking for this molecule. 

Recent works in the topic \cite{Armenise2013,Kustova2014,Heijkers2015,SahaiThesis,Terraz2019} make use of rates determined using the Schwartz-Slawsky-Herzfeld (SSH). The SSH theory is a first-order perturbation theory (FOPT) which relies on the scaling of lower-level state-to-state (STS) reaction rate coefficients to obtain higher-level rates. These may lead to non-physical values for collision probabilities at high-temperatures, which are pervasive in atmospheric entry flows. Additionally, SSH models are limited to single quantum jumps. An alternative to the Schwartz-Slawsky-Herzfeld (SSH) theory would be to apply more accurate trajectory methods (quasi-classical or quantum) over proposed Potential Energy Surfaces (PES) to compute ro-vibrational energy exchange probabilities. However, PES methods have yet to be applied to large scales and have fallen short of a description up to the dissociation limit of \ce{CO2} due to the inherent computational cost of such sophisticated methods. Increased complexity is still not a realistic option, however a more sophisticated modelling of \ce{CO2} kinetics is still desirable. As such, we propose to use the Forced Harmonic Oscillator (FHO) theory to model \ce{CO2} vibrational state-to-state kinetics. Since the FHO theory is the extension to higher order terms of the same kinetic theory as SSH, it maintains an affordable computational budget while keeping the results physically consistent. Additionally, a more physically consistent treatment of dissociation is achieved by acknowledging the different pathways to \ce{CO2} decomposition, including the enhancement of this process in the presence of \ce{O} atoms. Coupled with a more adequate treatment of the dissociation pathways, this presents a step up from current vibrational state-to-state kinetic models of \ce{CO2}. 

This work will be structured as follows: Section \ref{sec:stateoftheart} will present the state-of-the-art of \ce{CO2} heavy-impact kinetic modeling in low and high-temperature gases and plasmas. In section \ref{sec:theory} the theoretical framework of this work will be presented covering the governing equations, the determination of the manifold of levels, the interactions between singlet and triplet states of \ce{CO2}, the calibration with experimental data, the \ce{CO} and \ce{O} thermochemistry that was added to the model, and culminating with a brief description of underlying assumptions and model flaws. Section \ref{sec:results} will discuss some aspects of the state-specific kinetic datased produced in this work, and will present some theoretical test-cases for a dissociating and a recombining pure \ce{CO2} flow. A further comparison of dissociation times is carried out against available shock-tube experiments. Section \ref{sec:discussion} discusses the key findings of our new model, highlighting the importance of the key \ce{CO2 + O <-> CO + O2} reaction and discussing the large uncertainty that exists at lower temperatures, possible future improvements to the model, including the influence of radiative losses, and possible strategies for reducing the computational overhead of our model, which remains intractable for complex multidimensional applications, having over 20,000 rates. This section closes with an extensive discussion on the theoretical developments that are lacking for correcting the theoretical flaws of current state-to-state \ce{CO2} models (including ours), henceforth introducing the concept of \textit{vibrational chaos} for higher-lying levels of \ce{CO2}. We then summarize a possible roadmap for novel experiments.

\section{State-of-the-art}
\label{sec:stateoftheart}

As discussed in the introduction, \ce{CO2} is a molecule that plays a key role in applications such as combustion, atmospheric radiative transfer, atmospheric entry of Spacecrafts (since Venus and Mars are composed of 97\% \ce{CO2} and 3\% \ce{N2}), stratospheric processes in the atmosphere of Venus, Earth, and Mars, \ce{CO2} lasers, and plasma discharges, particularly for plasma reforming applications and SynGas production. \ce{CO2} is further a molecule with high societal impact, since it is chiefly responsible for the global warming trends of Earth's atmosphere \cite{LinodaSilva2020}.

Since our work is focused on  nonequilibrium heavy-impact kinetic processes of \ce{CO2}, we will restrict the scope of our discussion of the state-of-the-art to 1) high-temperature dissociation processes of \ce{CO2}, studied in shock-tube facilities; and 2) low-temperature nonequilibrium processes in \ce{CO2}, for the case of \ce{CO2} lasers, stratospheric processes, and plasma reforming of \ce{CO2}. While radiative heat transfer of \ce{CO2} impacts nonequilibrium excitation and dissociation processes of \ce{CO2}, for the sake of compactness we do not provide a discussion on the state-of-the-art on this topic and the reader may be referred to our recent work \cite{Vargas2020} where this discussion is carried out.

\subsection{Thermal dissociation}
\label{sec:stateofthearthighT}

Shock-tube investigations of thermal dissociation rates for \ce{CO2} were initiated in the mid-60's of the past Century, in support of Mars/Venus exploration missions. 

Brabbs \cite{Brabbs1963} carried out shock-tube experiments for a highly diluted \ce{CO2}/\ce{Ar} mixture around 2,500-2,700 K, using two complementary techniques: 1) a single pulse technique where the gas is shocked and successively quenched by a rarefaction wave, freezing the chemistry and allowing chemical analysis; 2) measurement of \ce{CO2} near-UV chemiluminescence bands\footnote{\ce{CO + O + M <-> CO2^* + M} followed by \ce{CO2}$^*$ \ce{-> CO2} + $h\nu$.}. Both measurement techniques provided similar results. Davies carried out shock-tube experiments in high dilution \ce{CO2}/\ce{Ar} mixtures, with post-shock temperatures in the range of 3,500-6,000 K \cite{Davies1964} and 6,000-11,000 K \cite{Davies1965}. The first work probed the emission of the IR vibrational bands of \ce{CO2} at $2.7/4.3$ $\mu$m whereas the high-temperature study additionally measured the \ce{CO2} chemiluminescence bands. The comparison between both measurements showed an order of magnitude difference, providing evidence that dissociation/recombination processes of \ce{CO2} might be a multi-step process. Michel \cite{Michel1965} measured reflected shockwaves IR radiation at 4.3 $\mu$m in diluted \ce{CO2}/\ce{Ar} and \ce{CO2}/\ce{N2} mixtures, in the range T=2,800-4,400 K. These studies were followed through by Fishburne \cite{Fishburne1966} and Dean \cite{Dean1973} who measured IR radiation (at 2.7/4.3 $\mu$m for Fishburne, 4.3 $\mu$m for Davis) in 3,000-5,000 K shock-tube flows, for \ce{CO2}/Ar mixtures (and \ce{CO2}/\ce{N2} mixtures for Fishburne), again with a high dilution ratio. Generalov and Losev \cite{Generalov1966} measured the rate of dissociation for \ce{CO2} without dilution from other mixtures, probing the \ce{CO2} near-UV chemiluminescence bands through absorption spectroscopy, in the 3,000-5,500 K range. Kiefer \cite{Kiefer1974} carried out shock-tube experiments for a highly diluted \ce{CO2}/\ce{Kr} mixture in the 3,600-6,500 K range using a laser schlieren technique. Hardy \cite{Hardy1974} and Wagner \cite{Wagner1974} measured reflected shockwaves IR radiation at $4.3$ $\mu$m and \ce{CO + O} chemiluminescence in the near-UV in un-diluted \ce{CO2} mixtures, in the T=2,700-4,300 K and T=3,000-4,000 K ranges respectively. Finally, Ebrahim \cite{Ebrahim1976} complemented the work of Generalov and Losev, determining the rate of dissociation of pure \ce{CO2} in the 2,500-7,000 K range, using a Mach--Zender interferometry technique\footnote{A common characteristic of most of these early studies is the presence of \ce{CO2} in a highly diluted gas bath of either \ce{Ar}, \ce{Kr}, or \ce{N2} so as to avoid the establishment of significant endothermic reactions as, for example, Ar has a rather large ionization energy of 15.76 eV and \ce{N2} a rather large dissociation energy of 9.79 eV, hence keeping a high post-shock-temperature.}.

These initial measurements showed a significant scattering between the different dissociation rate values, but also in the dissociation activation energy, which ranged from 2.9 to 4.6 eV \cite{Ebrahim1976} \footnote{The bond dissociation energy of \ce{CO2} to \ce{CO}($^1\Sigma$)\ce{+ CO}($^3$P) being 5.52 eV.}. Potential explanations for these large uncertainties were attributed to the effects of impurities in the gas \cite{Clark1971,Zabelinskii1985}, or to a failure in reducing the systematic errors in the measured quantities \cite{Zabelinskii1985}. Another potential source of uncertainty \cite{Clark1969,Baber1974} stems from the fact that dissociation products (atomic oxygen) may react with \ce{CO2}, further dissociating it (reaction \ce{O + CO2 <-> CO + O2}). This means that the measured dissociation rate of \ce{CO2} (assessed through the measurement of \ce{CO2} concentrations) is in fact an effective rate resulting from a two-step process (dissociation of \ce{CO2} and dissociative recombination of O and \ce{CO2}). Clark further studied the dynamics of \ce{O + CO2} interactions using different isotopes of atomic oxygen \cite{Clark1970a}.

Another proposed explanation was that dissociation proceeded through a two step mechanism\footnote{Since the dissociation of the ground electronic state of \ce{CO2} to the ground states of CO and O is spin-forbidden.} consisting of the collisional excitation of \ce{CO2} to an upper electronic state, followed by dissociation:
\begin{align}
\textrm{CO}_2(^1\Sigma)+\textrm{M}&\rightarrow \textrm{CO}_2^*+\textrm{M}\\
\textrm{CO}_2^*&\rightarrow \textrm{CO}(^1\Sigma)+\textrm{O}(^3\textrm{P})
\end{align}
with Brabbs proposing that \ce{CO2}$^*$ corresponds to \ce{CO2}($^3\Pi$) \cite{Brabbs1963}, and Fishburne proposing that \ce{CO2}$^*$ corresponds to either \ce{CO2}($^3$B$_2$) or \ce{CO2}($^1$B$_2$) \cite{Fishburne1966}. However, such propositions were not entirely satisfactory, since this implied potential curve crossings below the dissociation limit, and that further energy was still required for dissociation 
\cite{Fujii1989}. This led to the understanding that detailed knowledge on the potential energy surfaces for the excited electronic states of \ce{CO2} was needed.

To allow improved insights on the dynamics of thermal \ce{CO2} dissociation, later studies focused on measuring the concentrations of atomic oxygen through Atomic Resonance Absorption Spectroscopy (ARAS) at 130 nm. This allowed avoiding ambiguities derived from the measurement of \ce{CO2} concentrations, such as having additional processes that lead to the depletion of \ce{CO2} (e.g. the aforementioned \ce{O + CO2 <-> CO + O2} reaction). Fuji \cite{Fujii1989} used an ARAS technique for reflected shockwaves in diluted \ce{CO2}/Ar mixtures, in the 2,300-3,400 K range. The determined rate had an activation energy of 3.74 eV. Burmeister \cite{Burmeister1990} used the same technique to determine \ce{CO2} dissociation rates in diluted \ce{CO2}/\ce{Ar} mixtures, in the temperature range 2,400--4,400 K. The activation energy was found to be 4.53 eV. Eremin \cite{Eremin1993} investigated the chemiluminescence radiation in pure \ce{CO2} shocked flows and developed a simplified kinetic model that allowed concluding that above 3,000 K, part of the dissociation products were composed of excited oxygen \ce{O}($^1$D), and not only \ce{O}($^3$P). In a later experiment, Eremin \cite{Eremin1996} carried out ARAS measurements in shocked \ce{CO2}/Ar diluted mixtures in the 4,100-6,400 K temperature range to determine the fraction of excited oxygen \ce{O}($^1$D) produced during \ce{CO2} dissociation. He found out that \ce{O}($^1$D) amounted to 0-10\% of the dissociation products. Park \cite{Park1993}, in the scope of the development of his Mars entry kinetic model, considered the results from Davies \cite{Davies1964,Davies1965}, fitting them to an Arrhenius expression where the activation energy was constrained to the accepted \ce{CO2} dissociation energy, and the pre-exponential factor set to $n=1.5$. Ibragimova \cite{Ibragimova2000} carried out a critical review of past works, proposing a \ce{CO2} dissociation rate valid in the 300-40,000 K temperature range, for different colliding partners (\ce{Ar}, diatomic molecules, \ce{CO2}, \ce{C}, \ce{N}, and \ce{O}). She also proposed a dissociation coupling factor $Z(T,T_v)$ for accounting for thermal nonequilibrium conditions during dissociation. Saxena \cite{Saxena2007} claimed that the anomalous low activation rates were the result of rapid secondary loss of \ce{CO2} via the reaction \ce{O + CO2 <-> CO + O2}. Activation rates of more recent experiments, discarding this effect, are systematically above 4.3 eV. A pressure dependence on the dissociation rate, noticeable at higher presures, was also found. Jaffe \cite{Jaffe2011} claimed that the explanation for the low activation energy laid in the transition from the $^1$X to $^3$B$_2$ state being the rate-determining step for \ce{CO2} dissociation. In a follow-up work, Xu et al. \cite{Xu2017} reviewed the past works and proposed an effective rate in the range 2,000-10,000 K.

\subsubsection{Vibrational relaxation and dissociation incubation times}

Vibrational relaxation and dissociation incubation times for the dissociation of \ce{CO2} have been derived by some authors from the aforementioned shock-tube dissociation measurements. These imply firstly developing analytical master equation models and then applying them to the measurements of time-dependent data from shock-tube experiments.

Weaner \cite{Weaner1967} compared the density relaxation times $\tau_E$ through Mach--Zender  interferometry  and the v$_3$ mode relaxation time through measurement of the 4.3 $\mu$m bands,  Weaner considered that the energy in the v$_3$ mode is less than 8\% of the other v$_1$ and v$_2$ modes and even less at lower temperatures, and that as such $\tau_E=\tau_{\text{v}_{1,2}}$. The obtained values for $\tau_E$ and $\tau_{\text{v}_3}$ were equivalent in the temperature range 450--1,000K, leading to the conclusion that all the vibrational modes of \ce{CO2} relax at the same rate.

Simpson \cite{Simpson1970} studied vibration relaxation in shocked flows in the range T=360-1,500K, for pure \ce{CO2} gases, and for \ce{CO2} diluted in nitrogen, deuterium and hydrogen, with an emphasis on the deactivation of the bending mode. The obtained results put into evidence different behaviours at low and high temperatures, wherein a strong dependence from the collisional partner reduced mass exists at high temperatures, but not a low temperatures.

Oehlschlaeger \cite{Oehlschlaeger2005,Oehlschlaeger2005a} carried out several shock-tube experiments, investigating the dissociation of \ce{CO2} behind reflected shock waves at temperatures in the 3,200--4,600 K range and pressures in the 45--100 kPa range. This has been done probing the \ce{CO2} UV bands (B$\rightarrow$X transition) in absorption using an UV laser, with a sampling rate in the millisecond range. Oehlschlaeger also carried out an extensive review of previous works regarding dissociation incubation times and reproduced his experiments through a simplified one-dimensional energy-grained master equation model. The obtained results highlight the importance for knowing the average energy $\langle\Delta E\rangle$ transferred through a collision, for the whole, $p$, $T$ range. Unfortunately, no shock speed values are reported, preventing the selection of his experimental work as a suitable test-case.

The incubation times obtained by Oehlschlaeger were deemed too high by Saxena \cite{Saxena2007}, who carried similar experiments using a laser Schlieren experimental technique, and found much lower incubation times in the temperature range T=4,000-6,600K, for \ce{CO2-Kr} mixtures. Saxena found that full vibrational equilibrium was reached before the onset of dissociation.

\subsection{Low-temperature nonequilibrium processes}
\label{sec:stateoftheartlowT}

\begin{figure}[!htbp]
\centering
\includegraphics[width=.7\textwidth]{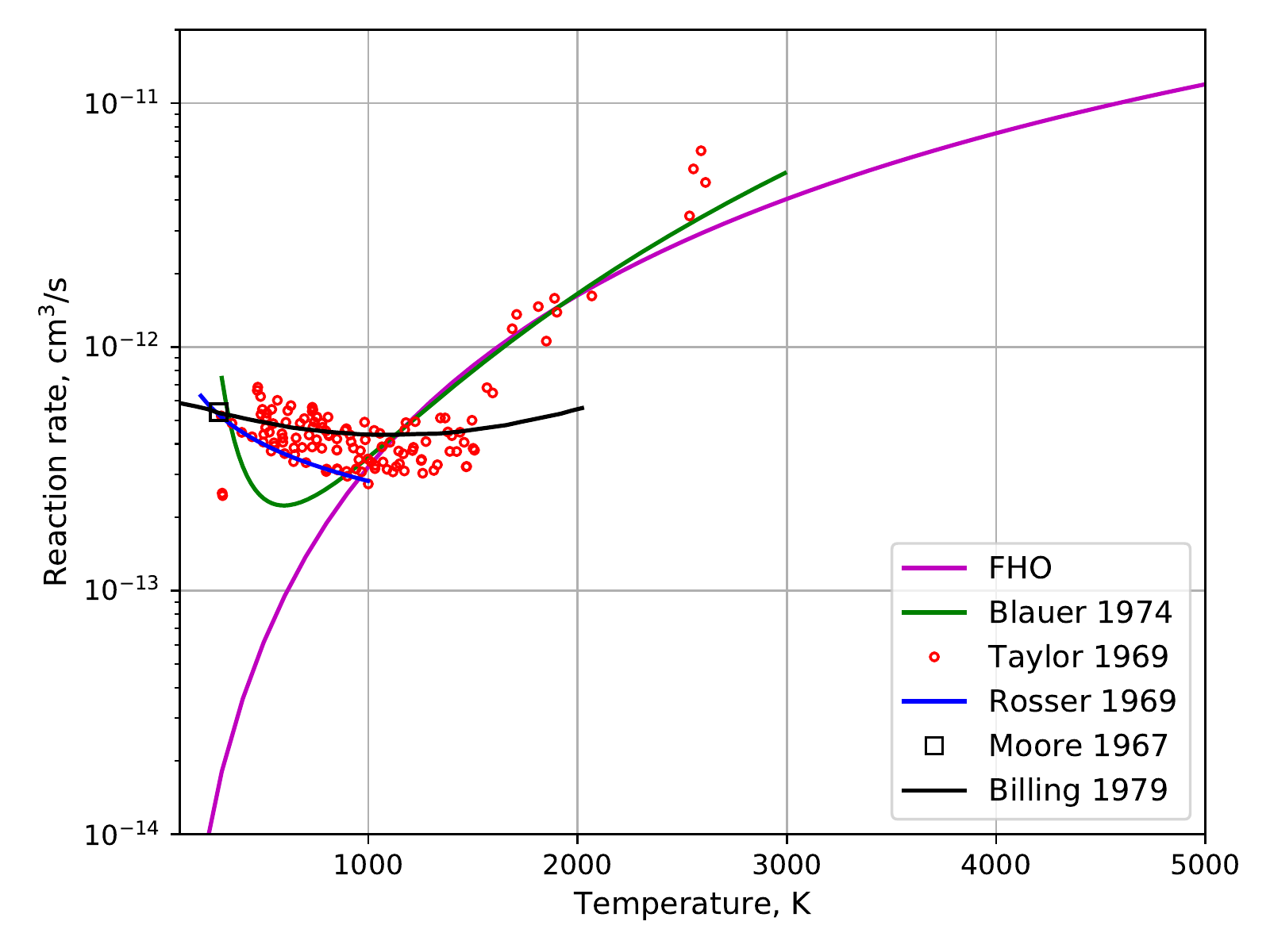}
\caption{Comparisons of rate coefficients\cite{Taylor1969,Rosser1969,Blauer1974,Moore1967,Billing1979} for the reaction \ce{CO2(001) + N2(0) <-> CO2(000) + N2(1)}.}%
\label{fig:CO2-N2-resonant}
\end{figure}

Room-temperature nonequilibrium processes in \ce{CO2} were the subject of a large amount of research studies at about the same time, in the wake of the development of the first \ce{CO2} lasers \cite{Patel1964}. Namely, the vibrationally-specific resonant reaction
\begin{align}
\ce{CO2(001) + N2(0) <-> CO2(000) + N2(1) + 18}\text{ cm}^{-1}
\label{eq:CO2v3N2vresonant}
\end{align}
was acknowledged as key to the inversion of the upper populations in \ce{CO2}-\ce{N2} lasers \cite{Moore1967,Sobolev1966,Sobolev1967-1,Sobolev1967-2,Gordiets1968,Taylor1969}. Interestingly, several measurements evidenced that a minimum of the rate for this process was reached at around 1,000 K, with a rate increase both at lower and higher temperatures. This behavior could not be expected to be explained by standard kinetic models like the Landau-Teller theory, which predicts increasing rates with increasing temperature, no matter the temperature range. To account for this low-temperature behavior, Sharma and Brau suggested \cite{Sharma1969} that the rate increase at lower temperatures could be attributed to long-range attractive forces between the dipole moment of \ce{CO2}(v$_3$) and the quadrupole moment of \ce{N2}. On this basis, they proposed the Sharma--Brau perturbation theory that has been very successful at reproducing this class of kinetic processes, predicting a temperature dependence of the order $T^{-1/2}$ in the low-temperature region, as opposed to the $T^{3/2}$ temperature dependence in the high-temperature range, attributable to short-range (nonadiabatic) repulsive forces. Fig. \ref{fig:CO2-N2-resonant} presents a list of the measured rates for process \ref{eq:CO2v3N2vresonant}, alongside with predicted rates from the SSH, FHO, and Sharma--Brau theory.

In the wake of these experimental and theoretical developments, a significant number of vibrationally-specific exchange rates have been measured for the three vibrational modes of \ce{CO2} and a great deal of collisional partners (both atomic and polyatomic \cite{Losev1976-1,Losev1976-2}).

%---------------------------- CO2 lasers -----------------------------

%----------------------------- Cooling of upper atmospheres ------------------------------

In the late 70's of the last Century, a new venue for the investigation of low-temperature vibrational nonequilibrium processes in \ce{CO2} was opened by the investigation of the key vibrational excitation processes:

\begin{align}
%\textrm{CO}_2(000)+\textrm{O}\rightarrow \textrm{CO}_2(\textrm{v}_2^{*})+\textrm{O}
\label{eq:CO2v2O-1}
\textrm{CO}_2(000)+\textrm{O}&\rightarrow \textrm{CO}_2(0\textrm{n}0)+\textrm{O}\\
\label{eq:CO2v2O-2}
\textrm{CO}_2(0\textrm{n}0)&\rightarrow \textrm{CO}_2(000)+h\nu
\end{align}

which have been recognized \cite{Bougher1994}\cite{LopezPuertas2001} to be the main contributor to the cooling of upper planetary atmospheres (Venus, Earth, Mars) trough the removal of translation energy from oxygen atoms by the bending mode of \ce{CO2} (eq. \ref{eq:CO2v2O-1}), and the subsequent radiation of energy away from the atmospheric layer (either towards the ground or towards Space eq. \ref{eq:CO2v2O-2}). This has led to an extensive number of theoretical and experimental works that have put a strong emphasis on the improvement of low-temperature, neutral kinetic models for \ce{CO2}.
 
%----------------------------- Cooling of upper atmospheres ------------------------------
%----------------------------- Plasma Reforming of CO2 ------------------------------

More recently, plasma reforming of \ce{CO2} into Syngas has been proposed as a potentially viable option for producing \ce{CO2}-neutral fuels \cite{Azizov1983,Fridman2008,Bogaerts2015,Capitelli2016}. This is achieved in microwave plasma sources, through the electron-impact excitation of the vibrations of \ce{CO2}. This has led to renewed studies on the modeling of low-temperature \ce{CO2} plasmas, with the development of detailed kinetic models for this specific application \cite{Kozak2014,DeLaFuente2016,Berthelot2016,Silva2018a}, the investigation of alternative high-power microwave \cite{Kwak2015}, or gliding arc \cite{Heijkers2017} plasma sources, the modeling of complex hydrodynamic effects \cite{Belov2018}, or the investigation on improving the efficiencies of this technique \cite{VanRooij2015}. Last but not least, experimental techniques have been developed up to a point where time-dependent populations for the lower-vibrational levels of \ce{CO2} can be measured in the discharge and post-discharge regions through Fourier Transform Infrared Spectroscopy (FTIR) \cite{Klarenaar2017a,Urbanietz2018,Stewig2020}.

As it can be inferred from these recent works, mostly published in the last decade, the modeling of \ce{CO2} nonequilibrium processes appears to be a contemporary "hot-topic". One may trace this to the seminal book by Fridman \cite{Fridman2008} which extensively reported and discussed past topical theoretical and experimental works, mostly of USSR origin. The theoretical methodologies discussed in this book were all but adopted by the contemporary plasma chemistry community, working as a \textit{de-facto} standard approach. In parallel, much effort was put into reproducing the experimental work discussed in this work, with an emphasis on reaching energy efficiencies for \ce{CO2} dissociation as high as reported therein. It is relevant to review both of these aspects, in an effort for understanding the current \textit{status-quo} on this topic.

The proposed pathways for \ce{CO2} dissociation by means of vibrational excitation in nonequilibrium plasmas start with an electron energy input in the range 1--2 eV. This provides excitation of low vibrational levels of \ce{CO2} which then excite higher levels through V--V ladder-climbing processes. Finally a non-adiabatic transition to the excited electronic state X$^1\Sigma\rightarrow^3$B$_2$ takes place, followed by dissociation. Based on the works by Rusanov et al.\cite{Rusanov1985b} and Rusanov, Fridman, \& Sholin, \cite{Rusanov1986}, Fridman considers that only the assymetric mode v$_3$ contributes for this up-pumping process, as the other mode (lumped bending and symmetric $bs$) has small enough energy spacings that V--T de-excitation processes dominate over V--V up-pumping processes, since typically $T<T_{v}$.

The \ce{CO2} conversion efficiencies of several classes of experiments in achieving such dissociation processes are extensively reviewed by Fridman. The maximum theoretical efficiency for dissociation in thermal plasmas is claimed not to exceed 43--48\%\cite{Levitsky1978a,Levitsky1978b,Butylkin1978,Butylkin1979}, with quasi-equilibrium arc discharges achieving efficiencies of no more than 15\%\cite{Polak1977}. Fridman points out that increased efficiency may be achieved for non-thermal plasmas, with efficiencies up to as 30\% for electron impact dissociation, achieved through plasma radiolysis using high-current relativistic electron beams at atmospheric pressure\cite{Legasov1978a,Legasov1978b,Legasov1978c,Legasov1978d,Vakar1978a,Vakar1978b}. If the pressure is reduced, the combination of electronic and vibrational excitation allows conversion efficiencies in the range of 20--50\% for plasma-beam experiments\cite{Ivanov1978,Nikiforov1979}. Fridman concludes that even higher efficiencies may be achieved in nonequilibrium discharges at moderate pressures, with conversion efficiencies of 60\% for a pulsed microwave/radiofrequency discharges in a magnetic field n the conditions of electron cyclotron resonance\cite{Asisov1977,Butylkin1981}. Finally, for flowing plasmas, conversion efficiencies as high as 80\% and 90\% are reported for subsonic\cite{Legasov1978a,Legasov1978b,Legasov1978c,Legasov1978d} and supersonic\cite{Asisov1981a,Asisov1981b,Asisov1983} flows respectively. These high efficiencies are claimed to be reached for input energies in the range 0.8--1.0 eV/molecule, and may be explained by the decrease of the gas translational temperature through the subsonic and supersonic expansions, hindering the V--T de-excitation processes. In particular, the maximum efficiency of 90\% is achieved for near-hypersonic plasma flows with Mach numbers in the range $M=3-5$.

All the sources cited by Fridman correspond to works carried out in the later 70's/early 80's
USSR. Such high efficiencies have to date not been reproduced, with a review by Ozkan \cite{Ozkan2015} finding that no atmospheric plasma sources achieve a energy efficiency above 15\%. Also of importance is the fact that no atmospheric plasma source can reach a fraction of converted \ce{CO2} higher than 25\%, meaning that plasma reforming of \ce{CO2} in large quantities is not yet a viable economic endeavor. More recently, Bongers \cite{Bongers2016} carried out a set of experiments on a vortex‐stabilized, supersonic 915MHz microwave plasma reactor, achieving \ce{CO2} conversion efficiencies in the range of 47--80\% and energy efficiency in the range of 35--24\% for an energy input in the range 10.3--3.9 eV/molecule. Upon increasing the mass flow of \ce{CO2} (from 11slm to 75slm), the energy input may be reduced to the range 1.9--0.6 eV/molecule, which is deemed as optimal by the Fridman review. In this case the energy efficiency increases, showcasing a range of 51--40\%, however the \ce{CO2} conversion output decreases significantly, to the range 11--23\%. Further analysis were carried out for a subsonic 2.45GHz microwave plasma reactor, wherein a maximum energy efficiency of 46\% was achieved. Of particular interest is a comparison of these results with the ones presented by Fridman, wherein the obtained efficiencies stayed well below the ones reported by Fridman, even for similar experience conditions.

Recent modeling works have further laid the theoretical foundations \cite{Kozak2014,Kustova2014} to which most of the subsequent published numerical models for \ce{CO2} adhere. Reference transitions are considered for the first vibrational levels of \ce{CO2} \cite{Blauer1974,Achasov1986,Joly1999a}, and SSH scaling laws are applied for determining transition rates for higher vibrational levels. A few sample rates include\cite{Armenise2013,Kustova2014}

\begin{equation}
k^{\textrm{CO}_2}_{\textrm{v}_2+1\rightarrow \textrm{v}_2}=k^{\textrm{CO}_2}_{010\rightarrow 000}(\textrm{v}_2+1)
\end{equation}

for V--T processes;

\begin{equation}
k^{\textrm{CO}_2}_{\textrm{v}_1+1,\textrm{v}_2+1,\textrm{v}_3\rightarrow \textrm{v}_1,\textrm{v}_2,\textrm{v}_3+1}=k^{\textrm{CO}_2}_{110\rightarrow 001}(\textrm{v}_1+1)(\textrm{v}_2+1)(\textrm{v}_3+1)
\end{equation}

for V--V--T processes, among others.

Energy levels are approximated by a simple anharmonic oscillator \cite{Suzuki1968}, up to the dissociation limit, which is considered to lie in the crossing to the $^3$B$_2$ state at about 5.56 eV. Dissociation occurs when a molecule in the v$_3$ level jumps to a v$_3+1$ level above this energy limit, dissociating with a probability of one. Typically the maximum quantum number for v$_3$ is considered to lie at 21 \cite{Kozak2014}.

\section{Theory}
\label{sec:theory}

In this section we describe our theoretical methods and models for determining the vibrational state-to-state rates of \ce{CO2}. Firstly, the governing equations for the flow are described followed by brief description of the \ce{CO2} molecule in terms of it's vibrational modes. Secondly, the determination of level energies is described followed by the Potential Energy Surface (PES) crossings between ground state and electronically excited \ce{CO2}. A detailed description of the Forced Harmonic Oscillator (FHO) is passed over in favor of a more qualitative description of the types of reactions this theory is applied to followed by a brief justification of the use and description of the Rosen-Zener theory for spin-forbidden interactions. The addition of macroscopic chemistry mechanisms to the kinetic scheme is performed on a step-by-step approach culminating with the addition of the \ce{CO} thermochemistry of Cruden \textit{et al.} \cite{Cruden2018} and a recap of the full dataset. Finally, a brief discussion on the flaws and limitations of this model is performed, focusing on future updates to the database which will correct some of these limitations, but also detailing the more underlying assumptions of the model. A more in-depth discussion on the possible venues for raising such limitations is carried out in section \ref{sec:discussion}.

\subsection{Governing Equations}

We use our in-house code SPARK (Simulation Platform for Aerodynamics, Radiation, and Kinetics), which is an object-oriented multiphysics code written in Fortran and capable of solving a set of Ordinary Differential Equations (ODE) for 0D/1D geometries, or a set of Partial Differential Equations (PDE) for 2D axyssimetric geometries, solving the Navier--Stokes equations with macroscopic and/or state-to-state chemistry, high-temperature thermodynamic models, multicomponent transport, and plasma effects (ambipolar diffusion, etc...) \cite{SPARK}. The code has been primarily applied to the simulation of atmospheric entry flows\cite{Lopez2014,Fernandes2019}, but is also being recently tailored for the simulation of atmospheric pressure plasma jets (APPJ's)\cite{Goncalves2020}. We have used the ODE setup for SPARK in this work, and the corresponding governing equations are shortly summarized here.

Considering mass, momentum and energy conservation equations, the governing equations are written:
\begin{align}
\frac{\partial \rho c_i}{\partial t} + \nabla\cdot(\rho \boldsymbol{v}c_i) &= \dot{\omega}_i \\
\frac{\partial \rho \boldsymbol{v}}{\partial t} + \nabla\cdot\left(\rho\boldsymbol{v}\boldsymbol{v} \right) &= -\nabla p \\
\frac{\partial \rho E}{\partial t} + \nabla\cdot(\rho H \boldsymbol{v}) &= 0
\end{align}
where $c_i$  and $\dot{\omega}_i$ are the mass fraction and mass production rate respectively of species $i$, $\rho$  and $p$ the gas density and pressure, $\boldsymbol{v}$ the velocity vector of the gas, $E$ and $H$ the total energy and enthalpy of the gas. The system of equations above are the inviscid Euler equations. A temporal relaxation system of equations may be obtained by setting the spatial derivatives to zero and assuming a calorically perfect gas $E = \sum_i c_iC_{V,i}T$:
\begin{align}
\frac{\partial c_i}{\partial t} &= \frac{\dot{\omega}_i}{\rho} \\
\frac{\partial T}{\partial t} &= -\frac{\sum\limits_i \varepsilon_i\dot{\omega}_i}{\rho C_{V}^f}
\end{align}
where the gas specific heat $C_V^f = \sum_i c_i C_{V,i}$ and $\varepsilon_i$ is the internal energy of species $i$. The system has been formulated in terms of primitive variables $c_i$ and $T$ leading to the number of ODE equal to the number of species/states plus one. The space relaxation system is obtained similarly by setting the time derivative to zero:
\begin{align}
\frac{\partial c_i}{\partial x} &= \frac{\dot{\omega}_i}{\rho v} \\
\frac{\partial v}{\partial x} &= \frac{c_1b_2 - c_2b_1}{a_1b_2-a_2b_1} \\
\frac{\partial T}{\partial x} &= \frac{c_1a_2-c_2a_1}{a_1b_2-b_1a_2}
\end{align}
where the $a$, $b$ and $c$ coefficients are defined thusly:
\begin{align}
a_1 = \frac{\rho v^2}{p} -1 \text{,} &\quad a_2 = \rho v^2 \\
b_1 = \frac{v}{T} \text{,} &\quad b_2 = \rho v C_p^f \\
c_1 = -\frac{M}{\rho}\sum_i \frac{\dot{\omega}_i}{M_i} \text{,} &\quad c_2 = - \sum_i h_i\dot{\omega}_i
\end{align}
where $M_i$ is the molar mass of species $i$ and $M$ is the molar mass of the gas averaged by the molar fraction. The spatial relaxation scheme leads to a number of ODEs equal to the number of species/states plus two. In parts of this work there will be an assumption of a shock wave passing through the gas. The shock wave is treated as a discontinuity and properties across the shock are transformed through the known Rankine-Hugoniot conditions.

\subsection{The \ce{CO2} molecule}

\ce{CO2} is a linear triatomic molecule. It has three vibrational modes, symmetric stretch which corresponds to the equal stretch of the \ce{C-O} bonds in the molecule, is usually denoted as v$_1$, $s$ or $ss$. The bending mode, which is considered doubly degenerate when there is vibrational angular momentum present ($l_2 > 0$). The bending mode corresponds to the deformation of the linearity of the molecule and is usually denoted as v$_2$, $b$ or $be$. The third and final mode corresponds to the compression of a \ce{C-O} bond and the stretch of the other \ce{C-O} mode, the so-called asymmetric stretch which is usually denoted as v$_3$, $a$ or $as$. A vibrational level of \ce{CO2} might be annotated as v$_1$v$_2$v$_3$ or v$_1$v$_2^{l_2}$v$_3$ when the vibrational angular momentum number is specified. Many works also use an additional assignment number $r$ so called ranking number. It is used when authors prefer to fix v$_2 = l_2$ and it provides a convenient way to identify vibrational levels which may be grouped by Fermi resonance. In this work we will not consider the $l_2$ structure of bending levels and will assume that v$_2 = l_2$ instead of the possible range of $l_2=\text{v}_2,\text{v}_2-2,\text{v}_2-4,...,1$ or $0$. As such, all bending levels have a degeneracy of $2$ instead of $g=\text{v}_2+1$. Additionally, using $g=\text{v}_2+1$ would be assuming that the average energy for v$_2$ levels (considering the possible $l_2$ manifold) would be at v$_2 = l_2$, which is significantly higher than the actual average at high v$_2$. Some tests were also performed comparing both models for degeneracies and big differences in the macroscopic and microscopic quantities were not detected.

\subsubsection{The question of the Fermi resonance}

Fermi resonance is an effect in which vibrational levels with the same molecular symmetry and a small energy gap are observed shifted in the spectrum, with a bigger energy gap and with line intensities different than those expected. This is usually interpreted as a coupling between the symmetric and bending modes of \ce{CO2} \cite{Kozak2015}. While citing this phenomena, it is a widespread approximation to justify the use of a single temperature to characterize the symmetric and bending modes. In this work, no \textit{a priori} coupling between the symmetric and bending mode can be admitted. Since the included bending levels are considered to be v$_2 = l_2$ there are no symmetric levels with the same molecular symmetry and therefore, no mode coupling is considered in this work. Furthermore, it cannot be reasonably assumed that the the small energy gap condition for Fermi Resonance to occur is maintained higher in the vibrational ladder where anharmonicity leads to a wider gap between the "would-be" resonant states. As such, resonances between levels higher in the vibrational ladder are accidental which ties in to the concept of \textit{vibrational chaos}. These concepts and other possible approaches for the modelling of higher levels of \ce{CO2} are discussed more in depth in section \ref{sec:discussion}. It is also worth mentioning that Fermi resonances in the level energies do not always translate to equiprobable levels populations for the resonant states \cite{Rosser1972,Losev1976-1,Allen1980,Millot1998,Joly1999a}, and that the fitting of high-resolution FTIR spectra in the 4.3$\mu$m region (as discussed in section \ref{sec:stateoftheartlowT}) is very insensitive to the v$_1$ and v$_2$ level populations, which means that the error bars in the fits allow for a great latitude of interpretation of the level populations for symmetric and bending states, which may be considered with separate temperatures ($T_{v_1}\neq T_{v_2}$) \cite{Urbanietz2018} or with equivalent temperatures ($T_{v_1}=T_{v_2}$) \cite{Stewig2020} indistinguishably \cite{Urbanietz2020}.

\subsection{Level energies}
\label{sec:levels}

\begin{figure}[ht]
\centering
\includegraphics[width=0.49\linewidth]{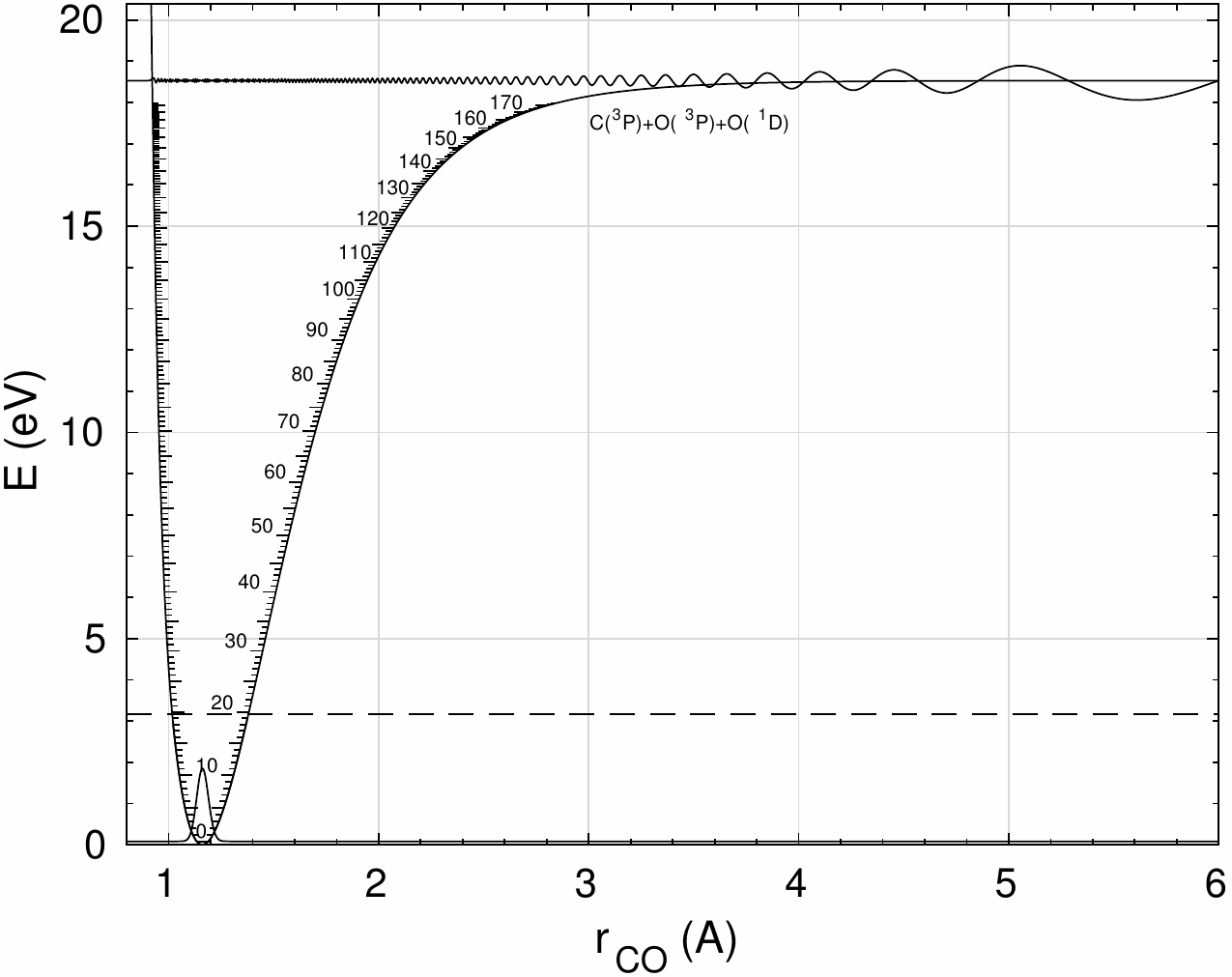}
\includegraphics[width=0.49\linewidth]{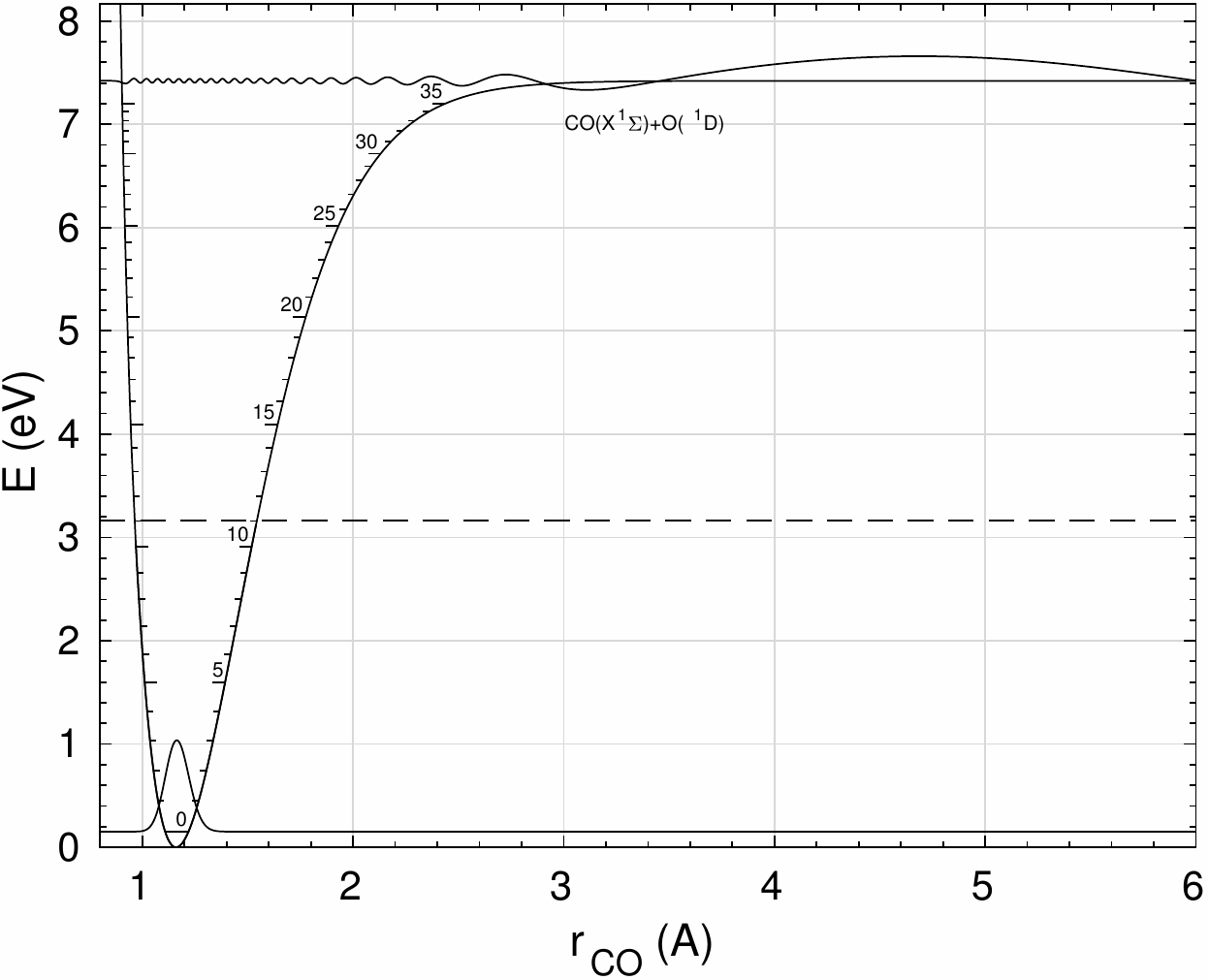}
\caption{Symmetric (left) and asymmetric (right) 1D potentials of the \ce{CO2} ground state with extrapolation to correct asymptotic limit of potential by the RKR method. The extrapolation is done above the dashed line, below which the data belongs to the NASA--Ames--2 PES.}
\label{fig:CO2Xpotential_sa}
\end{figure}

\begin{figure}[ht]
\centering
\includegraphics[width=0.49\linewidth]{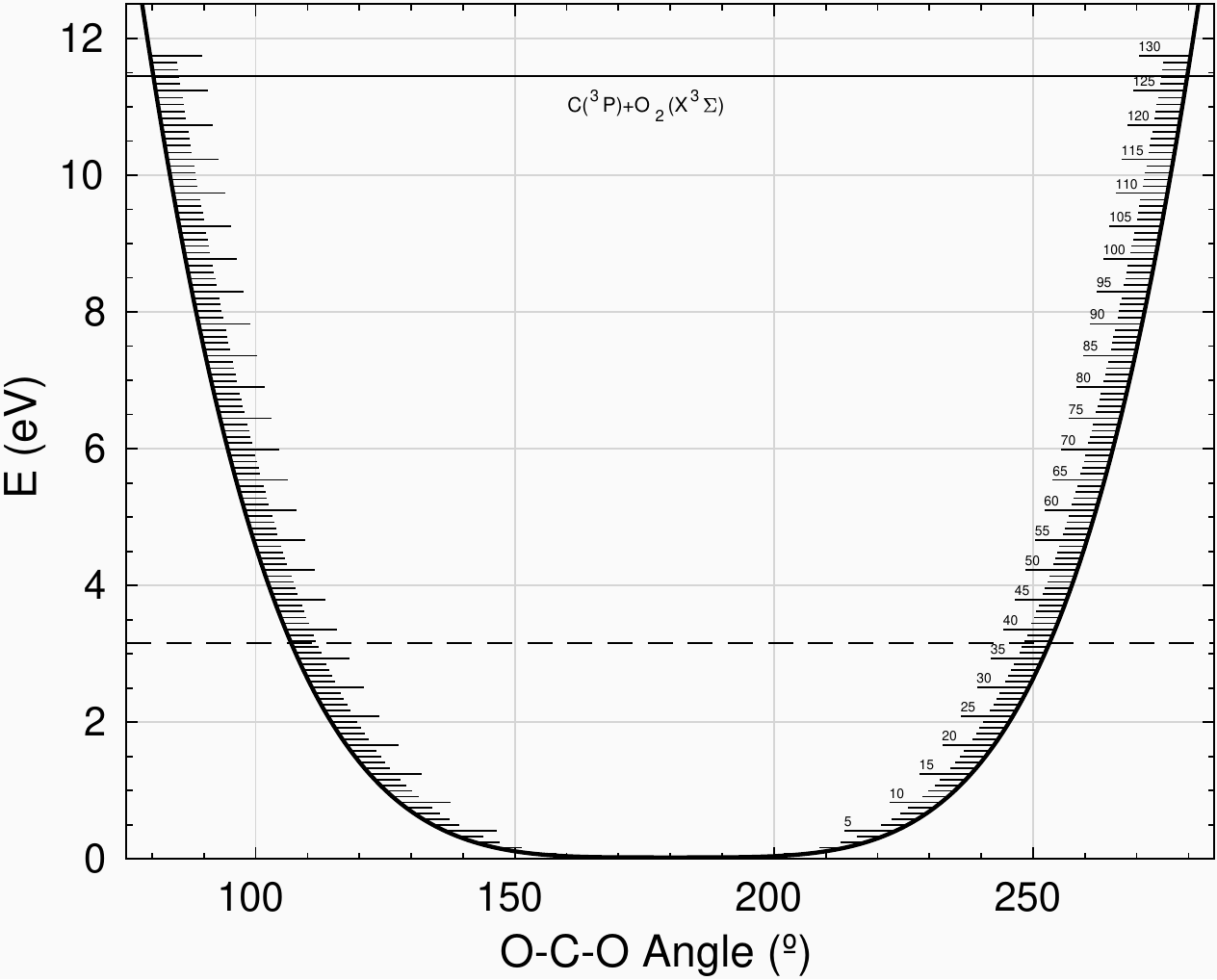}
\caption{Bending angular potential of the \ce{CO2} ground state. Above the dashed line extrapolation of the potential is performed by the expression $ax^2 + bx^4$. Below the dashed line is the data extracted from the NASA--Ames--2 PES.}
\label{fig:CO2Xpotential_b}
\end{figure}

The first step in the generation of new reaction rates is to determine a manifold of level energies. In this work, this is performed for the ground state and the electronically excited state of \ce{CO2} denoted $^3$B$_2$. We start with the ground state. Since there are no ground state Potential Energy Surfaces (PES) that are accurate up to the dissociation energy of \ce{CO2}, the asymptotic limits of dissociation must first be established. These must be different for each mode as each breaks apart in a different configuration. The dissociation energy can be computed by the balance of the enthalpy of formation for the products of dissociation. These are as follows:
\begin{itemize}
\item Symmetric stretch: \ce{CO2}(X$^1\Sigma$) + 18.53 eV $\rightarrow$ C($^3$P) + O($^3$P) + O($^1$D),
\item Bending: \ce{CO2}(X$^1\Sigma$) + 11.45 eV $\rightarrow$ C($^3$P) + O$_2$(X$^3\Sigma$),
\item Asymmetric stretch: \ce{CO2}(X$^1\Sigma$) + 7.42 eV $\rightarrow$ CO(X$^1\Sigma$) + O($^1$D).
\end{itemize}
Having the asymptotic behaviour of each mode allows the extension of a PES to the near-dissociation limit and thus well behaved a 1D potential curve for each mode is obtained. This process is the known Rydberg--Klein--Rees (RKR) method which is detailed in \cite{LinodaSilva2008}. Upon obtaining a well-behaved potential curve, the radial Schr\"{o}dinger's equation may be solved and a manifold of levels can be determined. The NASA--Ames--2 PES by \citet{Huang2012} (kindly shared by Dr. Huang) is used up to $25,500$ cm$^{-1}$ and extended to the respective dissociation limit in the long range for the symmetric and asymmetric stretch modes by Hulburth \cite{Hulburt1941} and Rydberg \cite{LinodaSilva2008} potentials respectively. In the short range, the potential is extended by a repulsive of the form ($a/x^b$). Thus, the 1D potentials of the symmetric and asymmetric stretch modes are obtained and are reported in fig. \ref{fig:CO2Xpotential_sa}. The dotted line in each figure represents the limit to which the NASA--Ames--2 PES is used, above which the potentials are extrapolated. The eigenvalues obtained from the radial Schr\"{o}dinger's equation are also plotted along the potential curve. A full line at $18.53$ and $7.42$ eV represents the asymptotic limit for the potential curve of each mode, along with the last bound solution of Schr\"{o}dinger's equation. The bending mode requires a different treatment. The symmetry of the bending mode potential excludes the possibility of using the same treatment described above as there is no expectation for the shape of the potential near the asymptotic limit. Furthermore, it is of no benefit to model such extreme states close to the dissociation limit of "pure" bending of the molecule. Nevertheless due to the symmetry of the bending mode, the potential may be fitted to a polynomial expression $ax^2 + bx^4$ as described in \cite{Quapp1993} \footnote{excluding perturbations from other states}. As for the energy levels, and once more acknowledging the symmetry of the potential, these may be extrapolated from a polynomial expression with a greater degree of confidence. In this work the Ch\'{e}din polynomial fit \cite{Chedin1979} is used for this purpose. Figure \ref{fig:CO2Xpotential_b} shows the symmetry of the bending mode along with the energy levels obtained from the Ch\'{e}din fit. In the same figure the dashed line represents the threshold above which the potential is extrapolated from the NASA--Ames--2 PES. A solid line represents the asymptotic limit which cannot be captured by the extrapolation of the employed $ax^2 + bx^4$ polynomial.

\begin{figure}[ht]
\centering
\includegraphics[width=0.6\linewidth]{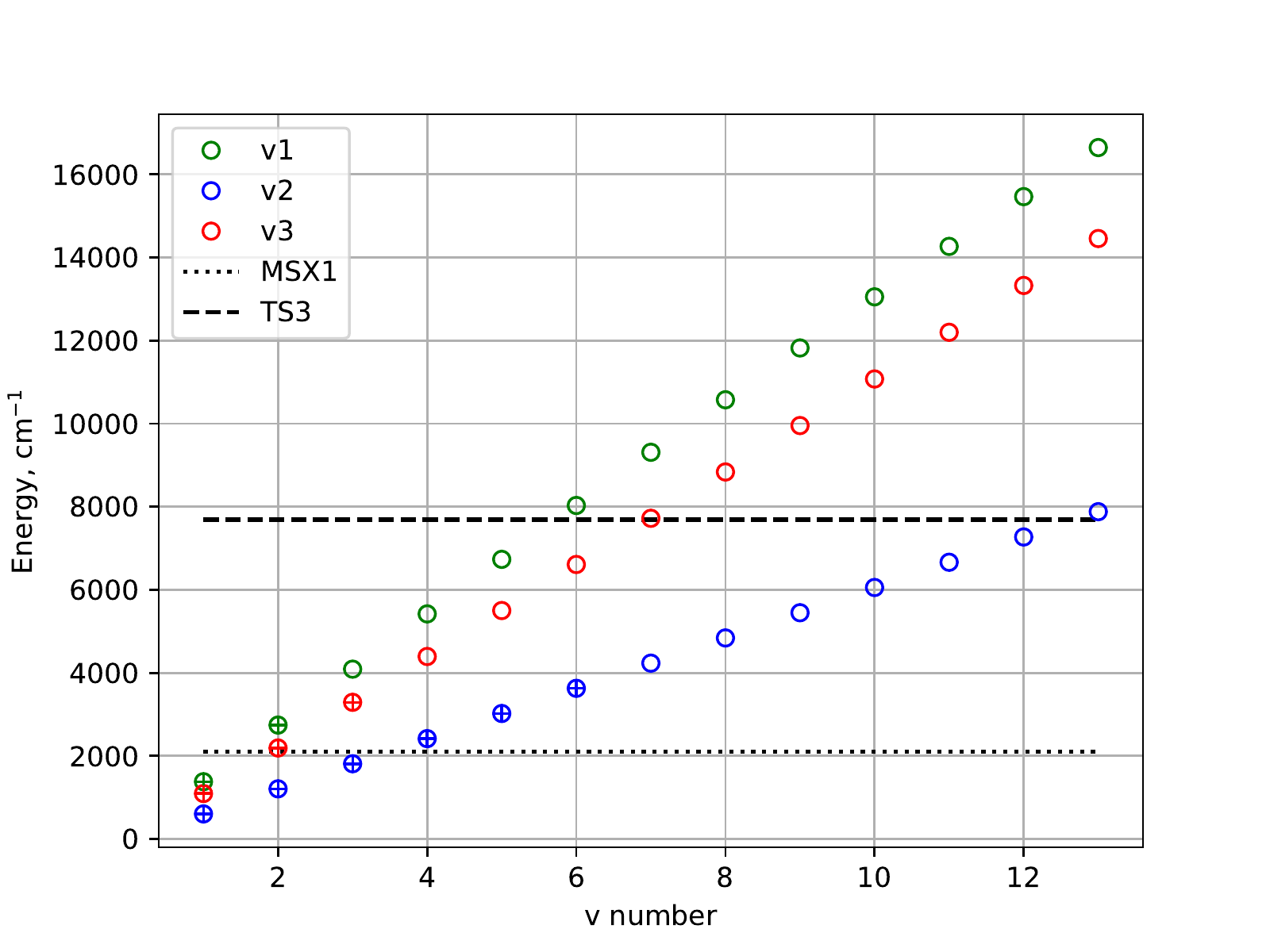}
\caption{Manifold of vibrational levels of \ce{CO2}$(^3\text{B}_2)$ used in this work. These levels were determined with expression \ref{eq:fit} and with the coefficients in table~\ref{tab:3B2levs}. MSX1 refers to the seam of crossing energy of the \ce{CO2}$(^3\text{B}_2)$ with the ground state, TS3 refers to the dissociation energy of \ce{CO2}$(^3\text{B}_2)$.}
\label{fig:3B2manifold}
\end{figure}

Electronically excited \ce{CO2} in the $^3$B$_2$ state is a bent molecule in its equilibrium configuration with the same vibrational modes as the ground state \ce{CO2}. Although there are some available PES of this excited state, not having the symmetry of a linear molecule precludes the use of the same methods as in the fundamental state. Solutions to Schr\"{o}dinger's equation could possibly be found but the assignment of each solution to a state would be made a complex endeavour which requires a work of it's own. Instead we will use the values found and assigned in the work of Grebenshchikov \cite{Grebenshchikov2017}. These may be used to fit a polynomial based on the same expression as the Ch\'{e}din fit \cite{Chedin1979}. The resulting polynomial takes the shape:
\begin{equation}
\label{eq:fit}
E(\text{v}_1,\text{v}_2,\text{v}_3) = \sum_{i=1,2,3} \omega_i \text{v}_i + \sum_{i=1,2,3} x_{ii} \text{v}_i^2 + x_{12} \text{v}_1 \text{v}_2 + x_{13} \text{v}_1 \text{v}_3 + x_{23} \text{v}_2 \text{v}_3.
\end{equation}
The coefficients found by fitting this polynomial are in table~\ref{tab:3B2levs}. Figure~\ref{fig:3B2manifold} presents the levels found by fitting the values in \cite{Grebenshchikov2017} to equation~\ref{eq:fit}. The circles with a cross inscribed are the values found in \cite{Grebenshchikov2017} and the other are energy levels found from extrapolating the fit. The fine dotted line, labeled MSX1 is the seam of crossing between the ground state of \ce{CO2} and the $^3$B$_2$ state. The line labeled as TS3 is the dissociation energy of the $^3$B$_2$ state. Usually, the spacing of levels will be lowered as the dissociation limit is approached. This is not the case in the asymmetric stretch mode of the $^3$B$_2$ state as the extrapolation of a polynomial expression does not allow replicating this behaviour. However as a first approximation it is a reasonable enough estimation. The line labeled as TS3 is the dissociation limit of the $^3$B$_2$ state. We assume that as in the case of the ground state the asymptotic limit for each mode will be different and as such the symmetric and bending modes are not asymptotically limited by TS3.

This concludes the determination of the level manifold used in this work. The number of levels used in this work is summarized in tab.~\ref{tab:manifold}. The number of levels in the v$_3$ mode for both electronic states is fixed since dissociation occurs through this mode. A v$_3$ level above the dissociation limit is considered to be quasi-bound (q.b.) and dissociates with probability 1 \cite{LinoDaSilva2007,Adamovich1995a}. The number of levels in the other modes may differ. We have elected to use $59$ and $100$ levels since these correspond to the same energy chosen arbitrarily, roughly $\approx9$ eV. No study was conducted to verify the sensitivity of the number of levels in the other modes. However, it was verified that using a smaller amount of levels in the ground state would lead to greater numerical instability in the code. With the levels included in tab.~\ref{tab:manifold} and the ground state of each electronic level there is a total of 245 vibrational levels in the model.

\begin{table}[ht]
\centering
\caption{Coefficients obtained from fitting the expression in equation~\ref{eq:fit} to the energy levels found in \cite{Grebenshchikov2017} in units of cm$^{-1}$.}
\label{tab:3B2levs}
\footnotesize
\bgroup
\def\arraystretch{1.5}
\begin{tabular}{ccccc}
\hline
$\omega_1$ & $\omega_2$ & $\omega_3$ & $x_{11}$ & $x_{22}$ \\ 
\hline
1.387e+3 & 6.031e+2 & 1.092e+3 & -8.254e+0 & 2.311e-1 \\
\hline
$x_{33}$ & $x_{12}$ & $x_{13}$ & $x_{23}$  & \\
\hline
1.501e+0 & -1.358e+1 & -6.422e+1 & -1.808e+1 & \\
\hline
\end{tabular}
\egroup
\end{table}

\begin{table}[ht]
\centering
\caption{Number of levels in the manifold used in this work for each mode of vibration and each electronic mode of \ce{CO2}.}
\label{tab:manifold}
\footnotesize
\bgroup
\def\arraystretch{1.5}
\begin{tabular}{cccc}
\hline
\ce{CO2} & v$_1$ & v$_2$ & v$_3$ \\ \hline
X$^1\Sigma$ & 59 & 100 & 41 \\
$^3$B$_2$ & 12 & 25 & 6 \\ \hline
\end{tabular}
\egroup
\end{table}

\subsection{PES Crossings}
\label{sec:PEScrossings}

\begin{figure}[ht]
\centering
\includegraphics[width=0.70\linewidth]{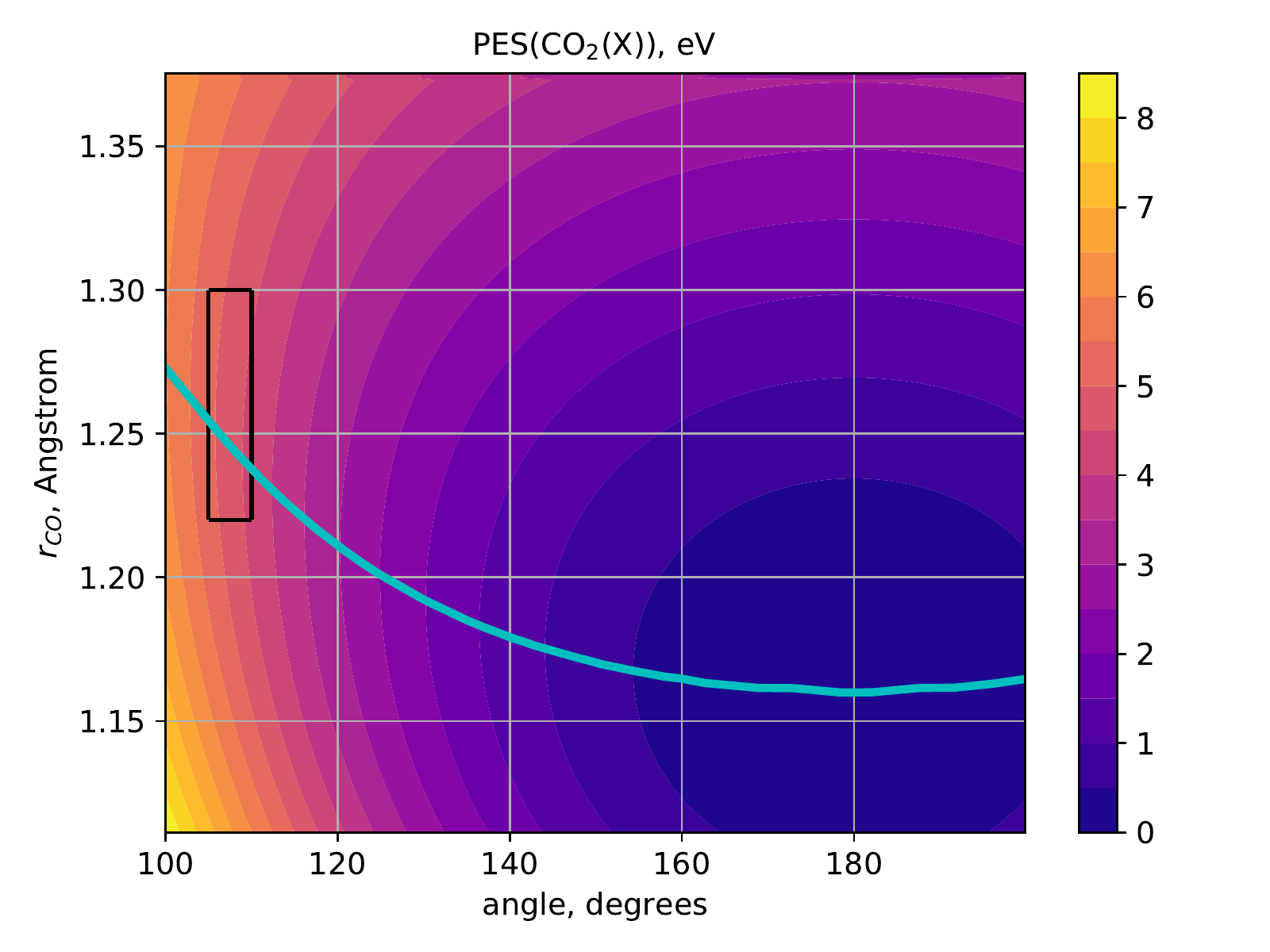}
\caption{Color map of the PES of the ground state of \ce{CO2} with $r_{\text{CO1}}=r_{\text{CO2}}$. The cyan line which crosses the color map is the equilibrium $r_{\text{CO}}$ distance given a specific angle. The black box on the left side of the figure is the probable configuration space for the crossing between the \ce{CO2} ground and triplet state to occur.}
\label{fig:firstX}
\end{figure}

In this work we aim to account for the different pathways to dissociation of \ce{CO2}. For this, the configuration of the interactions between the ground state and electronically excited states of \ce{CO2} needs to be defined. The work of Hwang \& Mebel \cite{Hwang2000} provides a basis for the configuration of these interactions. There are two seams of crossings between the ground and $^3$B$_2$ states of \ce{CO2}. The first takes place at approximately $4.99$ eV when the \ce{CO2} molecule is bent close to the equilibrium configuration of the $^3$B$_2$ state. The exact configuration of the crossing will depend on the calculation method of calculation used. Looking at the range of values proposed in \cite{Hwang2000} we define an approximate region of the exact configuration, between 1.22--1.30 \r{A} and 105.0--110.4º, assuming the \ce{C}--\ce{O} bonds to be the same length. This region is plotted in  fig.~\ref{fig:firstX} as a black box. Fig.~\ref{fig:firstX} also presents a 3D color map of the PES of the ground state of \ce{CO2} considering both \ce{C}--\ce{O} bonds at the same length and varying the angle between them. The cyan line in the aforementioned figure is the position of minimum potential at each angle. The cyan line intersects the region where the exact configuration of the crossing is most likely to be. This indicates that the crossing may occur purely through the ground state with v$_2$ excitation, as this mode will play the main role interacting with the $^3$B$_2$ excited state. Only a few vibrational levels close to the crossing will interact with the bending levels of the ground state. The interacting $^3$B$_2$ levels have been defined as the ones within 2000 cm$^{-1}$ of the crossing (MSX1). These coincide with the levels listed in the work of Grebenshchikov \cite{Grebenshchikov2017}.

The second crossing takes place at $5.85$ eV in a linear configuration of ground state \ce{CO2} but with the bonds at very different lengths. In this crossing, the determined configuration in \cite{Hwang2000} has one bond length at approximately $1.9$ \r{A} and the other bond at $1.3$ \r{A}. This suggests that the mode which interacts the most in this crossing is the asymmetric stretch, v$_3$. The excited state in this crossing is in a repulsive configuration and being excited to this state will lead to immediate dissociation. As we don't have the functional form for the triplet state there is no possibility to compute the energy levels of the quasi-bound states. As such, we assume there is a level at the energy of the crossing of $5.85$ eV to which the v$_3$ levels of the ground state of \ce{CO2} will dissociate to. 

We have now defined which levels are interacting in which crossings. The theory which is usually applied in vibrational state-to-state collisions is the Landau-Zener theory. Since these crossings are singlet-triplet interactions, this theory cannot be straightforwardly applied in this case. The next subsection discusses some of these details. There are other singlet and triplet electronically excited states of \ce{CO2} in a configuration alike to the $^3$B$_2$ level which was discussed in this section. These other states are not considered in this work due to the scarcity of published data on crossings between the ground state and other electronically excited states of \ce{CO2}.

\subsection{Vibration-Translation Processes}
Here we summarize the key equations for this theory and its extension to triatomic molecules such as \ce{CO2}. The inner details of the FHO theory and its different generalizations are not described here and may be consulted in \cite{STELLAR,LinoDaSilva2007}.\\

The FHO model computes Vibrational-Translational (VT) energy exchanges:
\begin{equation}
\label{eq:VT}
\ce{CO2}(i) + \ce{M <-> CO2}(f) + \ce{M}
\end{equation}
where $i$ and $f$ are the vibrational levels for the same mode and \ce{M} is a generic collision partner. In these reactions the electronic state can be the ground or $^3$B$_2$. The corresponding FHO transition probability is \cite{Kerner1958,Treanor1965}:
\begin{align}
P(i\rightarrow f,\varepsilon)=i!f!\varepsilon^{i+f}\exp\left(-\varepsilon\right)%
\left|\sum_{r=0}^{n}\frac{(-1)^{r}}{r!(i-r)!(f-r)!\varepsilon^{r}}\right|^{2}% 
\label{eq:PVT}%
\end{align}
with $n=min(i,f)$.

For transitions involving larger vibrational number changes and at higher-vibrational numbers, it is no longer possible to accurately compute transition probabilities using exact FHO factorial expressions. When such a computation is not possible it is instead replaced by the approximations suggested by Nikitin and Osipov \cite{Nikitin1977} that make use of Bessel functions. Details of these approximations are found in \cite{STELLAR,LinoDaSilva2007}. Dissociation (Vibration--Dissociation reactions, VD) may occur through the asymmetric stretch mode when the level in the products of the reaction is a quasi-bound (q.b.) level:
\begin{equation}
\ce{CO2}\text{(v}_3) + \ce{M <-> CO2}\text{(q.b.)} + \ce{M <->[{P = 1}] CO + O + M}
\end{equation}
where the dissociation products are \ce{CO}(X$^1\Sigma$) and \ce{O}($^1$D) in the case the reactant is the ground state, and \ce{CO}(X$^1\Sigma$) and \ce{O}($^3$P) in the case the reactant is the $^3$B$_2$ state. Dissociation is considered to be the only possible outcome in a quasi-bound state, and the recombination reaction is computed through detailed balance. A VT reaction may also occur when the initial and final level are not in the same mode (Intermode Vibration--Translation, IVT). In that case the VT reaction is considered to be two VT reactions and the total probability is considered to be the product of these two collisions:
\begin{equation}
\label{eq:IVT}
\ce{CO2}(i) + \ce{M <-> CO2}(0) + \ce{M <-> CO2}(f) + \ce{M}
\end{equation}
where \ce{CO2}(0) is the fundamental vibrational state of either the ground electronic state or the $^3$B$_2$ state. We note that this classical approximation will lead to a vanishingly small transition probabilities at higher v's\footnote{unless the translational temperature is very high}. This might not be necessarily the case since many \textit{accidental resonances} may occur for these higher levels, which means our approach will be underestimating IVT processes for these higher energy levels. Approaches to address this shortcoming will be discussed more ahead in section \ref{sec:discussion}. Finally, a Vibrational-Vibrational-Translational (VVT) reaction may also be modeled. In these reactions, two molecules with the same vibrational level can collide in an almost resonant fashion such that at the end of the process, one has gained a quanta and the other lost one quanta:
\begin{equation}
\label{eq:VVT}
\ce{CO2}(\text{v}) + \ce{CO2}(\text{v}) \ce{<-> CO2}(\text{v}+1) + \ce{CO2}(\text{v}-1).
\end{equation}
The corresponding FHO transition probability is  \cite{Zelechow1968}:
\begin{align}
&P(i_{1},i_{2}\rightarrow f_{1},f_{2},\varepsilon,\rho)=\left|\sum_{g=1}^{n}(-1)^{(i_{12}-g+1)}\right.\notag\\
&\times C_{g,i_{2}+1}^{i_{12}}C_{g,f_{2}+1}^{f_{12}}\varepsilon^{\frac{1}{2}(i_{12}+f_{12}-2g+2)}\exp\left(-\varepsilon/2\right)\notag\\
&\times\sqrt{(i_{12}-g+1)!(f_{12}-g+1)!}\exp\left[-i(f_{12}-g+1)\rho\right]\notag\\
&\left.\times\sum_{l=0}^{n-g}\frac{(-1)^{l}}{(i_{12}-g+1-l)!(f_{12}-g+1-l)!l!\varepsilon^{l}}\right|^{2}
\label{eq:PVVT}%
\end{align}
with $i_{12}=i_{1}+i_{2},\ f_{12}=f_{1}+f_{2}$ and $n=min(i_{1}+i_{2}+1,f_{1}+f_{2}+1)$. In these reactions, a residual amount of energy is transferred to the translational movement of the molecules due to the anharmonicity of the vibrational ladder. 

\begin{figure}[!htbp]
\centering
\includegraphics[width=.6\textwidth]{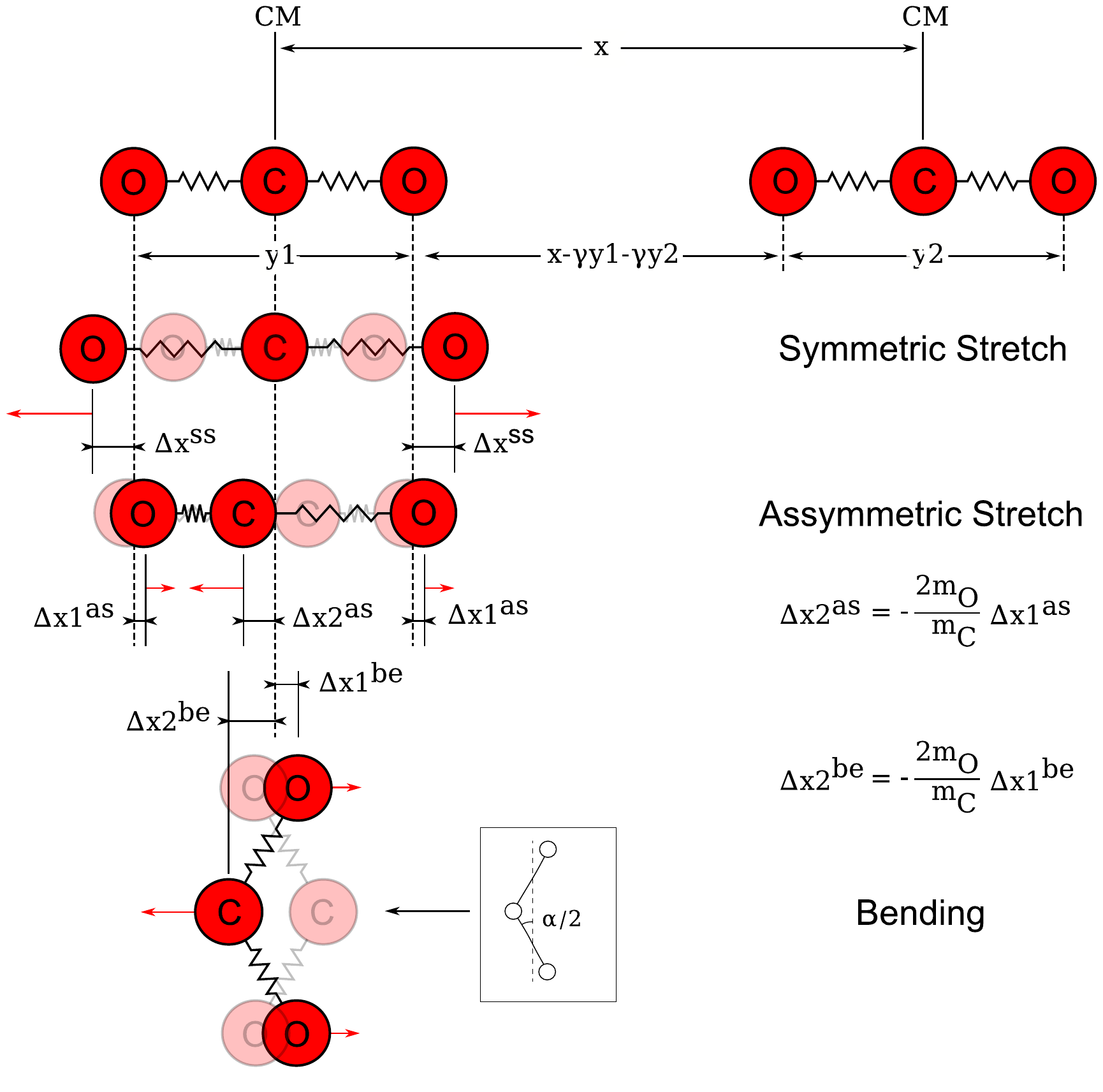}
\caption{CO$_2$ mass coordinates for the symmetric stretch, asymmetric stretch and bending modes}%
\label{fig:CO2-modes}
\end{figure}

For eqs. \ref{eq:PVT} and \ref{eq:PVVT}, $\varepsilon$ and $\rho$ are related to the two-state First-Order Perturbation Theory (FOPT) transition probabilities, with  $\varepsilon=P_{\textrm{FOPT}}(1\rightarrow0)$, $\rho=\left[4\cdot P_{\textrm{FOPT}}(1, 0\rightarrow0, 1)\right]^{1/2}$, and $C_{ij}^{k}$ is a transformation matrix \cite{LinoDaSilva2007}. Certain parameters in the model must be adjusted such that experimental measurements may be effectively reproduced by the calculations. As such, a so called semi-empirical adjustment of collision parameters takes place. Most notably, the inter-molecular potential, which is taken as a Morse potential with shape $V=E\left\{1-\exp{\left[-\alpha(r-r_{eq})\right]}\right\}^2$, and the steric factors, which are corrective factors to account for the isotropy of collisions when these are computed in a 1D framework, will be adjusted such that calculations will match experimental rate coefficient measurements in the best possible way. The expressions for $\varepsilon$ and $\rho$ are given by Cottrell \cite{Cottrell1955}, and Zelechow \cite{Zelechow1968}:
\begin{align}
\varepsilon&=\frac{8\pi^{3}\omega\left(\tilde{m}^{2}/\mu\right)\gamma^{2}}{\alpha^{2}h}
\frac{\cosh^{2}\left[\frac{\left(1+\phi\right)\pi\omega}{\alpha\bar{v}}\right]}{\sinh^{2}\left(\frac{2\pi\omega}{\alpha\bar{v}}\right)}\\
\phi&=(2/\pi)\tan^{-1}\sqrt{\left(2E_{m}/\tilde{m}\bar{v}^{2}\right)}\notag\\
\rho&=2\left(\tilde{m}^{2}/\mu\right)\gamma^{2}\alpha\bar{v}/\omega.
\end{align}
The corresponding mass parameters have been obtained for the three motions of \ce{CO2} (symmetric stretch, assymetric stretch and bending, see fig. \ref{fig:CO2-modes}). These are reported in Table 
\ref{tab:CO2-phi}. The $^3$B$_2$ state is bent in its equilibrium configuration but the same parameters in table \ref{tab:CO2-phi} are considered valid for the time being.

\begin{table}[!htbp]
\caption{Mass parameters for the three vibrational modes of CO$_2$}%
\label{tab:CO2-phi}%
\centering%
\begin{tabular}{l c c}
\hline
osc. mode & reduced mass $\mu$ & mass param. $\gamma$\\
\hline
sym. stretch & $m_{\textrm{O}}$ & 1/2 \smallskip\\
asym. stretch & $\dfrac{m_{\textrm{C}}m_{\textrm{O}}}{m_{\textrm{C}}+2m_{\textrm{O}}}$ & 1/2 \smallskip\\
bending & $\dfrac{m_{\textrm{C}}m_{\textrm{O}}}{2(2m_{\textrm{O}}+m_{\textrm{C}})}$ & 1/2 \smallskip\\
\hline
\end{tabular}
\end{table}

\begin{figure}[ht]
\centering
\includegraphics[width=0.49\linewidth]{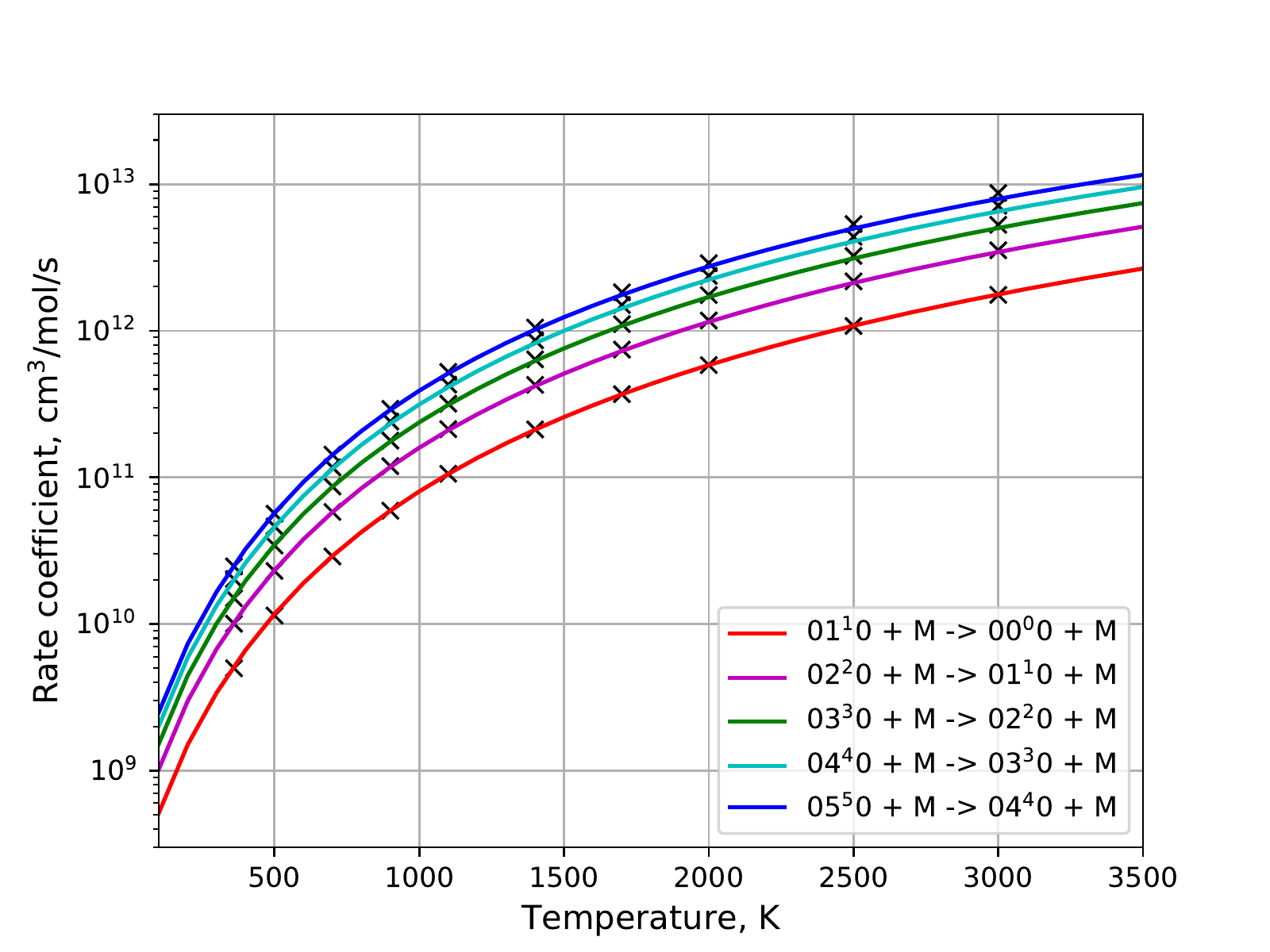}
\includegraphics[width=0.49\linewidth]{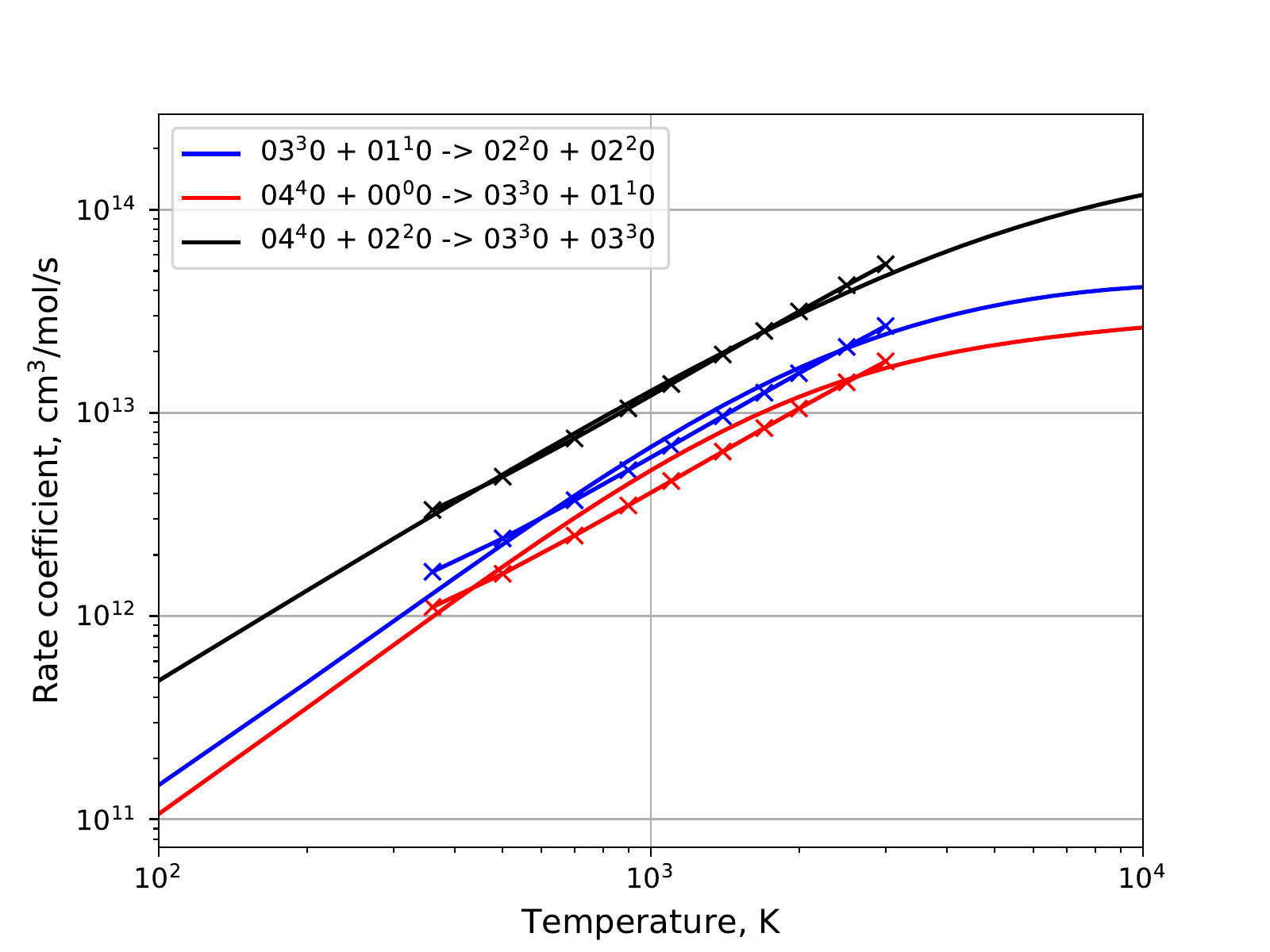}
\caption{Examples of adjustment of FHO computed rates to SSH published VT rates (left) and experimentally determined VVT rates (right). The used collision parameters may be found in tab.~\ref{tab:semiempirical}. More examples of these adjustments are found in \cite{STELLAR}.}
\label{fig:fitexample}
\end{figure}

Tab.~\ref{tab:semiempirical} presents the semi-empirical coefficients obtained in the adjustment. The first two columns are for pure VT and VVT reactions as described in equations~\ref{eq:VT} and~\ref{eq:VVT}. The last three columns are specific for IVT reactions such as the ones in equation~\ref{eq:IVT}. There are no experimental measurements specific for IVT v$_2$v$_3$ reactions and as such the coefficients are considered to be the same as for VT reactions. All experimental measurements are made for the ground state of \ce{CO2}, and we assume they are the same for \ce{CO2}($^3$B$_2$) since in this case there is no experimental data available for calibrating the FHO model. Some of the adjustment results are presented in fig.~\ref{fig:fitexample} where examples of the VT and VVT reaction rates found in the literature my be recovered after adjusting the semi-empirical coefficients accordingly. More examples can be found in \cite{STELLAR}.

\begin{table}[ht]
\centering
\caption{Collision parameters used in the FHO rate calculation processes in this work.}
\label{tab:semiempirical}
\footnotesize
\bgroup
\def\arraystretch{1.5}
\begin{tabular}{cccccc}
\hline
 & VT$_{\text{s,b,a}}$ & VVT$_{\text{s,b,a}}$ & IVT$_{\text{s,b}}$ & IVT$_{\text{s,a}}$ & IVT$_{\text{b,a}}$ \\ \hline
S$_{VT}$ (10$^{-4}$) & 6 & - & 2 & 1 & 6 \\
S$_{VVT}$ (10$^{-3}$) & - & 6.5 & - & - & - \\
$\alpha$ (\r{A}$^{-1}$) & 4.3 & 4.3 & 3.0 & 4.3 & 4.3 \\
E (K) & 650 & 650 & 300 & 650 & 650 \\ \hline
\end{tabular}
\egroup
\end{table}

\subsection{Vibrational-Electronic Processes}

In earlier works \cite{Jaffe2011,Xu2017} studying the same crossings mentioned in section \ref{sec:PEScrossings} it has been proposed for studying the aforementioned spin-forbidden interactions with the Landau-Zener theory. In these works, the authors proposed to use the off-diagonal terms of the Hamiltonian of the system, which is non-zero for the singlet-triplet interaction, to obtain the probability of the crossing. The difficulty would then lie in determining the off diagonal term which coincides with the spin-orbit coupling term. Though this is mathematically sound, the Landau-Zener theory was developed assuming a constant off-diagonal term \cite{Nikitin1999}. The Rosen-Zener theory is more appropriate in the case of spin-forbidden interactions, although we haven't verified the region of applicability of this theory in this case. The preference of this theory over the Landau-Zener theory can be justified thus: "\textit{The Rosen-Zener model, [...] can be associated with the one-dimensional motion of a system featuring the exponential off-diagonal coupling between the zero-order states of constant spacing [...].}" as can be read in the review of Nikitin \cite{Nikitin1999}. It continues: "\textit{In a way, this model [Rosen-Zener] is the opposite to the avoided crossing Landau-Zener model, for which the spacing between the zero order states features the crossing while the off-diagonal term is constant}". In other words, the crossing of singlet-triplet interactions is described by the off-diagonal terms in the Hamiltonian of the system. The Landau-Zener theory assumes a description of the crossing contained in the diagonal terms, while the Rosen-Zener theory assumes this description to be featured in the off-diagonal term. This is the case in the crossings that are dealt with in this work and thus, the use of the Rosen-Zener theory is justified.

\begin{figure}[ht]
\centering
\includegraphics[width=\linewidth]{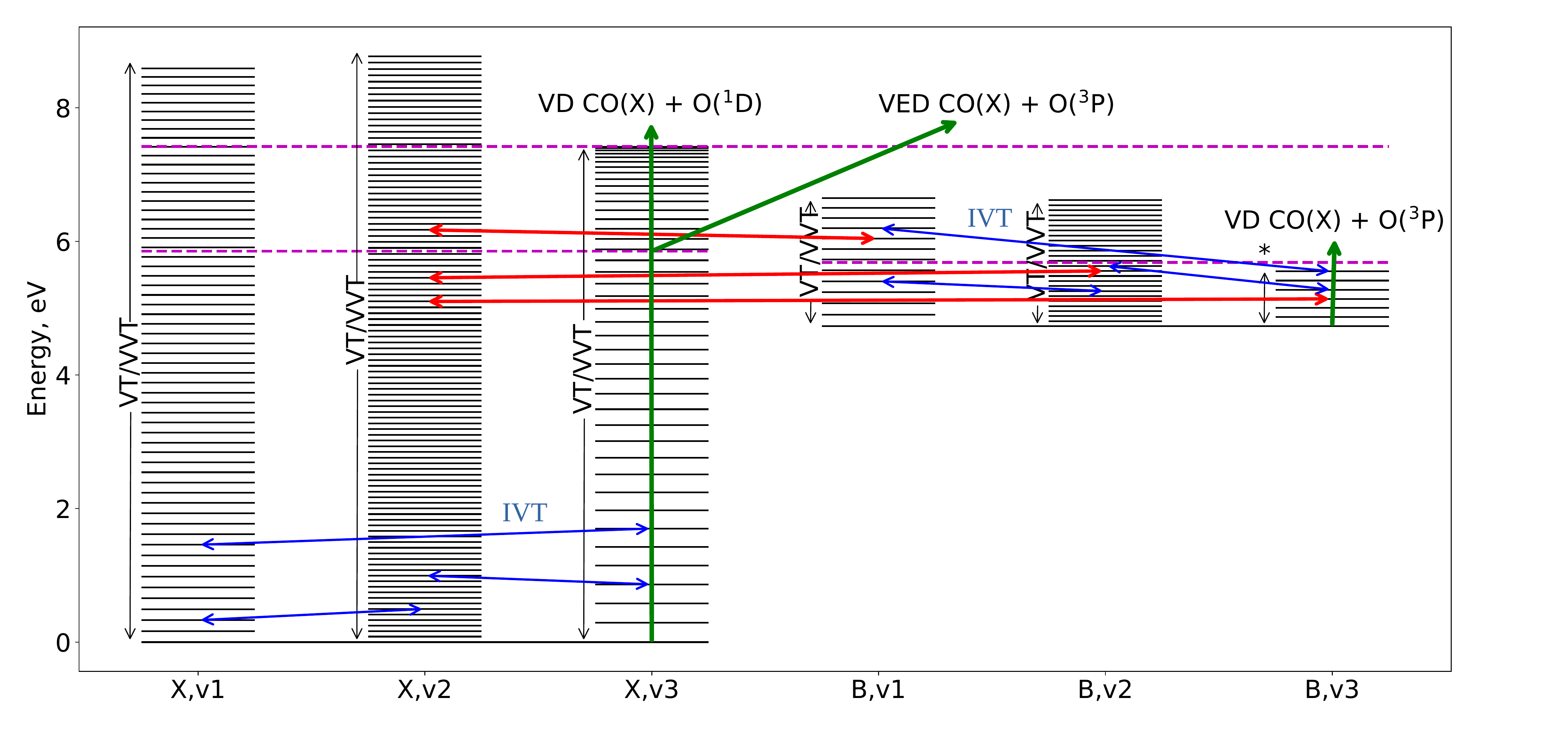}
\caption{Schematic of kinetic model with included levels and types of interactions between levels. Dissociation reactions indicate the products of dissociation and the dashed lines represent crossings or dissociation limits for certain configurations of \ce{CO2}. The exact spacings between adjacent vibrational levels are reported in the figure.}
\label{fig:model}
\end{figure}

The Rosen-Zener probability may be written as a function of velocity $v$,
\begin{equation}
\label{eq:RZformula}
P(v) = \left[1+\exp\left(\frac{\pi\Delta E}{\hbar\alpha v}\right)\right]^{-1}.
\end{equation}
The probability is dependent on the difference in energy of the interacting levels $\Delta E$, and the repulsive term of the interaction potential $\alpha$. With the current expression and with no data to calibrate the rate coefficients for these kind of reactions, we are forced to use the values as they are calculated. The expression above tends to $1/2$ in the high velocity limit. In the high-temperature regimes, this might lead to a somewhat higher than expected rate coefficient, comparable (but still lower) to the collisional rate coefficient. Thus the interaction between the singlet ground state of \ce{CO2} and the triplet $^3$B$_2$ state of \ce{CO2} is taken care of. A schematic of the overall model discussed in this work is presented in figure~\ref{fig:model}. The ground state is labeled as X, the excited triplet state as B, and their respective vibrational modes are presented. Arrows point and label to show what kind of interactions are present in this model. 

\subsection{Macroscopic Chemistry}
\label{sec:macroreac}

\begin{figure}[ht]
\centering
\includegraphics[width=0.7\linewidth]{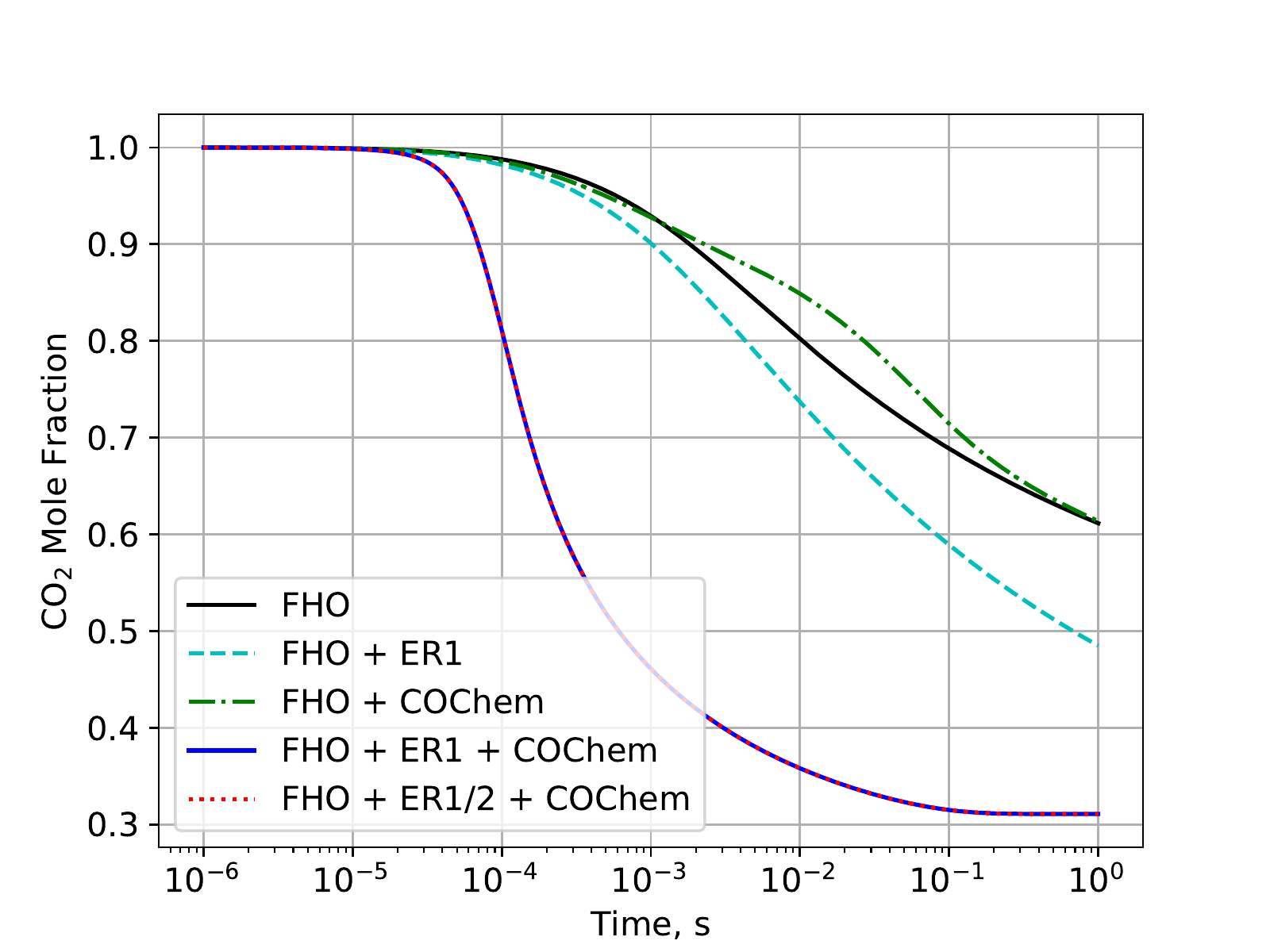}
\caption{\ce{CO2} simulated mole fraction in a $3690$ m/s shock at $133$ Pa in pure \ce{CO2}. The different lines correspond to the inclusion or exclusion of certain reactions. FHO corresponds to the model schematized in fig.~\ref{fig:model}. The inclusion of the exchange reaction \ce{CO2 + O <-> CO + O2} (Sharipov), the \ce{CO2 + C <-> CO + CO} and the \ce{CO} chemistry to the aforementioned FHO model are labeled ER1, ER2 and COChem respectively. The reaction \ce{CO2 + C <-> CO + CO} is seen to considerably enhance dissociation of \ce{CO2}, reducing the characteristic decomposition time to 10$^{-4}$--10$^{-5}$s}
\label{fig:dissociation}
\end{figure}

At this point, some simulations were conducted to test if the model with its current dissociation pathways is consistent with typical dissociation times and degrees for \ce{CO2}. A 1D shock wave at $3690$ m/s was simulated passing a \ce{CO2} gas at 1 Torr and 300K. In these conditions dissociation was expected to be noticeable on the $10^{-4}$ seconds scale and the gas would be nearly in equilibrium under a second. These are general expectations based on simulations using the macroscopic models of \citet{Park1994} and \citet{Cruden2018}. We are not expecting to reproduce the aforementioned models but rather to obtain similar results within limits of reasonability provided by the macroscopic models. Figure~\ref{fig:dissociation} presents in blue the \ce{CO2} mole fraction of this first simulation with the model described so far. This result fails to meet the expected macroscopic physical behaviour for these conditions as the flow has not arrived to equilibrium after 1 second. As such, other mechanisms important for \ce{CO2} decomposition should be added to further complement the model. Usually these chemical processes are not available in a state-to-state form. A redistribution of a reaction rate may be carried out to transform a macroscopic reaction:
\begin{align}
\ce{CO2 + X <-> AB + CD}
\end{align}
into a set of vibrational state-to-state rates of the same reaction:
\begin{align}
\ce{CO2}(\text{X}^1\Sigma\text{,v})\ce{ + X <-> AB + CD}.
\end{align}
The redistribution employed in this work is identical to the one found in \cite{AnnaloroPhDThesis}. A brief description is warranted here. Redistributing a macroscopic reaction rate to a state to state reaction rates relies on the balance of energy of a reaction. If the reagents have energy above the activation energy $E_a$, then the reaction is exothermic, if not the reaction is endothermic. $E_a$ is taken as the balance of the enthalpies of formation between the products and reagents of the reaction or alternatively with the third Arrhenius coefficient $\theta k_B$ converted to a suitable unit. We assume the shape of the exothermic state to state reactions to be:
\begin{equation}
K_v(T) = A_v T^n  \exp\left(-\frac{\theta}{T}\right)
\end{equation}
and the endothermic state to state reactions to be:
\begin{equation}
K_v(T) = B_v T^n.
\end{equation}
where $A_v$ and $B_v$ are vibrational-state specific coefficients. We further assume that $A_v$ has the shape,
\begin{equation}
A_v = b\frac{E_a-E_v}{E_a-E_{-}}
\end{equation}
and $B_v$ has the shape,
\begin{equation}
B_v = b\frac{E_{+}-E_a}{E_v-E_a},
\end{equation}
where $b$ is some coefficient dependent on temperature, $E_v$ is the energy of level $v$ and $E_{-}$ is the last level where the reaction is endothermic and $E_{+}$ the first level in which the reaction is exothermic. We then take a macroscopic reaction rate $K_{\text{macro}}$, and force this equality be true:
\begin{equation}
K_{\text{macro}}(T) = \sum_v K_v(T) \frac{g_v\exp{\left[-E_v/(k_BT)\right]}}{Q_v}
\end{equation}
where $g_v$ is the degeneracy of level $v$, $E_v$ the energy of level $v$ and $Q_v$ the vibrational partition function of the considered levels for redistribution. Following the above equation, we substitute the assumed shapes for the state to state reaction rates and the functions of $A_v$ and $B_v$ to solve for $b$,
\begin{equation}
b = \frac{K_{\text{macro}}Q_v}{\sum\limits^{E_{-}}_{E=0}\frac{E_a-E_v}{E_a-E_{-}}g_v\exp{\left(-T_a/T\right)} + \sum\limits^{E_{\text{max}}}_{E_{+}}\frac{E_{+}-E_a}{E_v-E_a}\exp{\left(-E_v/(k_B T)\right)}}.
\end{equation}
This yields a state to state reaction rate set that is self-consistent with the initial macroscopic reaction rate. In this work we only consider the vibrational manifold of the ground state of \ce{CO2} for redistribution, assuming that the triplet \ce{CO2} effects are negligible.

\subsubsection{\ce{CO2 + O <-> CO + O2} exchange processes}

\begin{figure}[ht]
\centering
\includegraphics[width=0.8\linewidth]{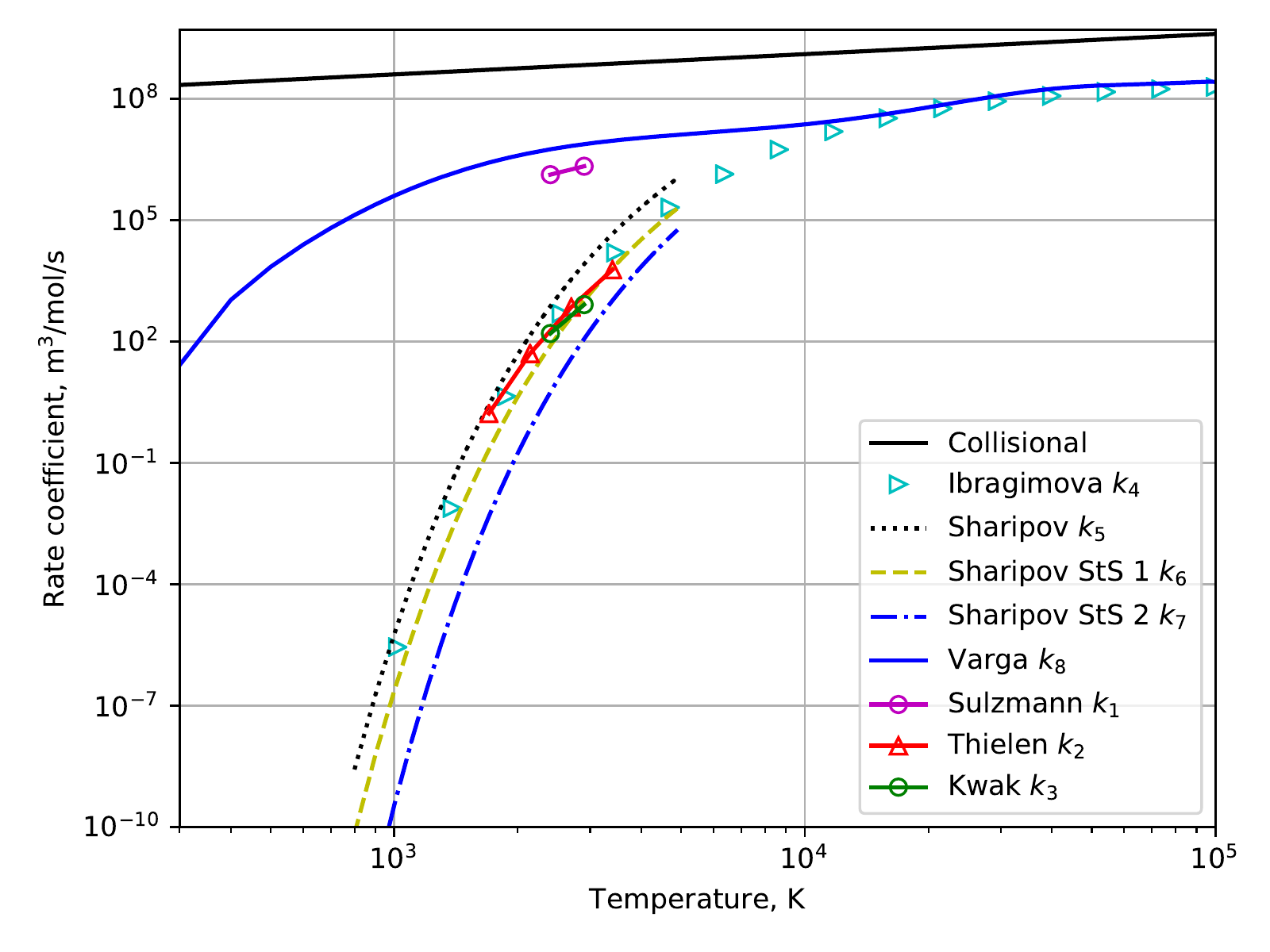}
\caption{Exchange reactions \ce{CO2 + O <-> CO + O2} found in some works in literature plotted against the collisional \ce{CO2 + O} rate coefficient.}
\label{fig:exreaction}
\end{figure}

It has been experimentally observed that the addition of \ce{O} atoms to the gas composition increases the dissociation rate of \ce{CO2} \cite{Clark1970a}. This is usually attributed to the exchange reaction \ce{CO2 + O <-> CO + O2}. Tab.~\ref{tab:excratereview} presents some of the different exchange reaction rates found in literature and considered for this work. Reactions $1$ and $2$ are taken from experimental studies \cite{Sulzmann1965, Thielen1983} of the inverse reaction \ce{CO + O2 <-> O2 + O}. The first reported source is the original experiment and the second is where the rate coefficient for the \ce{CO2 + O <-> CO + O2} reaction was reported. Reaction rate $3$ from \cite{Kwak2015} is mentioned as an adaptation from \cite{Schofield1967} but it is not made clear how this adaptation was performed. Rate $4$ was originally reported in \cite{Ibragimova1991}. The original document could not be found and as such it is not known whether this rate is experimental or calculated through other means. Rates $5$, $6$, and $7$ are presented as Arrhenius fit coefficients valid in the $800-5,000$ K region from QCT calculations carried out by Sharipov and Starik \cite{Sharipov2011} on the \ce{CO + O2} collision. Rate $5$ assumes a Boltzmann distribution of the internal states of all intervening chemical species. Rates $6$ and $7$ are state to state rates corresponding to different seams of crossing between the \ce{CO + O2} ground states intermolecular potential and \ce{CO2 + O} intermolecular potential. The first crossing allows only the production of ground state oxygen, \ce{O($^3$P)} while the second crossing also produces \ce{O($^1$D)}. The authors of \cite{Sharipov2011} decided not to branch the production of different \ce{O} levels as this reaction is negligible compared to others presented in the same work. An additional reaction is presented in \cite{Sharipov2011} where \ce{CO + O2(a)} produces \ce{CO2 + O} but the contribution of \ce{O2(a)} will be neglected in this work and considered only for future developments of this model. Finally, rate $8$ coefficients were obtained from sensitivity analysis of combustion experiments in Varga PhD Thesis \cite{VargaThesis2017} and reported to have low uncertainty. An inversion of rates $5$ to $8$ were performed using the SPARK code by computing the partition functions of intervening molecules and the equilibrium constant of the \ce{CO + O2 <-> CO2 + O} reaction. As such the presented $5$ to $8$ rates in table \ref{tab:excratereview} correspond to the \ce{CO + O2 -> CO2 + O} inverse reaction. The forward reaction rates were fitted through a 9th order polynomial as:
\begin{equation}
\label{eq:9thpolynomial}
K = \exp{\left(\frac{a}{\overline{T}^3} + \frac{b}{\overline{T}^2} + \frac{c}{\overline{T}} + d\log{(\overline{T})} + e + f\overline{T} + g\overline{T}^2 + h\overline{T}^3 + i\overline{T}^4\right)}.
\end{equation}
The coefficients for reactions $5$ to $8$ using the above equation are reported in tab. \ref{tab:forwardreactionrates}. Reaction rates $1$ to $5$ and $8$ were assumed to involve only the ground state of each chemical species. Reactions $6$ and $7$ were used to derive a reaction rate coefficient for the \ce{CO2 + O($^3$P) <-> CO + O2} and \ce{CO2 + O($^1$D) <-> CO + O2} reactions by assuming a $50:50$ branching for \ce{O} atoms in rate $7$. The addition of $k_6$ and half of $k_7$ are labeled as 'Sharipov StS 1' and half of $k_7$ is labeled as 'Sharipov StS 2' in figure~\ref{fig:exreaction} where these rate coefficients are plotted along with $k_1$ to $k_5$ and $k_8$ of table~\ref{tab:excratereview}. The collisional rate coefficient of the \ce{CO2 + O} collision is also plotted in black in the same figure. Back to fig.~\ref{fig:dissociation}, the curve with the redistributed \ce{CO2 + O} exchange reaction reported by Sharipov \cite{Sharipov2011} and denoted $k_5$ in table \ref{tab:excratereview}, is plotted with the label 'FHO + ER'. An improvement to the previous model is obtained but still far from an equilibrium state at the second time scale. 
\begin{table}[ht]
\centering
\caption{Reaction rates for the \ce{CO2 + O <-> CO + O2} (direct or inverse) found in literature and considered for this work. Reactions 1 to 4 are rates for \ce{CO2 + O -> CO + O2} and reactions 5 to 8 are for the inverse reaction.}
\label{tab:excratereview}
\footnotesize
\bgroup
\def\arraystretch{1.5}
\begin{tabular}{cccccc}
\hline
\# & A (cm$^3$/mol/s) & n & $\theta$ (K) & Notes & Source \\ \hline
$k_1$ & 2.11e+13 & 0.0 & 6,651 & 2,400-3,000 K & \onlinecite{Sulzmann1965},\onlinecite{Schofield1967} \\
$k_2$ & 2.10e+13 & 0.0 & 27,800 & 1,700-3,500 K at 1.8 bar & \onlinecite{Thielen1983},\onlinecite{Park1994} \\
$k_3$ & 2.14e+12 & 0.0 & 22,848 & Adapted from $k_1$ & \onlinecite{Kwak2015} \\
$k_4$ & 2.71e+14 & 0.0 & 33,800 & - & \onlinecite{Ibragimova1991},\onlinecite{Cruden2018} \\
$k_5$ & 4.32e+7 & 1.618 & 25,018 & inv. reac., 800-5,000 K & \onlinecite{Sharipov2011} \\ 
$k_6$ & 7.63e+6 & 1.670 & 26,950 & \ce{O($^3$P)}, see $k_5$ & \onlinecite{Sharipov2011} \\ 
$k_7$ & 5.18e+6 & 1.728 & 33,470 & \ce{O($^1$D)} and \ce{O($^3$P)}, see $k_5$ & \onlinecite{Sharipov2011} \\ 
$k_8$ & 2.88e+12 & 0.0 & 24,005 & inv. reac., set of exp. & \onlinecite{VargaThesis2017} \\ \hline
\end{tabular}
\egroup
\end{table}

\begin{table}[ht]
\centering
\caption{Coefficients for the \ce{CO + O2 -> CO2 + O } reactions $5$ to $8$ of table \ref{tab:excratereview} using equation \ref{eq:9thpolynomial} for fitting. Units are in cm$^3$/mol/s.}
\label{tab:forwardreactionrates}
\footnotesize
\bgroup
\def\arraystretch{1.5}
\begin{tabular}{cccccccccc}
\hline
 & a & b & c & d & e & f & g & h & i \\ \hline
$k_5$ & -4.071e-1 & 2.888e+0 & -3.693e+1 & -1.716e+0 & 3.578e+1 & 4.620e-1 & -7.840e-3 & 6.734e-5 & -2.220e-7 \\
$k_6$ & 2.643e-5 & 2.072e+2 & -3.115e+1 & 1.678e+0 & 2.969e+1 & -5.880e-4 & 9.232e-6 & -8.196e-8 & 2.877e-10 \\
$k_7$ & 2.377e-5 & 2.073e+2 & -3.767e+1 & 1.736e+0 & 2.970e+1 & -5.871e-4 & 9.217e-6 & -8.182e-8 & 2.872e-10 \\
$k_8$ & -4.071e-1 & 2.888e+0 & -1.194e+1 & -3.334e+0 & 3.571e+1 & 4.619e-1 & -7.840e-3 & 6.734e-5 & -2.219e-7 \\ \hline
\end{tabular}
\egroup
\end{table}

\subsubsection{\ce{CO2 + C <-> CO + CO} exchange reaction}

An important reaction in industrial processes is the Boudouard reaction \ce{2CO <-> CO2 + C}. In the aforementioned processes this usually involves a phase change of the product carbon and is one of the reactions responsible for the creation of soot. Therefore, most available data does not consider all chemical species in the gaseous phase. The NIST Chemical Kinetics Database contains two estimates for the reaction rate coefficient at $300$ K for the reaction \ce{CO2 + C -> 2CO}. We will take the lowest estimate which is also the most recent \cite{Husain1975} as $k_f = 6.022\times 10^{8}$ cm$^3$/mole/s. We give a temperature dependence to this reaction as:
\begin{equation}
\label{eq:extension}
k_f(T) = \frac{k_f(300\text{K})}{Z(300\text{K})}Z(T)
\end{equation}
where $Z(T)$ is the \ce{CO2 + C} collisional rate coefficient. This formulation may be interpreted as the total ratio of collisions \ce{CO2 + C} to collisions resulting in \ce{CO + CO} or other products is constant. After the expansion, a redistribution as described in this section is carried out. Despite being an important industrial process, in the gaseous phase this reaction is not expected to be important since \ce{C} atoms are very reactive and disappear fast in gaseous environments to form other compounds. It is included just for the sake of completeness. The inclusion of this reaction is represented in figure \ref{fig:dissociation} with the label 'FHO + ER1/2 + COChem'. As a relatively negligible process it did not justify a simulation by itself and the previously discussed mechanisms and as such was added later.

\subsubsection{Quenching of O atoms}

Quenching of atomic \ce{O} is also introduced in the model. Specifically, the reaction \ce{O($^1$D) + M <-> O($^3$P) + M} is addressed with data from literature. Before that, a brief discussion on the importance of this mechanism is carried here. The appendix of the work of Fox and Ha\'{c} \cite{Fox2018} provides an extensive review of cooling mechanisms of hot \ce{O} atoms. These hot atoms need not be electronically excited \ce{O}: even translationally excited \ce{O($^3$P)} atoms may redistribute their energy to the ro-vibrational modes of molecules. Important collision partners are \ce{CO}, \ce{O2} and \ce{CO2}. One contribution \cite{Dunlea2004} has even reported an efficient deposition of translational energy into the ro-vibrational modes of \ce{CO2}. Additionally, the importance of the excited \ce{O($^1$D)} atom cannot be understated as it's excess energy may be redistributed to vibrationally excited \ce{O2} and \ce{CO2} molecules through the reactions: \ce{O($^1$D) + CO2 <-> O2($\text{v,J}$) + CO + 1.63} eV and  \ce{O($^1$D) + CO2 <-> O($^3$P) + CO2($\text{v,J}$) + 1.97} eV. However, in this work we will only deal with quenching reactions like \ce{O($^1$D) + M <-> O($^3$P) + M}. Some measurements of the reaction rate coefficients at $300$ K are summarized in tab.~\ref{tab:OQuenching}. These reactions are given a temperature dependence according to eq.~\ref{eq:extension} and fitted to Arrhenius rates
\begin{equation}
k_f = AT^n\exp{\left(-\frac{\theta}{T}\right)}.
\end{equation}
A reaction rate coefficient with partner \ce{O($^3$P)} is also available from \cite{Yee1990} but with no temperature validity range, with the shape:
\begin{equation}
k_f = A + B\sqrt{T} + CT
\end{equation}
and coefficients $A = 7.66\times10^{-12}$, $B = 2.13\times 10^{-13}$ and $C = -1.84\times 10^{-15}$ in units of cm$^3$/part./s. An Arrhenius fit was performed on this reaction from $300$ up to $2500$ K and checked for consistency up to $100,000$ K. The Arrhenius coefficients of the fitted \ce{O} quenching reactions are found in table~\ref{tab:OQuenchingArrhenius}. 

\begin{table}[ht]
\centering
\caption{Quenching reaction rates of \ce{O($^1$D) + M <-> O($^3$P) + M} at $300$ K. Units are $10^{-10}$cm$^3$/part/s.}
\label{tab:OQuenching}
\footnotesize
\bgroup
\def\arraystretch{1.5}
\begin{tabular}{cccc}
\hline
M & \ce{CO2} & \ce{CO} & \ce{O2} \\ \hline
$k_f$ & 1.03 & 0.3 & 0.41 \\ 
Source & \onlinecite{Wine1981} & \onlinecite{Donovan1970} & \onlinecite{Davidson1976} \\ \hline
\end{tabular}
\egroup
\end{table}

\begin{table}[ht]
\centering
\caption{Quenching reaction rates of \ce{O($^1$D) + M <-> O($^3$P) + M} fitted by Arrhenius experessions.}
\label{tab:OQuenchingArrhenius}
\footnotesize
\bgroup
\def\arraystretch{1.5}
\begin{tabular}{ccccc}
\hline
M & \ce{CO2} & \ce{CO} & \ce{O2} & \ce{O($^1$D)} \\ \hline
A ($10^{12}$cm$^3$/mol/s) & 3.58 & 1.04 & 1.43 & 4.21 \\
n & 0.50 & 0.50 & 0.50 & 0.088 \\
$\theta$ (K) & 0.00 & 0.00 & 0.00 & 21.91 \\ \hline
\end{tabular}
\egroup
\end{table}

\subsubsection{Other rates}

Other processes, not directly related to \ce{CO2}, could still make an impact on the concentration of \ce{CO2}. As such we have included the thermochemistry for \ce{CO}, reported in the work of Cruden \textit{et al.} \cite{Cruden2018},into our kinetic model. In these reactions we have assumed that \ce{O} is in it's ground state. The rates we have included are presented in tab.~\ref{tab:otherrates}. In fig.~\ref{fig:dissociation} the label 'COChem' indicates the inclusion of the reactions in tab.~\ref{tab:otherrates} of the simulation. In the curve with label 'FHO + COChem' the \ce{CO2} mole fraction is closer to the base 'FHO' model than the curves with the exchange reaction \ce{CO2 + O <-> CO + O2} labeled 'FHO + ER1', 'FHO + ER1 + COChem' and 'FHO + ER1/2 + COChem'. The latter curves indicate that the inclusion of the \ce{CO2 + O} exchange reaction is essential to obtain a better dissociation trend for \ce{CO2}. Adding the thermochemistry in \cite{Cruden2018} provides a source of \ce{O} atoms which accelerate the \ce{CO2} decomposition.

\begin{table}[ht]
\centering
\caption{Reactions and Arrhenius coefficients proposed in \cite{Cruden2018}}
\label{tab:otherrates}
\footnotesize
\bgroup
\def\arraystretch{1.5}
\begin{tabular}{ccccc}
\hline
Reaction & A (cm$^3$/mol/s) & n & $\theta$ (K) & Source \\ \hline
\ce{CO + M <-> C + O + M} & $7.99\times10^{38}$ & -5.5 & 129,000 & \onlinecite{Hanson1974} \\
\ce{C2 + O <-> CO + C} & $3.61\times10^{14}$ & 0.0 & 0.0 & \onlinecite{Fairbairn1969} \\
\ce{C2 + M <-> C + C + M} & $1.82\times10^{15}$ & 0.0 & 64,000 & \onlinecite{Fairbairn1969} \\
\ce{CO + O <-> C + O2} & $3.9\times10^{13}$ & -0.18 & 69,200 & \onlinecite{Park1994} \\
\ce{O2 + M <-> O + O + M} & $1.2\times10^{14}$ & 0.0 & 54,246 & \onlinecite{Warnatz1984} \\
\ce{C + e- <->C+ + e- + e-} & $3.7\times10^{31}$ & -3.0 & 130,700 & \onlinecite{Park1994} \\
\ce{O + e- <-> O+ + e- + e-} & $3.9\times10^{33}$ & -3.78 & 158,500 & \onlinecite{park1989nonequilibrium} \\
\ce{CO + e- <-> CO+ + e- + e-} & $4.5\times10^{14}$ & 0.275 & 163,500 & \onlinecite{Teulet2009} \\
\ce{O2 + e- <-> O2+ + e- + e-} & $2.19\times10^{10}$ & 1.16 & 130,000 & \onlinecite{Teulet2009} \\
\ce{C + O <-> CO+ + e-} & $8.8\times10^{8}$ & 1.0 & 33,100 & \onlinecite{Park1994} \\
\ce{CO + C+ <-> CO+ + C} & $1.1\times10^{13}$ & 0.0 & 31,400 & \onlinecite{Park1994} \\
\ce{O + O <-> O2+ + e-} & $7.1\times10^{2}$ & 2.7 & 80,600 & \onlinecite{Park1993} \\
\ce{O2 + C+ <-> O2+ + C} & $1.0\times10^{13}$ & 0.0 & 9,400 & \onlinecite{Park1994} \\
\ce{O2+ + O <-> O2 + O+} & $2.19\times10^{10}$ & 1.16 & 130,000 & \onlinecite{Park1993} \\ \hline
\end{tabular}
\egroup
\end{table}

\subsection{Final dataset}
\label{sec:rates}

To summarize the model developed in this work, reaction labels and number of reactions are presented in tab.~\ref{tab:reactions}. This excludes the macroscopic reactions from \cite{Cruden2018}, since these are presented in tab.~\ref{tab:otherrates}. There are $11$ species in the model: \ce{CO2}, \ce{CO}, \ce{O2}, \ce{C2}, \ce{C}, \ce{O}, \ce{CO+}, \ce{O2+}, \ce{C+}, \ce{O+}, \ce{e-}, two of which are state specific, \ce{CO2} with $201$ vibronic levels and \ce{O} with $2$ electronic levels. The model contains a total of $22,569$ reactions, only $14$ of which are not STS.

\begin{table}
\centering
\caption{List of reactions included in our \ce{CO2} kinetic model. Not listed are the reactions taken from the kinetic scheme in \cite{Cruden2018} which describes \ce{CO} thermochemistry. These reactions can be found in table~\ref{tab:otherrates}.}
\label{tab:reactions}
\footnotesize
\begin{tabular}{lr@{}lcr}
\hline
  & \multicolumn{2}{c}{Name} & Type & \#Reac. \\ \hline
R1  & \ce{CO2}(X,v$^{\prime}_1$) + M &$\leftrightarrow$ \ce{CO2}(X,v$^{\prime\prime}_1$) + M & VT & 1770 \\
R2  & \ce{CO2}(X,v$^{\prime}_2$) + M &$\leftrightarrow$ \ce{CO2}(X,v$^{\prime\prime}_2$) + M & VT  & 5050 \\
R3  & \ce{CO2}(X,v$^{\prime}_3$) + M &$\leftrightarrow$ \ce{CO2}(X,v$^{\prime\prime}_3$) + M & VT  & 861  \\
R4  & \ce{CO2}(X,v$^{\prime}_1$) + \ce{CO2}(X,v'$_1$) &$\leftrightarrow$ \ce{CO2}(X,v$^{\prime}_1$+1) + \ce{CO2}(X,v$^{\prime}_1$-1) & VVT  & 58 \\
R5  & \ce{CO2}(X,v$^{\prime}_2$) + \ce{CO2}(X,v'$_2$) &$\leftrightarrow$ \ce{CO2}(X,v$^{\prime}_2$+1) + \ce{CO2}(X,v$^{\prime}_2$-1) & VVT  & 99 \\
R6  & \ce{CO2}(X,v$^{\prime}_3$) + \ce{CO2}(X,v'$_3$) &$\leftrightarrow$ \ce{CO2}(X,v$^{\prime}_3$+1) + \ce{CO2}(X,v$^{\prime}_3$-1) & VVT  & 41 \\
R7  & \ce{CO2}(X,v$^{\prime}_1$) + M &$\leftrightarrow$ \ce{CO2}(X,v$^{\prime\prime}_2$) + M & IVT  & 5900 \\
R8  & \ce{CO2}(X,v$^{\prime}_1$) + M &$\leftrightarrow$ \ce{CO2}(X,v$^{\prime\prime}_3$) + M & IVT  & 2478 \\
R9  & \ce{CO2}(X,v$^{\prime}_2$) + M &$\leftrightarrow$ \ce{CO2}(X,v$^{\prime\prime}_3$) + M & IVT  & 4200 \\
R10 & \ce{CO2}(B,v$^{\prime}_1$) + M &$\leftrightarrow$ \ce{CO2}(B,v$^{\prime\prime}_1$) + M & VT  & 78   \\
R11 & \ce{CO2}(B,v$^{\prime}_2$) + M &$\leftrightarrow$ \ce{CO2}(B,v$^{\prime\prime}_2$) + M & VT  & 325  \\
R12 & \ce{CO2}(B,v$^{\prime}_3$) + M &$\leftrightarrow$ \ce{CO2}(B,v$^{\prime\prime}_3$) + M & VT  & 21   \\
R13 & \ce{CO2}(B,v$^{\prime}_1$) + \ce{CO2}(B,v'$_1$) &$\leftrightarrow$ \ce{CO2}(B,v$^{\prime}_1$+1) + \ce{CO2}(B,v$^{\prime}_1$-1) & VVT  & 11 \\
R14 & \ce{CO2}(B,v$^{\prime}_2$) + \ce{CO2}(B,v'$_2$) &$\leftrightarrow$ \ce{CO2}(B,v$^{\prime}_2$+1) + \ce{CO2}(B,v$^{\prime}_2$-1) & VVT  & 24 \\
R15 & \ce{CO2}(B,v$^{\prime}_3$) + \ce{CO2}(B,v'$_3$) &$\leftrightarrow$ \ce{CO2}(B,v$^{\prime}_3$+1) + \ce{CO2}(B,v$^{\prime}_3$-1) & VVT  & 6  \\
R16 & \ce{CO2}(B,v$^{\prime}_1$) + M &$\leftrightarrow$ \ce{CO2}(B,v$^{\prime\prime}_2$) + M & IVT  & 300 \\
R17 & \ce{CO2}(B,v$^{\prime}_1$) + M &$\leftrightarrow$ \ce{CO2}(B,v$^{\prime\prime}_3$) + M & IVT  & 84  \\
R18 & \ce{CO2}(B,v$^{\prime}_2$) + M &$\leftrightarrow$ \ce{CO2}(B,v$^{\prime\prime}_3$) + M & IVT  & 175 \\
R19 & \ce{CO2}(X,v$^{\prime}_2$) + M &$\leftrightarrow$ \ce{CO2}(B,v$^{\prime\prime}_1$) + M & VE  & 103 \\
R20 & \ce{CO2}(X,v$^{\prime}_2$) + M &$\leftrightarrow$ \ce{CO2}(B,v$^{\prime\prime}_2$) + M & VE  & 311 \\
R21 & \ce{CO2}(X,v$^{\prime}_2$) + M &$\leftrightarrow$ \ce{CO2}(B,v$^{\prime\prime}_3$) + M & VE  & 163 \\
R22 & \ce{CO2}(X,v$^{\prime}_3$) + M &$\leftrightarrow$ \ce{CO + O}$(^1D)$ + M & VD  & 42  \\
R23 & \ce{CO2}(X,v$^{\prime}_3$) + M &$\leftrightarrow$ \ce{CO + O}$(^3P)$ + M & VE/VD  & 42  \\
R24 & \ce{CO2}(B,v$^{\prime}_3$) + M &$\leftrightarrow$ \ce{CO + O}$(^3P)$ + M & VD  & 7   \\
R25 & \ce{CO2}(X,v$^{\prime}_{1,2,3}$) + O($^3$P) &$\leftrightarrow$ \ce{CO + O2} & Zeldov. & 201 \\ 
R26 & \ce{CO2}(X,v$^{\prime}_{1,2,3}$) + C &$\leftrightarrow$ \ce{CO + CO} & Zeldov. & 201 \\ 
R27 & O($^1$D) + M &$\leftrightarrow$ O($^3$P) + M  & Quench. & 4 \\ \hline
\end{tabular}
\end{table}

\subsection{Underlying assumptions and restrictions}

\ce{CO2} is a triatomic molecule and consequently it has more degrees of freedom than a diatomic molecule. This induces complexities in the sense that modeling for such molecules needs to be tractable with a reasonable number of levels and rates, compatible with current-day computational resources. In this sense, much more restrictions and assumptions than in the case of diatomic molecules need to be brought. For example, diatomic molecules state-to-state models customarily assume a Boltzmann equilibrium for the rotational levels, solely modeling a reasonable number of vibrational and electronic states (in order of the hundred). For \ce{CO2} not only this has to be assumed, but furthermore additional restrictions have to be brought regarding the different vibrational degrees of freedom. These include:
\begin{itemize}
\item Full separation of the three vibrational modes of \ce{CO2}, only considering its so-called \textit{extreme states}. This means that for example the rate $\textrm{CO}_2(n',m',p')\rightarrow\textrm{CO}_2(n'',m',p')$ will be the same no matter what the $m$ and $p$ quantum numbers are. The calculations by Billing \cite{Billing1979} shown that differences from a factor of 5 (at room temperature) down to a factor of 1.5 (at 2000 K) exist for these rate coefficients with different $m$ and $p$. This is perhaps the most significant limitation of our model, with implications on the modeling of higher, near-dissociation levels, which we will discuss more ahead. Nevertheless, this allows us to achieve a computationally tractable model with about $N\approx250$ levels instead of the $N\approx10,000^{+}$ real ground state levels of \ce{CO2}.% With multi-quantum transitions we need a first approach to account for the $N^2$ V-T transitions and $N^4$ V-V-T transitions and this is simply not tractable for $N\approx 10,000$.

\item For the same reasons, there is no specific accounting of the $l_2$ bending quantum numbers, and transitions from any $v_2^{l_2}$ levels are assumed to have equivalent rates. Billing calculations from Ref. \cite{Billing1981} predict differences ranging from a factor of 5 to one order of magnitude in transitions from the same $v_2$ level, depending on the $l_2$ quantum number.

\item We extrapolate the \ce{CO2} PES in its 3 modes limit ($ss$, $be$, $as$) by a representative repulsive and near-dissociation potential. While this is not as accurate as defining a proper PES near-dissociation potential (which is not carried out in the NASA--Ames--2 PES), we should still be capable of providing correct near-dissociation trends, as compared with the usual extrapolation of polynomial expansions. Past similar approaches for diatomic molecules have provided quite accurate results \cite{LinodaSilva2008}.
\end{itemize}

In addition to those, other limitations currently exist, but could easily be waived in future works:

\begin{itemize}
\item Considering an isotropic Morse-like intermolecular potential, and assuming the collision as 1D with the application of a steric factor. Comparisons carried out with the FHO model against PES-based methods show that rates with the same order of magnitude are predicted. However the temperature dependence at low-temperatures is poorly reproduced by the FHO model, and attractive low-temperature effects should be modeled resorting to the Sharma-Brau \cite{Sharma1969} theory, and added to the rate provided by the FHO theory. Results in the higher temperature limit have a better agreement with PES results, as would be expected in the Landau-Teller limit (increasing $\log(K{_f})$ over T$^{-1/3}$). Regarding the scaling of rates to higher vibrational quantum levels, there is not enough PES-based data to provide a meaningful comparison. In any case, since this work is mostly concerned with mid to high-temperature regimes, we may safely neglect these low-temperature limits below room temperature.
\item The rates of collision are the same independently of the collisional partner, which is assumed to be \ce{CO2}. This assumption is temporarily used as a matter of convenience, since our worked examples are applied to pure \ce{CO2} flows. This assumption will need to be revisited for increasing the accuracy of the database or allowing for simulations of highly diluted \ce{CO2} flows (typically in Helium or Argon baths).
\end{itemize}

\section{Results}
\label{sec:results}

The results of the developed model are shown and discussed in this section. Firstly we present and discuss some of the calculated rates in tab.~\ref{tab:reactions}. Then, we showcase simulations of an isothermal gas with no dissociation, a \ce{CO2} gas excitation and relaxation with dissociation and  recombination simulations, which are presented and discussed with an emphasis on the behaviour of the modes of \ce{CO2} and mechanisms for the decomposition. Secondly, a comparison against available shock-tube data is performed.

\subsection{Rates Dataset}

This subsection will present some of the calculated rates presented in table~\ref{tab:reactions}. Notably, mechanisms R2, R7 and R23. Mechanisms R1--R3 to R10--R12 share the same functional form. The same can be said for mechanisms R7--R9 and R16--R18. Reactions that involve a spin-forbidden interaction such as R19, R20, R21 and R23 also share some similarities. 

\begin{figure}
\centering
\includegraphics[width=0.7\linewidth,keepaspectratio]{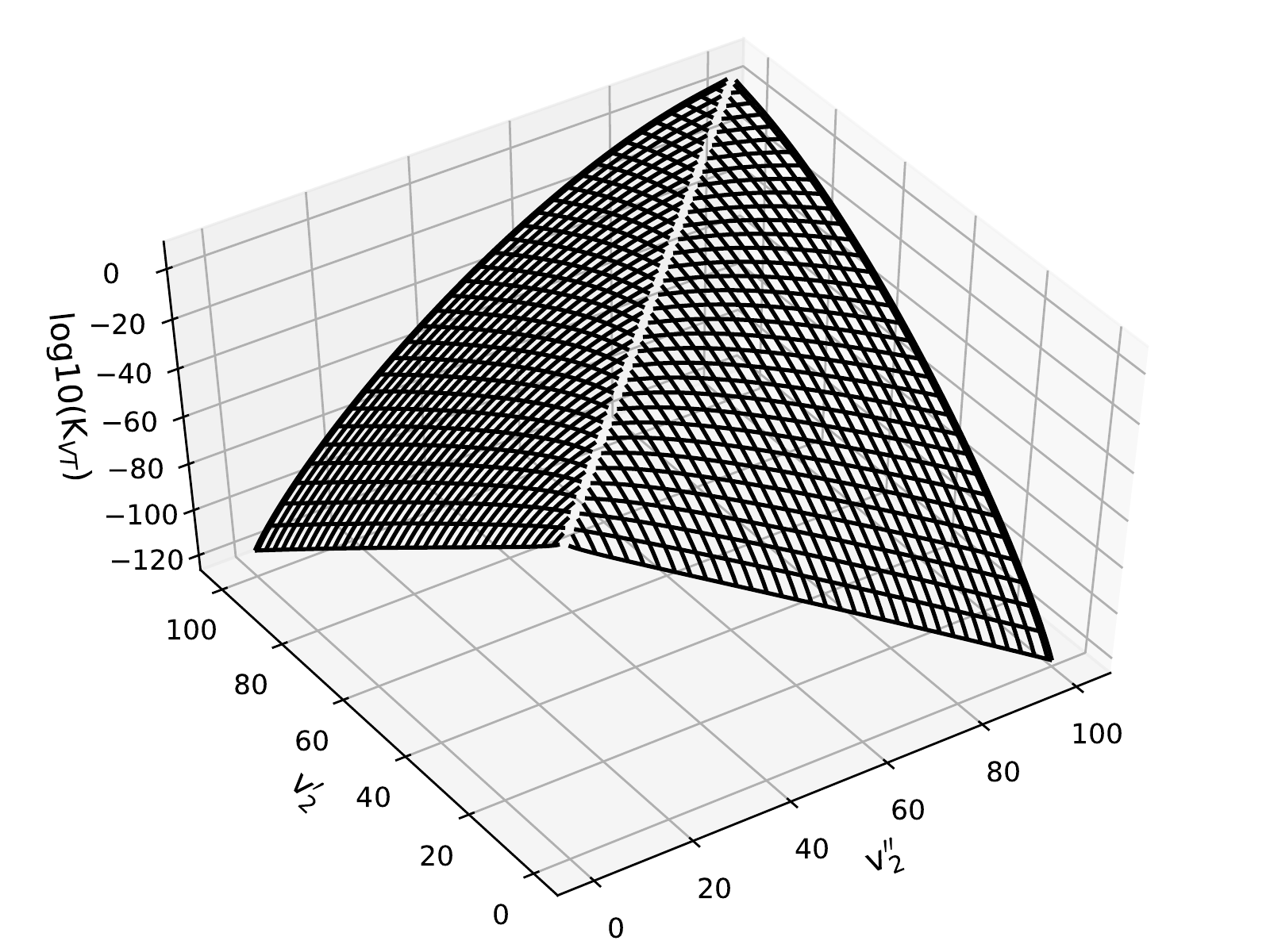}
\caption{Base 10 logarithm of VT rate coefficients at $5,000$ K for the bending mode of \ce{CO2}. As expected, large quantum jumps have lower probability than smaller jumps, with endothermic jumps falling out faster than the reverse exothermic jumps. VT rate coefficients of other modes have the same functional shape.}
\label{fig:wireframeVT}
\end{figure}

Firstly, in fig.~\ref{fig:wireframeVT} the $\log_{10}\left(K_{\text{VT}}\right)$ rate coefficients of the bending levels of the ground electronic state of \ce{CO2} are shown at $5,000$ K. These correspond to the rate coefficients of mechanism R2 in tab.~\ref{tab:reactions}. It is expected that transitions with small differences in the vibrational number should be stronger than transitions with greater $\Delta$v. This is observed as the rate coefficients tend to a maximum value around the plane where v$_2^{\prime} = \text{v}_2^{\prime\prime}$. The reactions with no change in vibrational number are not depicted as these correspond to no energy exchange happening. Two oblique surfaces corresponding to exothermic and endothermic reactions are observed. The exothermic reactions are slightly more likely than endothermic reactions and this is also observed by comparing the inclination of the surfaces against the $z$ axis scale in the planes $v_2^{\prime}=100$ and $v_2^{\prime\prime}=100$. Reaction mechanisms R1--R3 and R10--R12 are functionally the same as R2 with some deformations that might occur when the Bessel approximation \cite{Nikitin1977} is used.

\begin{figure}
\centering
\includegraphics[width=0.7\linewidth,keepaspectratio]{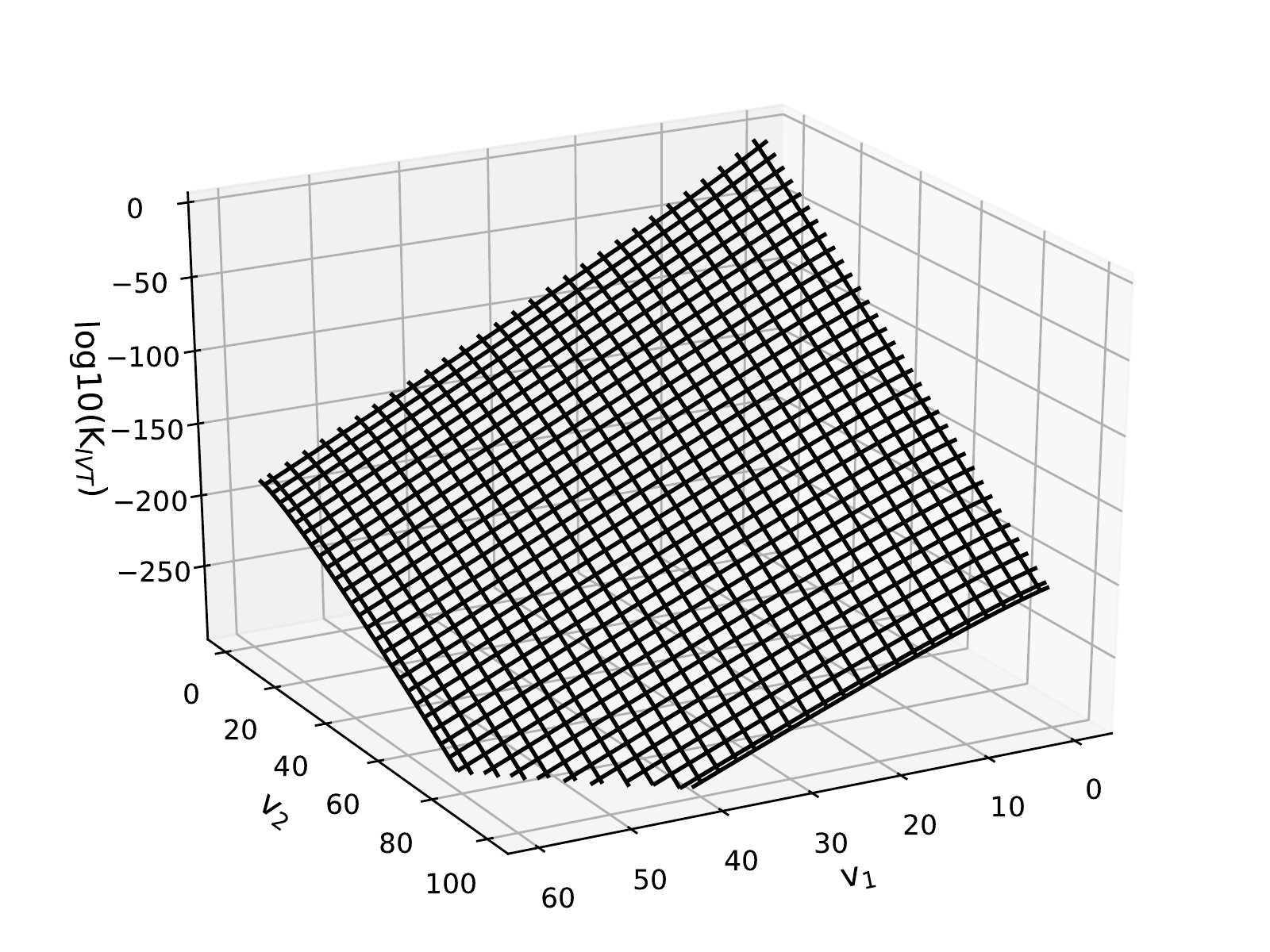}
\caption{Base 10 logarithm of IVT rate coefficients at $5000$ K between the symmetric and bending modes of \ce{CO2}. As expected, the probability of intermode energy exchange is lowest when the vibrational numbers are high and highest when the vibrational numbers are low. Other IVT rate coefficient sets have the same functional shape.}
\label{fig:wireframeIVT}
\end{figure}

Secondly, in fig.~\ref{fig:wireframeIVT} the $\log_{10}\left(K_{\text{IVT}}\right)$ rate coefficients between the bending and symmetric stretch levels of ground \ce{CO2} are shown at $5,000$ K. These processes correspond to mechanism R7 in table~\ref{tab:reactions}. As these transition probabilities are modeled as VT transitions with large changes in vibrational number, the rate coefficients drop very fast as the vibrational numbers increase. Reaction mechanisms R8, R9 and R16--R18 share the same functional form as mechanism R7.

\begin{figure}
\centering
\includegraphics[width=0.6\linewidth,keepaspectratio]{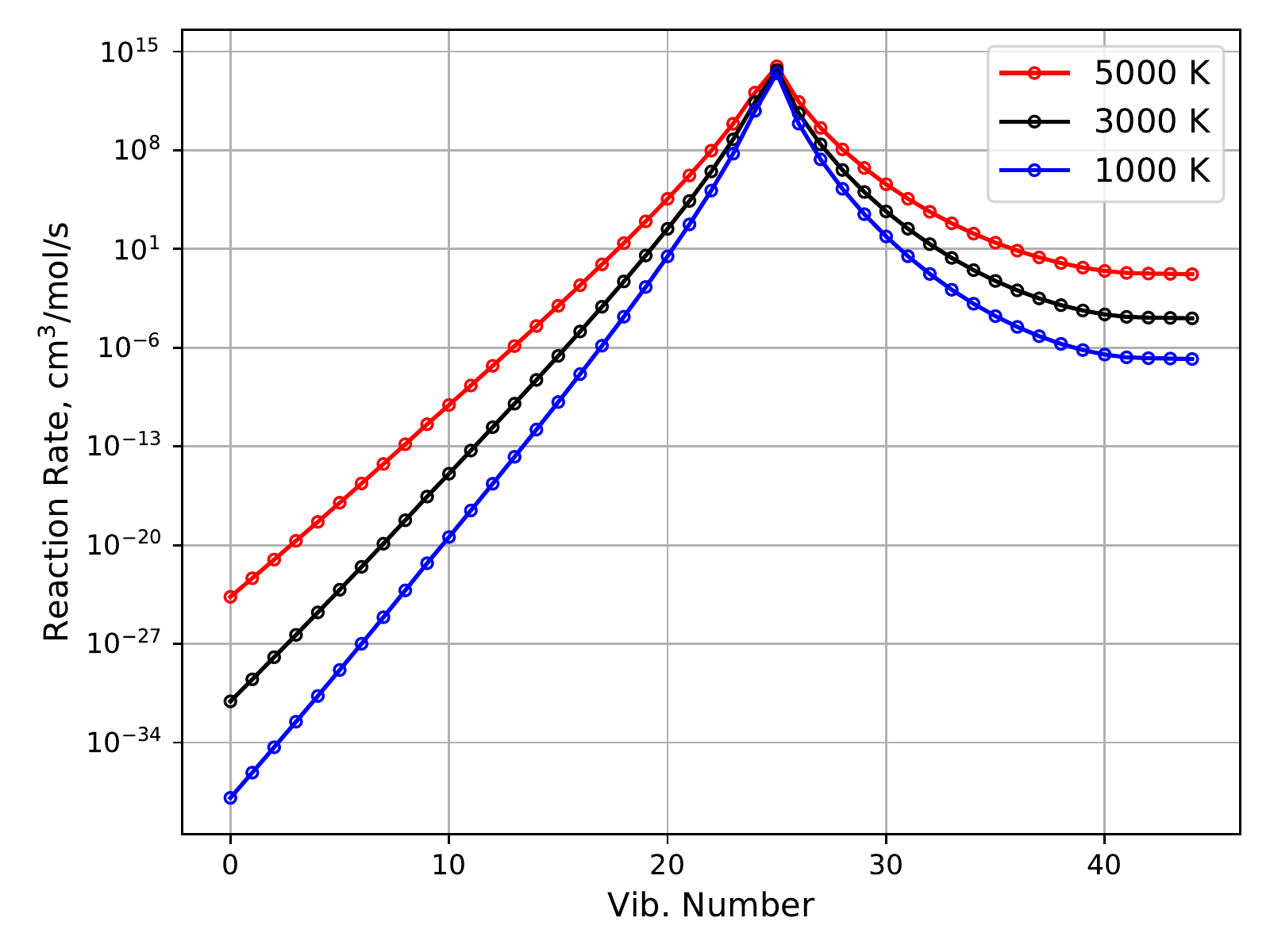}
\caption{Rate coefficients computed using the Rosen-Zener probability formula (eq.~\ref{eq:RZformula}) for the dissociation mechanism R23 in tab.~\ref{tab:reactions}. The rate coefficients are computed at $1,000$, $3,000$ and $5,000$ K and as expected, the greater transition probabilities lie close to the crossing point between the singlet and triplet \ce{CO2}.}
\label{fig:plotVED}
\end{figure}

Finally, in fig.~\ref{fig:plotVED} the rate coefficients of mechanism R23 in tab.~\ref{tab:reactions} are plotted at $1,000$, $2,000$ and $3,000$ K. A maximum for the rate coefficient is observed at the vibrational number which is closest to the crossing between the ground state of \ce{CO2} and the repulsive triplet state of \ce{CO2}. This is expected by the simple formulation of the Rosen--Zener probability formula in eq.~\ref{eq:RZformula}. Mechanisms R19, R20 and R21 are also computed through the same formula and may have different crossings which will correspond to horizontal shifts in the peak of the rate coefficients plotted in fig.~\ref{fig:plotVED}. Otherwise, the aforementioned mechanisms share the same functional form as R23.

\subsection{Theoretical test cases}

\subsubsection{\ce{CO2} isothermal}

\begin{figure}
\centering
\includegraphics[width=0.49\linewidth,keepaspectratio]{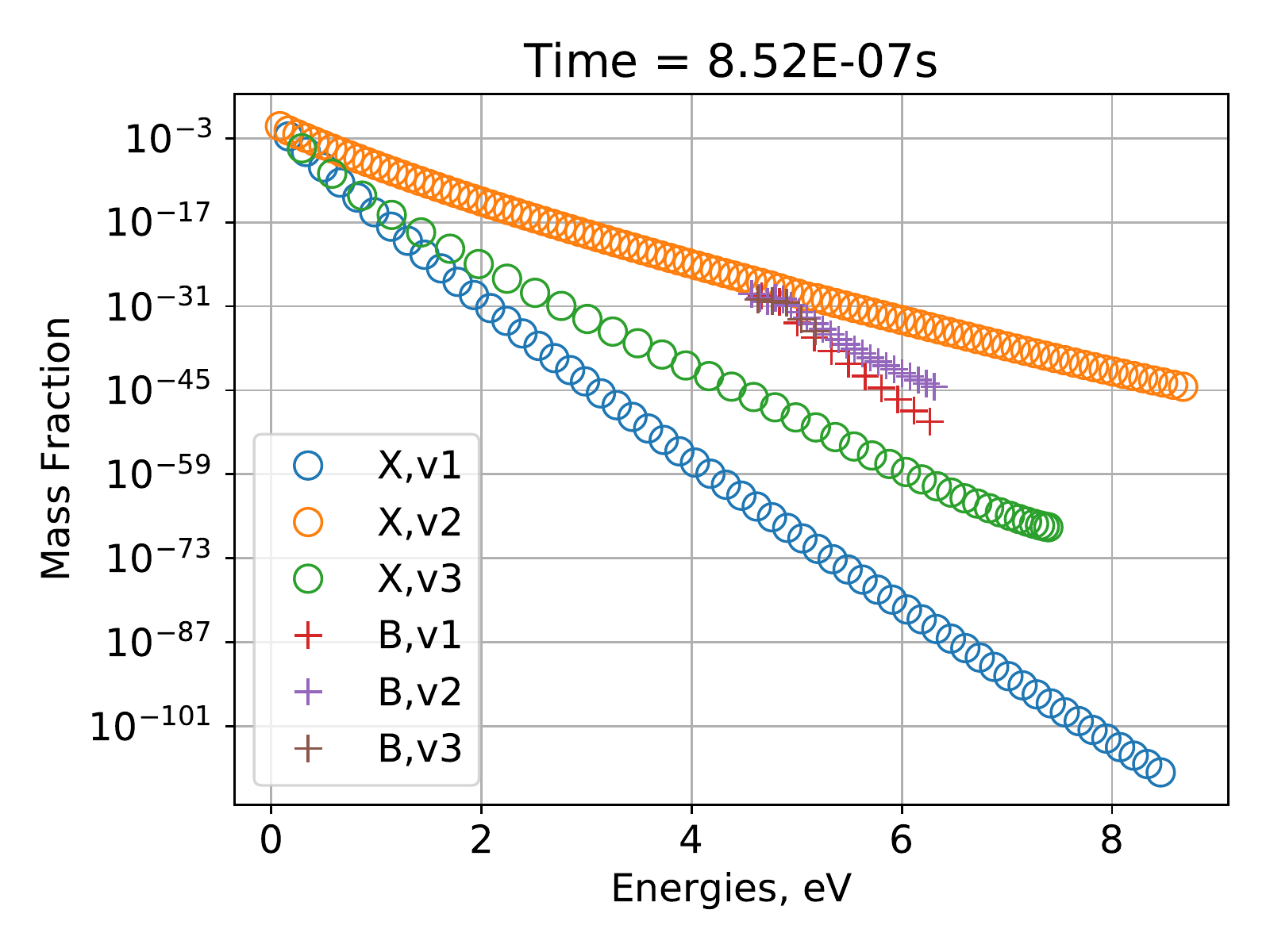}
\includegraphics[width=0.49\linewidth,keepaspectratio]{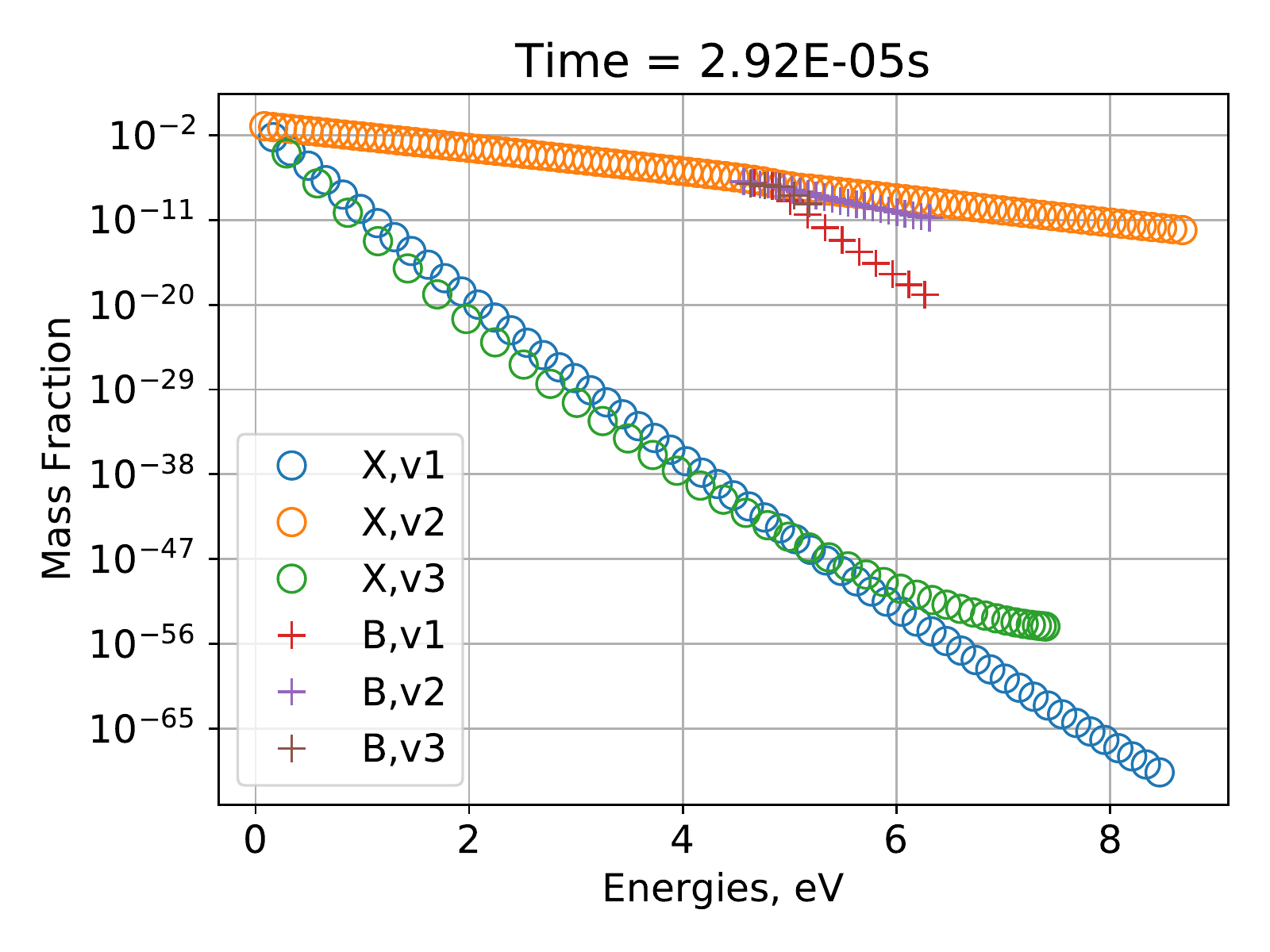}
\includegraphics[width=0.49\linewidth,keepaspectratio]{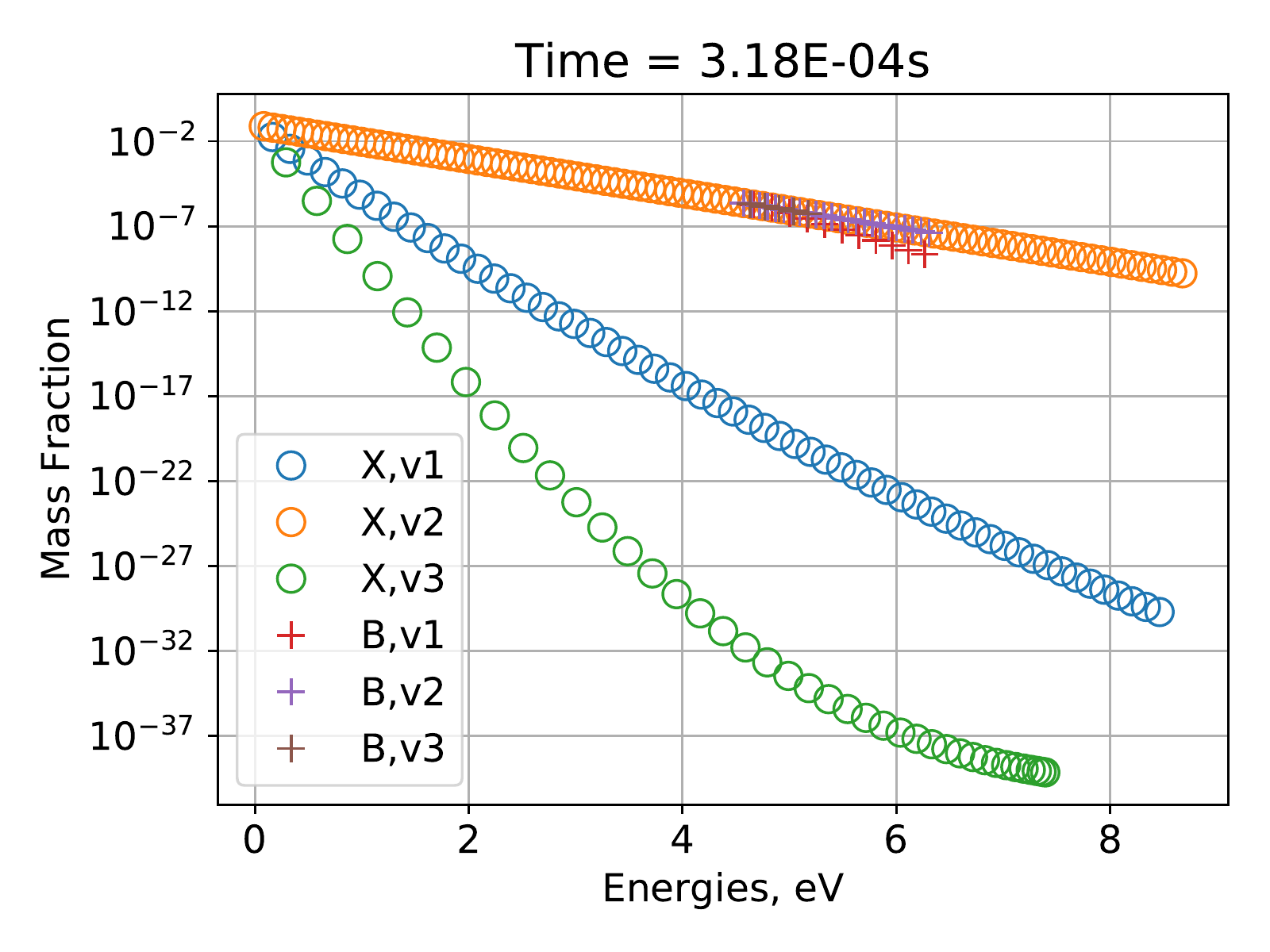}
\includegraphics[width=0.49\linewidth,keepaspectratio]{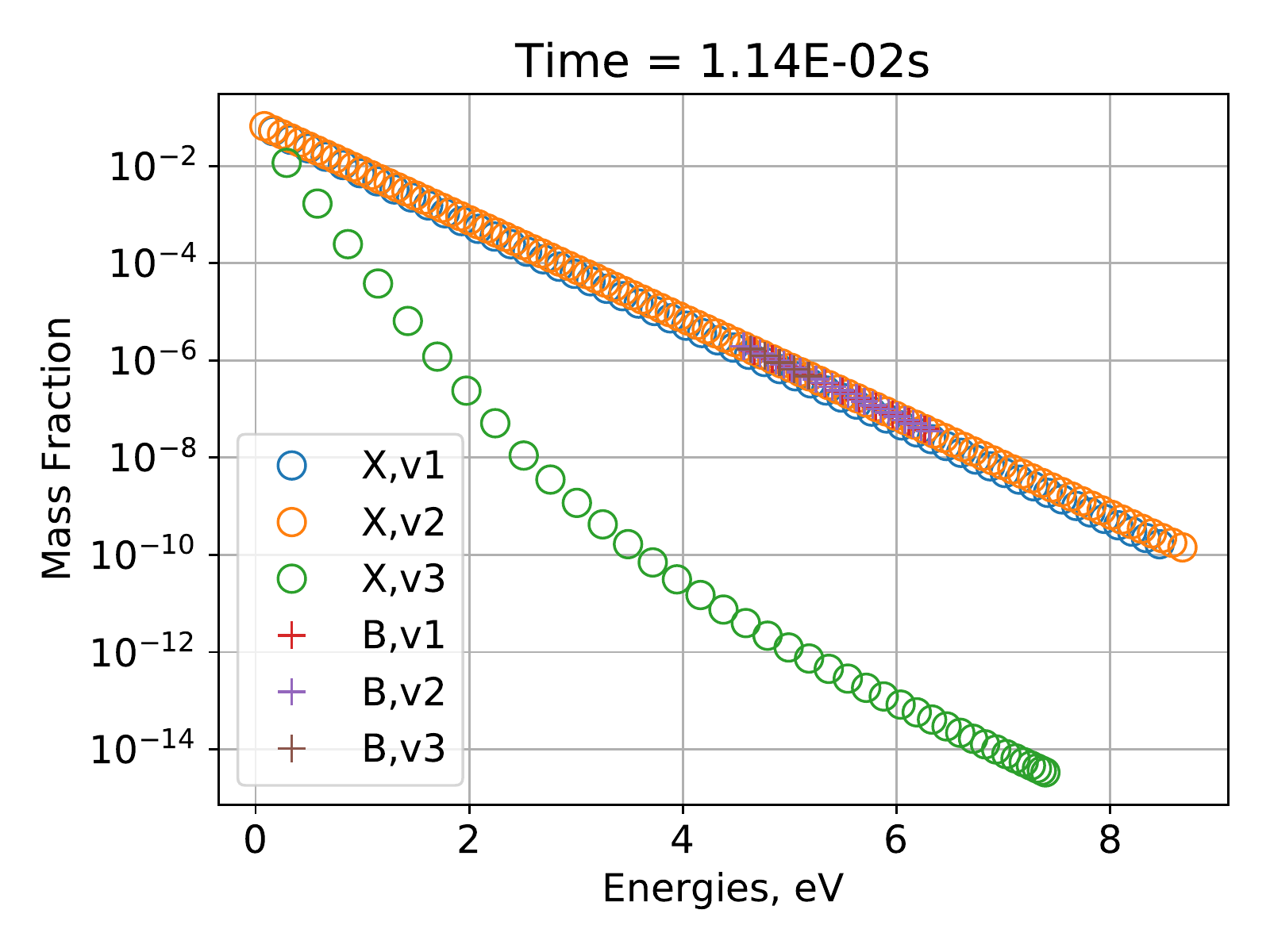}
\caption{Time snapshots of the \ce{CO2} vibrational distribution in a 0D Isothermal simulation for a pure \ce{CO2} gas at $2,000$ Pa initially at $300$ K and a final temperature of $5,000$ K.}
\label{fig:0DCO2only}
\end{figure}

An isothermal 0D simulation was performed for a pure \ce{CO2} gas, initially at $300$ K and $2,000$ Pa with the gas temperature suddenly raised to $5,000$ K. For this particular simulation we have decided not to use dissociation processes and consider only \ce{CO2} internal processes. Time snapshots of the mass fractions of \ce{CO2} are plotted in fig.~\ref{fig:0DCO2only}. In the top left figure, at $t=8.52\times10^{-7}$ seconds the bending mode is excited much faster than the other modes. This is expected since the spacing between consecutive vibrational levels is smaller. The electronically excited \ce{CO2} accompanies the excitation of the bending which gets more populated than the levels at the same energy of the asymmetric and symmetric stretch modes of the ground state of \ce{CO2}. Contrary to expectations the asymmetric mode is more populated than the symmetric stretch mode. That changes at $t=2.92\times10^{-5}$ seconds in the top right of figure~\ref{fig:0DCO2only}. At this time, the symmetric stretch mode overtakes the asymmetric stretch mode of \ce{CO2} ground state, except in the higher energy levels of the asymmetric stretch mode. At $t=3.18\times10^{-4}$ seconds the bending mode and the \ce{CO2} triplet state are almost in their equilibrium populations. At $t=1.14\times10^{-2}$ seconds all \ce{CO2} subpopulations are in equilibrium except the asymmetric stretch mode. From these simulations and considering the possible pathways for \ce{CO2} dissociation, it is expectable that the greatest contributor to \ce{CO2} dissociation is the pathway through the excited triplet state of \ce{CO2}.

\subsubsection{\ce{CO2} Dissociating flow}

\begin{figure}
\centering
\includegraphics[width=0.49\linewidth,keepaspectratio]{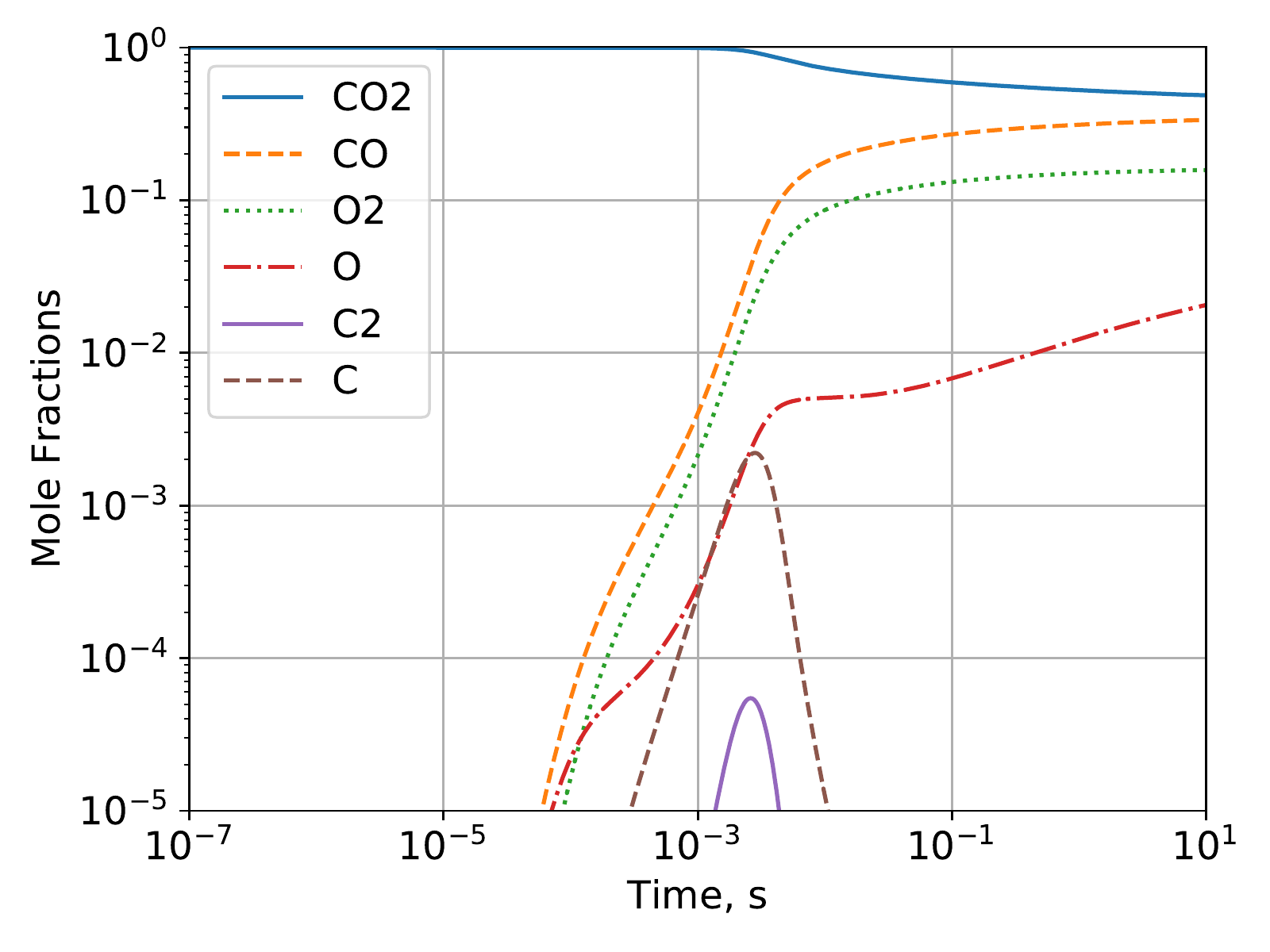}
\includegraphics[width=0.49\linewidth,keepaspectratio]{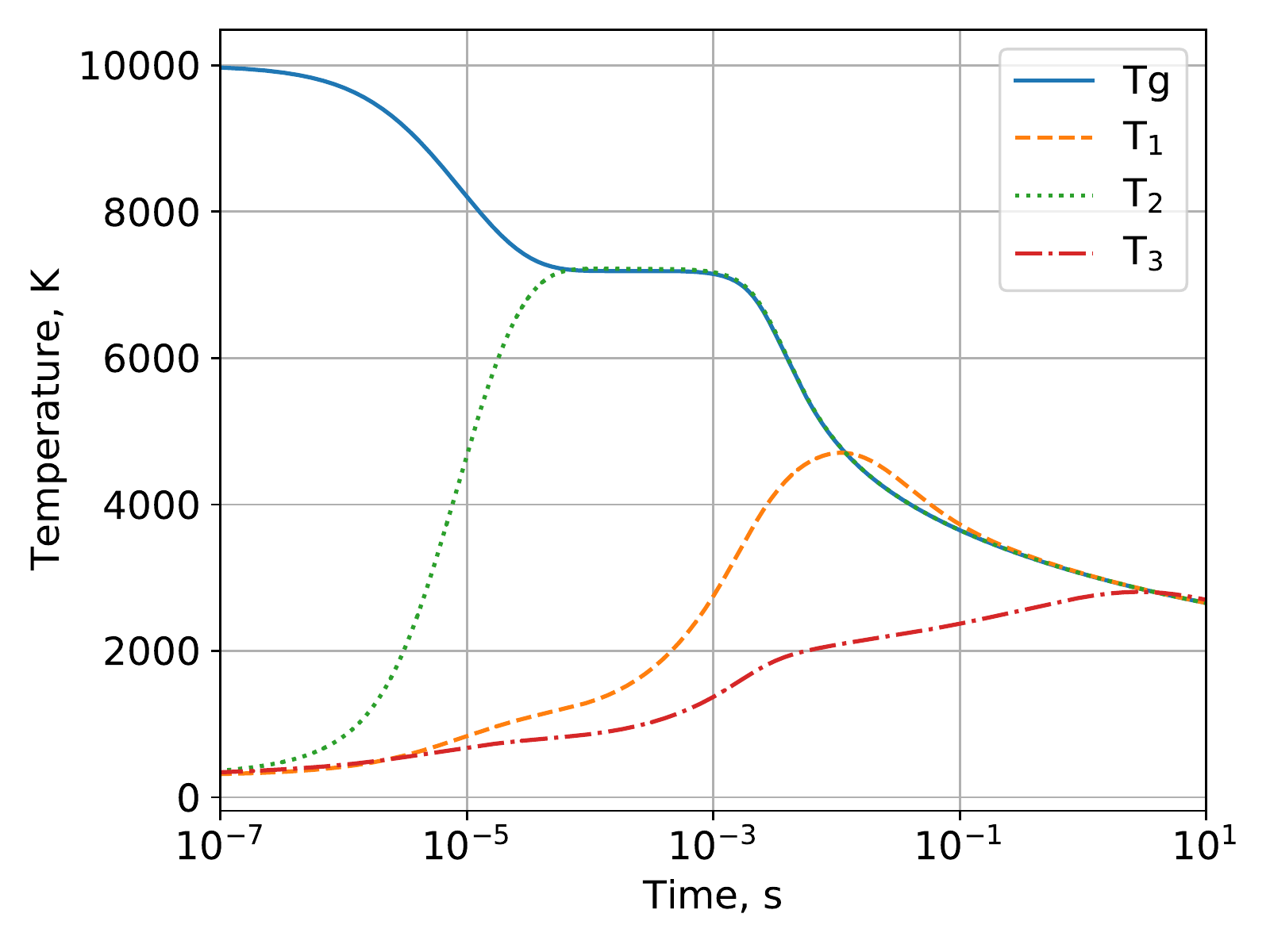}
\caption{Mole fractions (right) and temperatures (left) of \ce{CO2} gas initially at $300$ K and $2000$ Pa suddenly heated to $10,000$ K in a 0D simulation.}
\label{fig:0DCO2relax}
\end{figure}

Another 0D simulation was performed again in a pure \ce{CO2} gas initially at $300$ K and $2,000$ Pa which is suddenly heated to $10,000$ K and allowed to relax and dissociate. The mole fractions are plotted on the left in fig.~\ref{fig:0DCO2relax}. On the right, the gas temperature and the internal temperature obtained from fitting a Boltzmann distribution on the vibrational distribution of each mode of ground \ce{CO2} are presented. There are several features that should already be expected from the discussion of the previous test-case. Firstly, the relative distribution of vibrational modes of \ce{CO2} is in agreement with the distributions in fig.~\ref{fig:0DCO2only}, except now the gas is not isothermal and as such the temperature continuously drops until thermal equilibrium is achieved. Secondly, as seen in fig.~\ref{fig:dissociation}, the exchange reaction \ce{CO2 + O <-> CO + O2} dominates the decomposition of \ce{CO2} as the creation of \ce{CO} and \ce{O2} far outpaces the production of \ce{O} atoms which are created mostly through the VD reactions calculated through the FHO or Rosen--Zener dissociation models. Additionally, the initial temperature decay from $10^{-7}$ to $10^{-5}$ seconds stems mostly from the excitation of the bending mode and the exchange reaction \ce{CO2 + O <-> CO + O2} which is very efficient energy-wise as the gas temperature is mostly constant between $10^{-4}$ to $10^{-3}$ when the aforementioned reaction is most active. Beyond $10^{-3}$ seconds, chemistry becomes the most active process with dissociation of \ce{C2} molecules into \ce{C} atoms. Around $10^{-2}$ seconds \ce{CO2} vibrational modes are in equilibrium with each other.

\subsubsection{\ce{CO2} Recombining flow}

\begin{figure}[ht]
\centering
\includegraphics[width=0.7\linewidth]{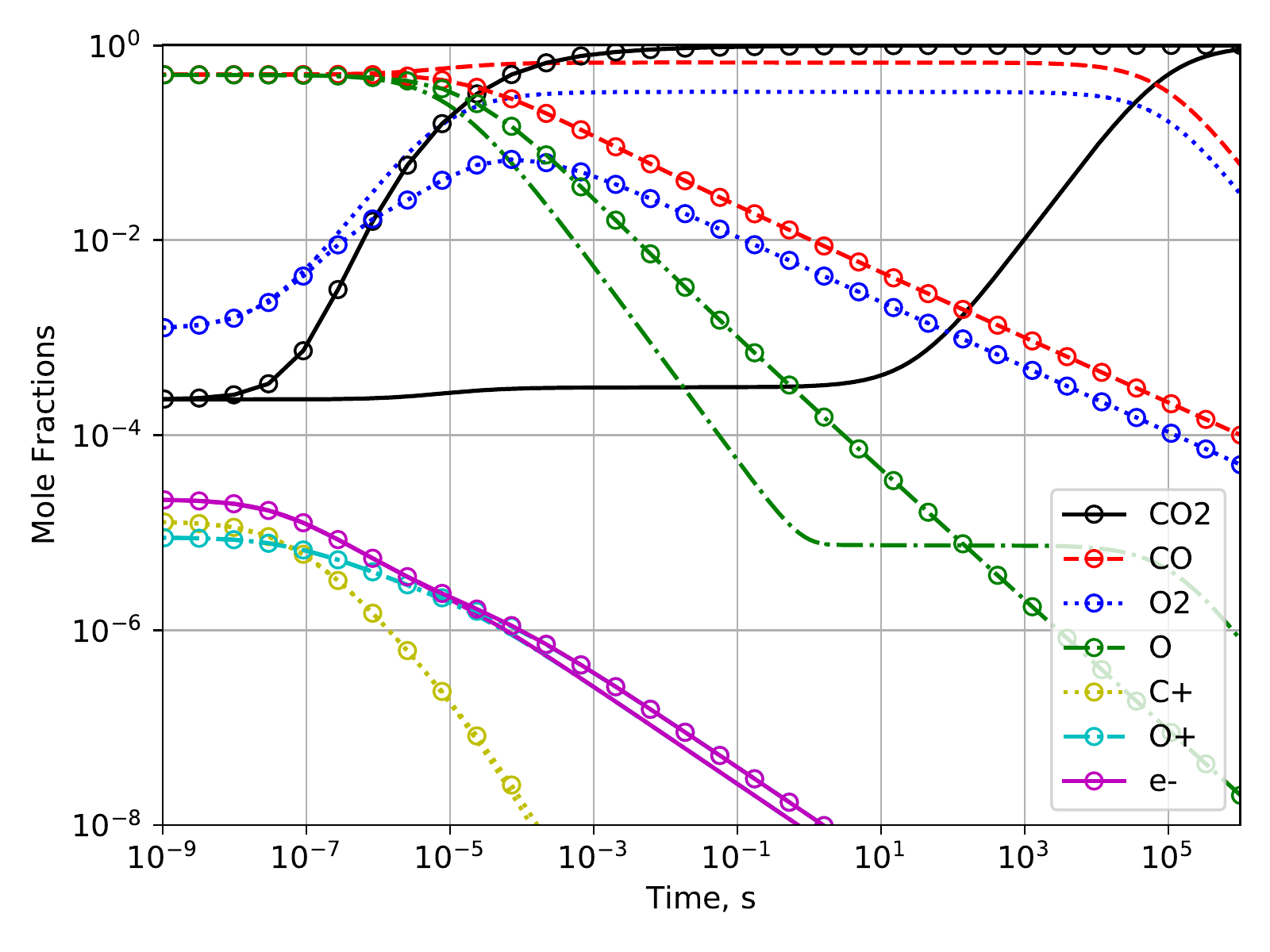}
\caption{Mole fractions for the \ce{C} and \ce{O} atom compounds and ions in a recombining isothermal gas kept at $1,000$ K and $1$ bar in a 0D simulation. The lines with circles correspond to the simulations using Varga \ce{CO2 + O <-> CO + O2} rate, those without circles to the simulations using Sharipov's rate.}
\label{fig:0Drecomb}
\end{figure}

A third and final 0D simulation attempts to capture the recombination dynamics of \ce{CO2}. The equilibrium composition of a \ce{CO2, CO, O2, C2, C, O} gas and their respective ions and electrons was determined at $5,000$ K using the SPARK code. A very similar mixture, to round the sum of all molar fractions to $1$, is determined and left at $1,000$ K and $1$ bar with the isothermal condition enforced. This simulation is performed twice with the \ce{CO2 + O <-> CO + O2} rate by Sharipov \cite{Sharipov2011} or Varga \cite{VargaThesis2017} respectively. This is thus since at $1,000$ K these reactions are over nine orders of magnitude different as can be respectively seen in fig. \ref{fig:exreaction}. It is also worth mentioning that, although most rate coefficients in fig. \ref{fig:exreaction} agree with the rate from Sharipov, the rate from Varga is obtained from a sensitivity analysis of experimental data of combustion experiments. As such, it is not a whimsical comparison but a demonstration of the disparity of the \ce{CO2 + O} reaction rate estimate at low temperatures. The mole fractions of this simulation are plotted in fig.~\ref{fig:0Drecomb} with the simple lines using the Sharipov rate and the lines with circles using the Varga rate. Very quickly, \ce{C} atoms and \ce{C2} molecules disappear from the gas and as such are not displayed in the figure. \ce{C+} atoms recombine slower, \ce{O+} and \ce{e-} take somewhat longer to recombine and their temporal variation is very similar. For the simulation using Sharipov's rate, \ce{O} atoms start recombining into \ce{O2} molecules at around $10^{-8}$ seconds and further combine with \ce{CO} to form \ce{CO2} at $10^{4}$ seconds. At around $10^{1}$ \ce{CO2} starts to recombine and is fully recombined around $10^{6}$ seconds. In contrast, the simulation using Varga's rate shows similar times for the recombination of \ce{O2} and \ce{CO2}. The \ce{O2} molecule concentrations diverge from the Sharipov simulation after $10^{-7}$ seconds to react with \ce{CO} to create more \ce{CO2}. The \ce{O} atoms are consumed more rapidly, either recombining into \ce{O2} or into \ce{CO2}, than \ce{CO} and \ce{O2} molecules recombining into \ce{CO2} which becomes close to fully recombined by $10^{-2}$ seconds. It is clear that the exchange reaction \ce{CO2 + O <-> CO + O2} is an essential mechanism in the recombination of \ce{CO2}. Its importance and the need to obtain accurate estimates of the reaction rate of this process at low temperatures should not be understated.

\subsection{Comparison with shock tube experiments}

\begin{figure}[ht]
\centering
\includegraphics[width=0.7\linewidth]{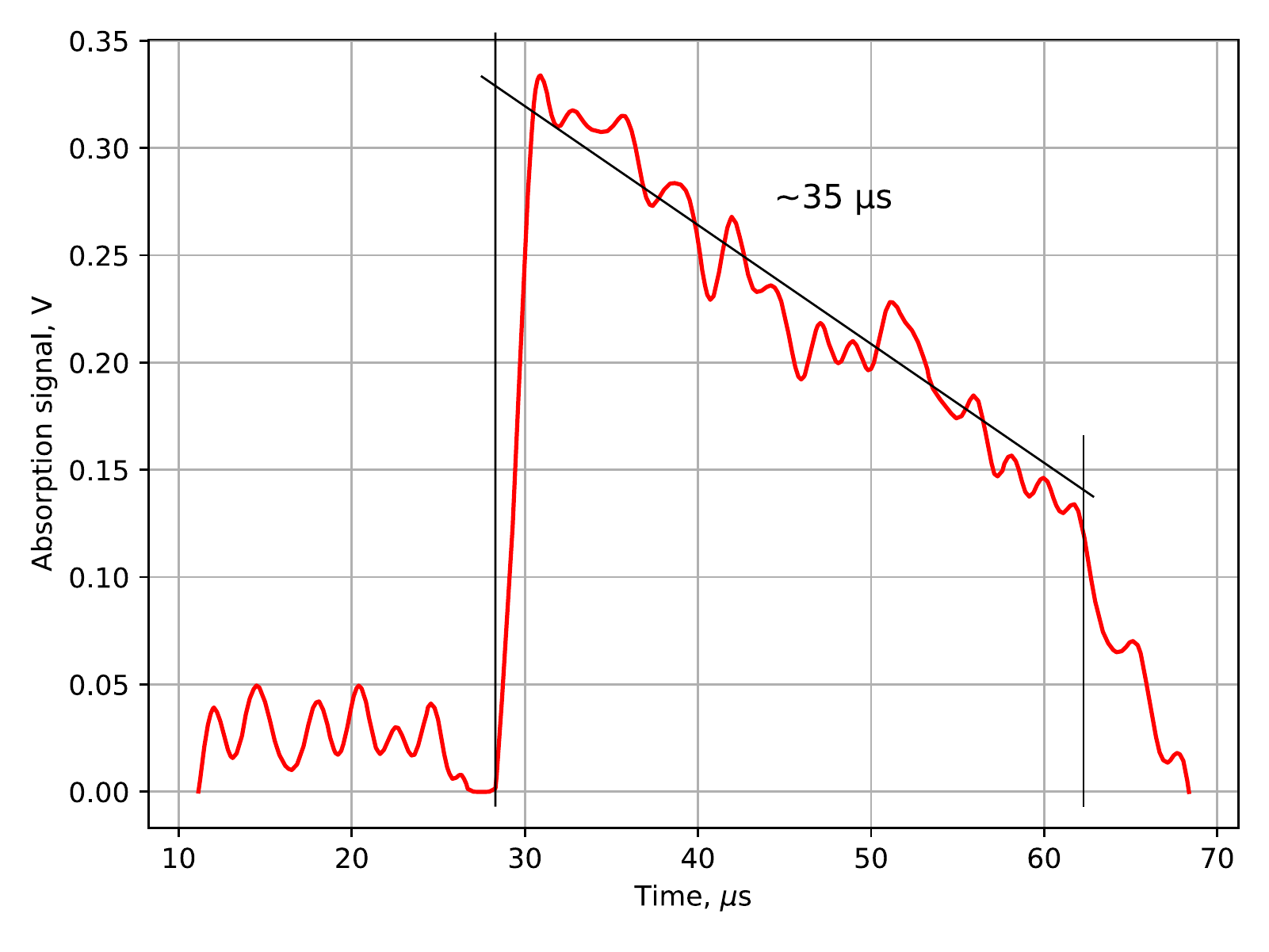}
\caption{Experimental measurement of \ce{CO2} absorbance in a shock tube. The shock velocity was $3,020$ m/s and the pressure in the tube $8.8$ Torr corresponding to the first shot in tab. \ref{tab:shockcharMIPT} and the left-most point in fig. \ref{fig:TimeDecomp}. Here the useful test time lies between 30 $\mu$s with the arrival of the shockwave, and 60 $\mu$s with the arrival of the contact wave. The black lines show how the time of decomposition can be estimated in these measurements.}
\label{fig:MIPTexamp}
\end{figure}

In 2008, a shock tube campaign was carried out in the Moscow Institute of Physics and Technology (MIPT) under a contract from the European Space Agency (ESA) \cite{SupportMIPTReport,ReportBeck2008,Reynier2011}. The objective of the study was to validate the existing CFD tools employed in the design of the EXOMARS mission. Among the employed diagnostics, a mercury lamp was used to measure the absorbance of the flow in the hot \ce{CO2}(B $\rightarrow$ X) UV band around 253.7 nm. This allows the measurement of the evolution of concentration of \ce{CO2} in the ground state and an estimation of the time of decomposition of the shock. Seven shots were performed using this diagnostic as per tab.~\ref{tab:shockcharMIPT} with initial temperatures of $T=300$ K and the test gas fully composed of \ce{CO2}. 1D simulations, using the \ce{CO2} FHO model described in this work, were carried out using the conditions of tab.~\ref{tab:shockcharMIPT} as the upstream conditions. Variations were performed by using the reaction rate sets of the different macroscopic coefficients for \ce{CO2 + O}. In a simulated profile, it is not as much straightforward to define the incubation time as it is in an experimental signal. In this work we have defined the following criterion for computing the incubation time: The dissociation time is the time at which the derivative of the molar fraction of \ce{CO2} is at it's lowest. In other words, the time of decomposition is when the molar fraction changes curvature which represents the moment where the flow is in a quasi-steady-state. In contrast, an example of the experimental measurements performed at MIPT is shown in fig. \ref{fig:MIPTexamp} with an estimation of the time of decomposition. There is a rise in the absorption signal of \ce{CO2} at 30 $\mu$s, which corresponds to the increase of density after the passage of the shockwave. At 60 $\mu$s, the arrival of the contact wave marks the end of the useful flow.

\begin{figure}[ht]
\centering
\includegraphics[width=0.8\linewidth]{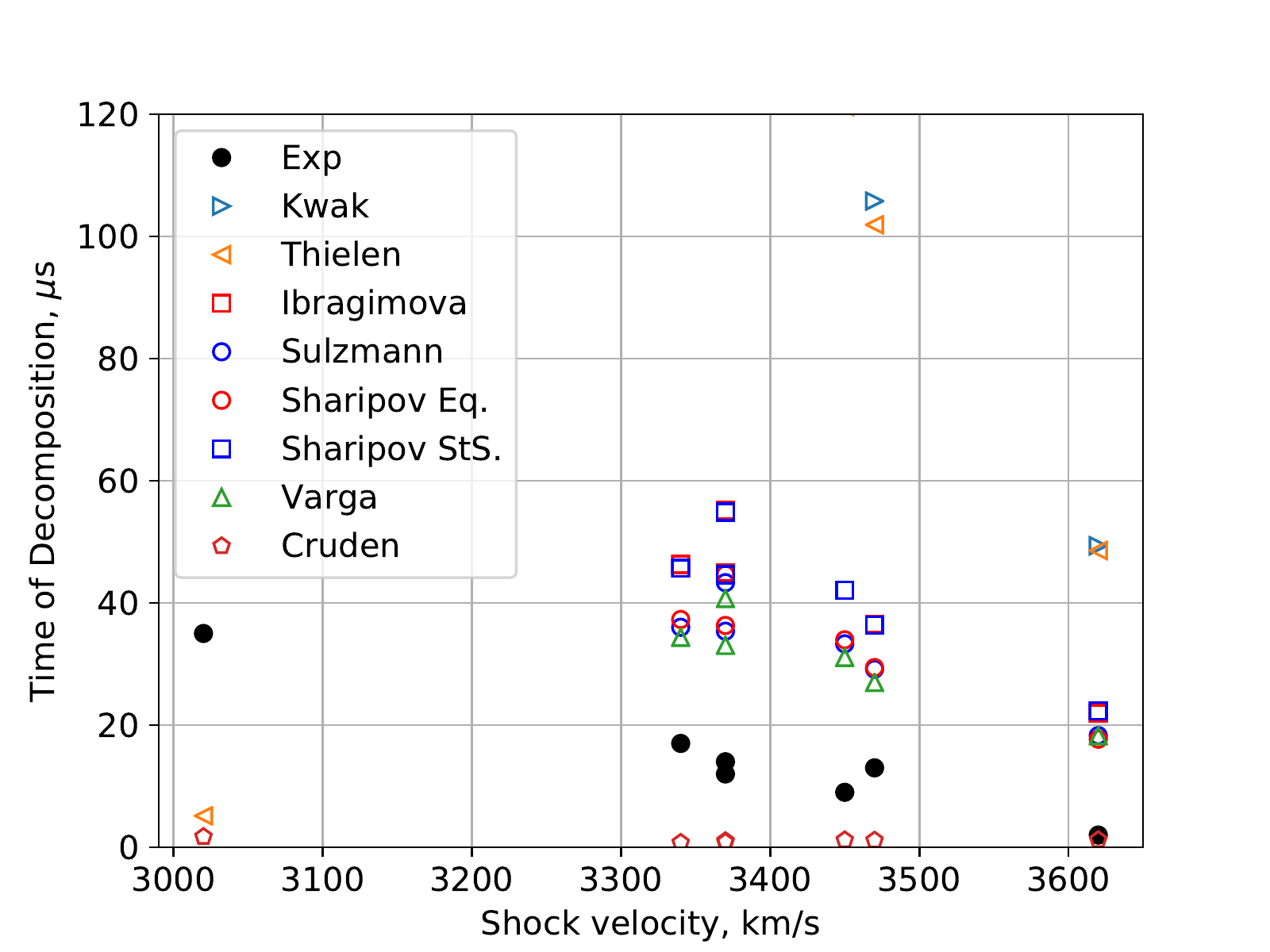}
\caption{Time of decomposition for several shocks in pure \ce{CO2}. Experimental points are plotted in full circles and calculated points with other symbols. A linear relationship between increasing shock speed and decreasing time of decomposition is observed in experimental and calculated points.}
\label{fig:TimeDecomp}
\end{figure}

The results of the simulations can be found in fig.~\ref{fig:TimeDecomp} with experimental points reported against the results of the FHO model with the exchange reactions described in section~\ref{sec:macroreac}. The lowest velocity point, near $3,000$ m/s is overestimated by all exchange rate coefficients except by Thielen. In fact, the simulations with this shock velocity do not have any physical significance as the model does not capture any significant dissociation. Therefore, we will ignore this point for the remainder of this analysis. As the shock velocity increases the time of decomposition decreases in a linear fashion independently of pressure. This trend is observed with whatever exchange reaction is used with the model, the inclination of the trend being what changes with the considered exchange reaction. The best results are achieved with the redistributed state-to-state reactions obtained from Varga \cite{VargaThesis2017} and good estimates are also achieved with the rates obtained from Sulzmannm Ibragimova and Sharipov \cite{Sulzmann1965,Ibragimova1991,Sharipov2011}. The results obtained from Thielen and Kwak \cite{Thielen1983,Kwak2015} do not provide a good estimate but also obtain a linear trend. Also plotted with these models is the macroscopic model presented by Cruden \textit{et al.} in \cite{Cruden2018}. This macroscopic model underestimates the time of decomposition and does not determine the same linear trend that is expected from the experiments and other models. This may be due more to the underlying assumption of a Boltzmann distribution for the internal states, rather than any inherent inadequacy of the proposed macroscopic kinetic rates.

\begin{table}
\centering
\caption{Shock characteristics of the experimental campaign carried out in MIPT.}
\label{tab:shockcharMIPT}
\footnotesize
\begin{tabular}{cccccccc}
\hline
Shot \# & 1 & 2 & 3 & 4 & 5 & 6 & 7 \\ \hline
Pressure (Torr) & 8.8 & 9.0 & 6.5 & 8.0 & 5.9 & 6.2 & 5.6 \\
Velocity (m/s) & 3,020 & 3,340 & 3,370 & 3,370 & 3,450 & 3,470 & 3,620 \\ 
Incubation Time ($\mu$s) & 35 & 17 & 12 & 14 & 9 & 13 & 2 \\ \hline
\end{tabular}
\end{table}

\section{Discussion}
\label{sec:discussion}

We have presented a vibrational state to state kinetic model for \ce{CO2} with a number of improvements over current state of the art models. Firstly, the Forced Harmonic Oscillator theory has been for the first time extended to linear triatomic molecules such as \ce{CO2} and deployed in a complete and self-contained state-to-state kinetic model tailored for heavy-impact reactions. This is a significant improvement compared to the traditional SSH scaling laws (discussed in section \ref{sec:stateoftheartlowT}). Instead of carrying a scaling of experimental rates, we estimate the different intermolecular potentials that approximates such experimentally-determined rates and then applied it over the whole vibrational level energies manifold. Furthermore, by using the general FHO theory instead of First Order Perturbation Theories such as the SSH model, we avoid obtaining transition probabilities above one in the high-temperature regime, and in the near-dissociation limit, where energy spacings become ever smaller. These are well known shortcomings of the SSH theory \cite{LinoDaSilva2007}. More importantly, we may now account for multiquantum transitions, which are well known to become important in higher temperature regimes \cite{LinoDaSilva2007}.

A second improvement against the current status-quo is provided by the detailed discussion on the crossings between the ground and electronic states of \ce{CO2} which is treated in a much more consistent fashion that in past works, which make the more na\"{i}ve assumption of an assymetric mode crossing from the linear ground-state configuration of \ce{CO2} to a near-dissociative vibrational level 70$^{\circ}$ bent excited state configuration around 5.56 eV. Instead, from the analysis of published PES, we find that PES crossings may instead arise from the bent v$_2$ mode at 4.99 eV, or trough the assymetric stretch mode v$_3$ at 5.85 eV. In this last case the excited state is repulsive and the \ce{CO2} molecule is allowed to dissociate immediately after the crossing.

Ultimately these two improvements are overshadowed by the findings that the \ce{CO2 + O <-> CO + O2} reaction is quintessential to both the dissociation and recombination dynamics of \ce{CO2}. Indeed, it has been found that without the inclusion of this rate, direct dissociation and recombination processes of \ce{CO2} become unrealistically slow, hinting that dissociation for \ce{CO2} is likely to follow a two-step process where direct dissociation of \ce{CO2} creates a first batch of atomic oxygen atoms \ce{O} which then induce further decomposition of \ce{CO2} into \ce{CO} and \ce{O2} products. This is consistent with the findings of our literature review in section \ref{sec:stateofthearthighT}.

\subsection{Possible short-term model developments}

There are certainly shortcomings in our model, which may immediately be pointed out. Nevertheless there are none which weren't also present in all previous models relying on simplified FOPT theories like SSH. Some of these flaws can be addressed in future works or updates to this database. 

\subsubsection{Inclusion of different collision partners and state-to-state kinetics for diatomic molecules}

The vibrational-translational effects in \ce{CO} and \ce{O2} atoms could be added for a further sophistication of the model. These rates have been reasonably well modeled in the past \cite{Adamovich1998,LinodaSilva2008b}. Other simple improvements would be to account for collisional partners other than \ce{CO2} in our model. This would require a straightforward review and calibration of the FHO model against further experimental rates, in the exact same fashion that was carried out in the development of our STELLAR-\ce{CO2} database \cite{STELLAR}. Nevertheless this will at most lead to a 5-fold increase of the number of rates in the model, so the real impact of these quality-of-life improvements in the model will have to be carefully weighted against its computational overhead.

\subsubsection{Improved modeling of intermode vibrational transitions}

The key shortcoming is undoubtedly the assumption for complete separability of the vibrational modes of \ce{CO2}. To include non-separated vibrational levels of \ce{CO2} into this model several obstacles have to be lifted. Firstly, a manifold of mixed levels must be found. This has been done partially in \cite{Vargas2020} where a vibrationally specific level database for \ce{CO2(X)} has been determined which reproduces the partition function of \ce{CO2} by direct summation up to $4000$ K within less than $1\%$ error. For \ce{CO2}($^{3}$B$_{2}$) the same polynomial expression determined in this work will have to suffice while novel PES-based works are not made available. Secondly, the reduced masses of non-separated modes have to be determined. We could overcome this difficulty by writing the reactions in such a way that only a single mode is interacting at any given reaction. These reactions could be written (as an example for symmetric stretch) thus \ce{CO2(\text{v}_1,\text{v}_2,\text{v}_3) + CO2(\text{v}_1 + 1, 0, 0) <-> CO2(\text{v}_1 + 1 ,\text{v}_2,\text{v}_3) + CO2(\text{v}_1 , 0, 0)}. While limiting, in the sense that only single quantum jumps can be modelled this way, it might become a first step in building a kinetic state to state model with no mode separability.

\subsubsection{Accounting for resonant processes}

By assuming full separability of the vibrational modes of \ce{CO2} we neglect a series of so-called \textit{accidental resonances} that arise in \ce{CO2} and which may effectively redistribute vibrational energy between the different modes of \ce{CO2}. However, even considering the \ce{CO2} \textit{extreme states}, one may find out that a quite large number of these states might be near-resonant (with energy spacings below 100 cm$^{-1}$). One may apply a simplified Landau--Teller model to yield an additional set of near-resonant rates and deploy these in our model, according to the expression \cite{Landau1969,Nevdahk2003,STELLAR}:

\begin{equation}
\textrm{K}_{res}'(T,\Delta E_v[\textrm{cm}^{-1}])=1.57\times10^{-11}\exp\left\{\left[-\frac{(\Delta E_v)^2}{T}\right]^{1/3}\right\}[\textrm{cm}^{3}/\textrm{part.}/\textrm{s}]
\label{eq:Landau-Teller1}
\end{equation}

While certainly not sufficient for accounting the full effects of these accidental resonance effects, this may be a better approximation that neglecting those processes altogether.

\subsubsection{Inclusion of radiative processes}

\begin{figure}[ht]
\centering
\includegraphics[width=0.7\linewidth]{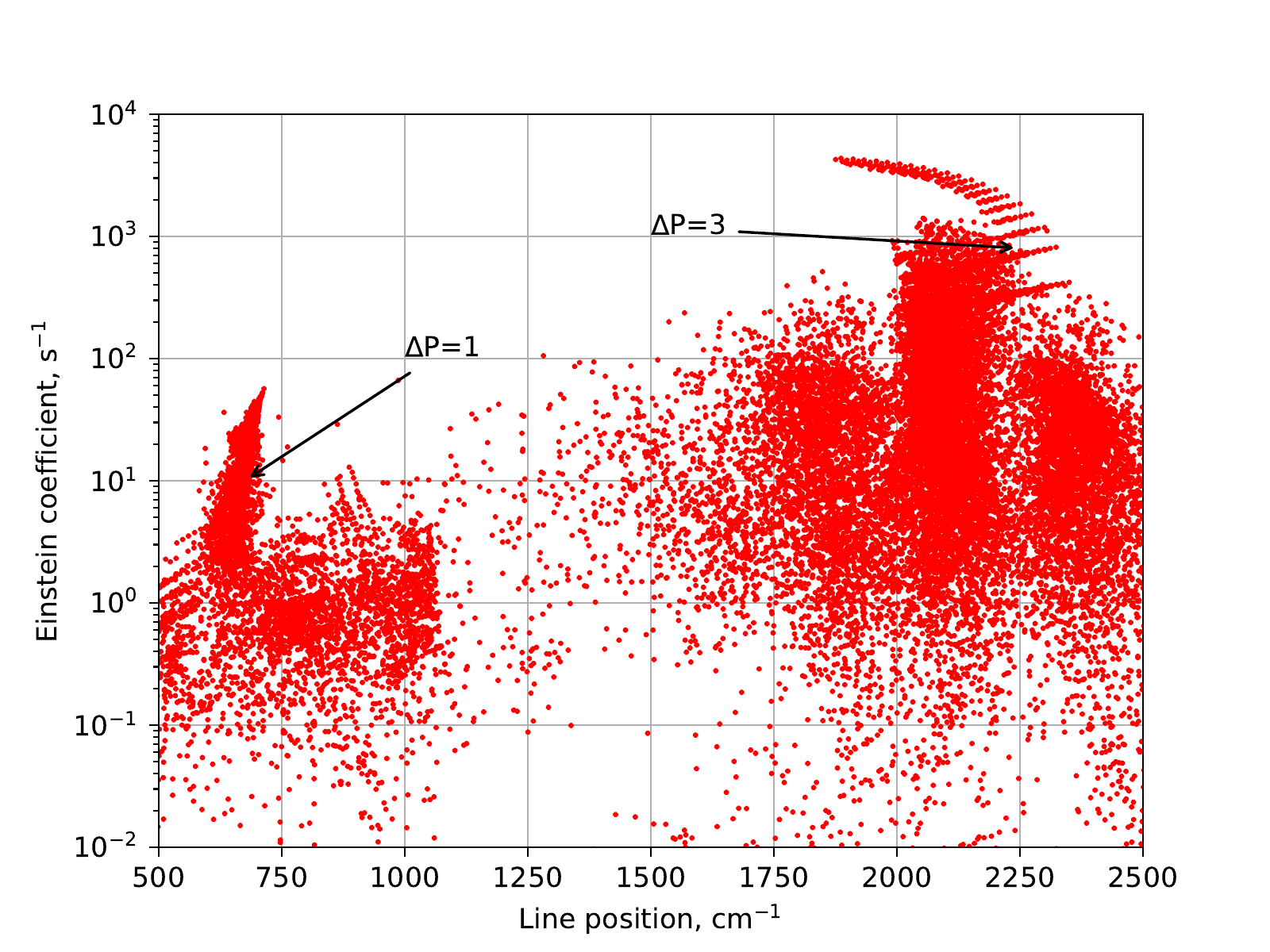}
\caption{Line positions and respective Einstein coefficients of vibrational bands of \ce{CO2} IR radiation. This figure focuses on the $15$ and $4.3$ $\mu$m regions of the \ce{CO2} IR spectra located around 600 and 2400 cm$^{-1}$ respectively. The arrows labeled $\Delta$P$=1$ and $\Delta$P$=3$ point towards the regions where polyad exchanges of that number are most common. The vibrational frequencies of \ce{CO2} are related by $\omega_2\approx\omega_3/3\approx\omega_1/2$, and a pseudo-vibrational number may be written from the sum of the vibrational numbers from the so-called polyad, P$=2\text{v}_1+\text{v}_2+3\text{v}_3$. Vibrational bands can be organized in the spectrum by their polyad variation as in this figure. The $\Delta$P$=1$ and $\Delta$P$=3$ regions indicate a radiative decay of $\Delta$v$_1$ and $\Delta$v$_3$ respectively. Particularly the $\Delta$v$_3$ decay might be an important inclusion in the continuity equation, owing to its high Einstein coefficients.}
\label{fig:posvsA}
\end{figure}

The inclusion of radiation of \ce{CO2} may also be achieved by using the vibrationally-specific database developed in~\cite{Vargas2020}. The database contains vibrational state-to-state Einstein coefficients which could be used by simply adding to the continuity equation the term $-A_{v^{\prime}v^{\prime\prime}}N_{v^{\prime}}$, which accounts for levels de-excitation through radiation losses. This would effectively model the escape of radiation from the gas. This application however, is not so straightforward as one might think. The use of extreme states in this work implies some sort of adaptation for the published Einstein coefficients that could be achieved by some sort of binning. Even after binning and distribution of spontaneous emission effects on the vibrational ladder, important Einstein coefficients of IR radiation in \ce{CO2} are in the order of $A_{v^{\prime}v^{\prime\prime}}\approx10^3$ s$^{-1}$ as in fig.~\ref{fig:posvsA}. The time-scale of interest for spontaneous emission effects is therefore higher than the vibrational kinetics time scale. As such, the inclusion of radiation might be moot for a lot of applications but significant in others. Appropriate methods for equating these general Einstein coefficients to the vibrational modes separability is currently underway.

\subsubsection{Uncertainty in the \ce{CO2 + O <-> CO + O2} process}

As discussed in previously, the \ce{CO2 + O <-> CO + O2} reaction is essential for the decomposition of \ce{CO2} and might as well also be critical in it's recombination through the inverse process. The rates plotted in fig.~\ref{fig:exreaction} are in qualitative agreement over $8,000$ K but this agreement does not exist in lower temperature ranges. Most reactions surveyed in this work, both experimental and numerical are in qualitative agreement in the $1,000$ to $5,000$ K range. Only two rates estimate a much higher rate coefficient between $1,000$ to $5,000$ K, the Sulzmann \cite{Sulzmann1965} and Varga \cite{VargaThesis2017} rates both experimental. At first one might consider these rates as outliers and trust the cluster of results that are in agreement. However, the rate obtained by Varga was obtained by numerical optimization of 39 datapoints and is reported to have low uncertainty. As such the relative agreement at lower temperatures for most of the reaction rates cannot be taken for granted. With a difference of over $9$ orders of magnitude to other rates at $1,000$ K, it may be that this reaction has been underestimated by several authors in the low-temperature ranges. Given its importance to \ce{CO2} chemical-kinetic processes, the emphasis on newer studies to obtain better estimates for this reaction should not be understated. Needless to say, a good quality PES-based quantum-chemistry analysis on this Zeldovich process would significantly advance the insight on \ce{CO2} dissociation processes.

\begin{figure}[ht]
\centering
\includegraphics[width=0.7\linewidth]{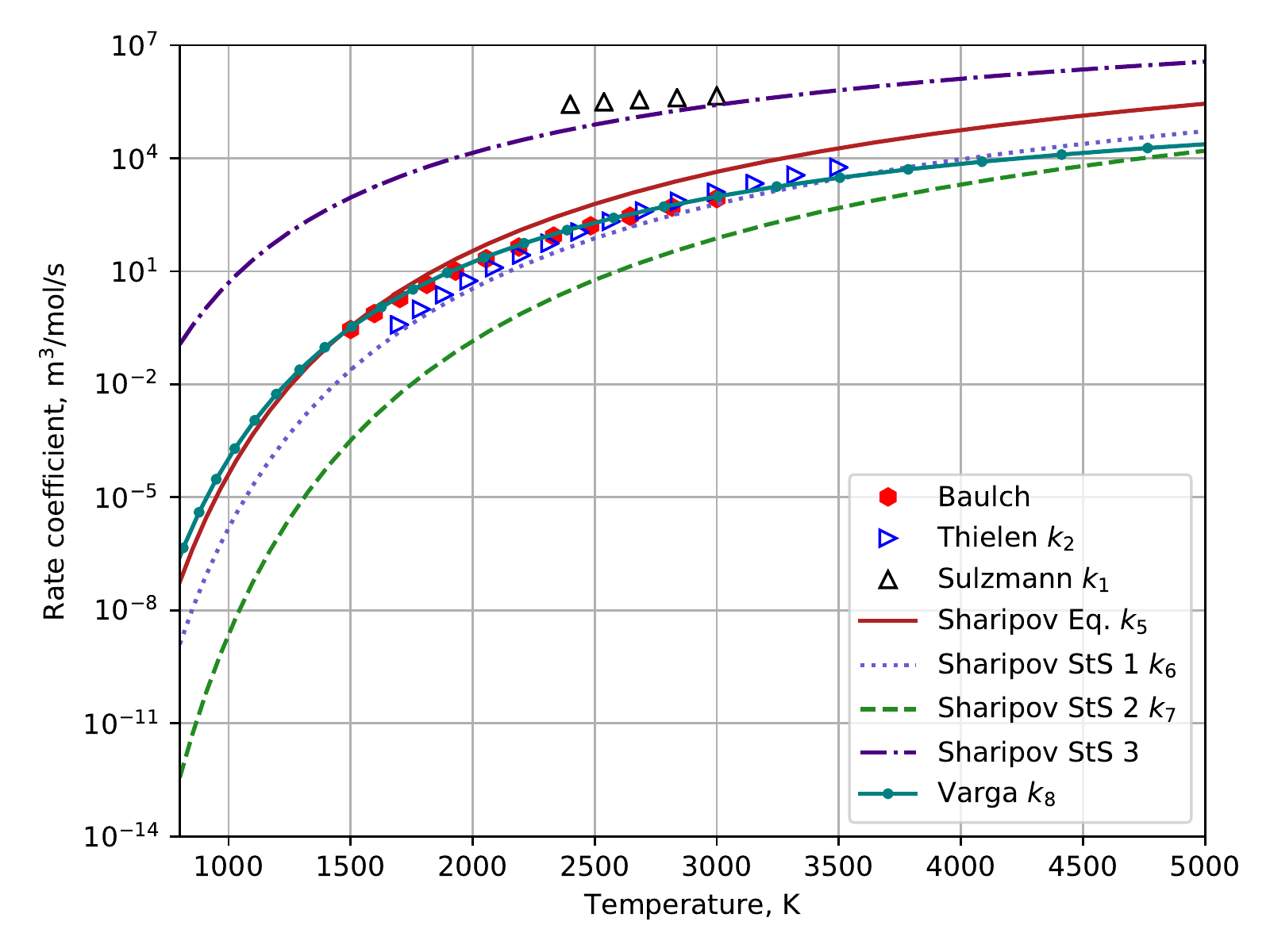}
\caption{Reaction rate coefficients of the reaction \ce{CO + O2 <-> CO2 + O}}
\label{fig:excinverse}
\end{figure}

Furthermore, the inverse reaction \ce{CO + O2 <-> CO2 + O} might hold the key for the recombination of \ce{CO2} at high-temperatures. In figure~\ref{fig:excinverse} the rate coefficients for the \ce{CO + O2 <-> CO2 + O} reaction are plotted. Reactions $k_1$, $k_2$ and $k_{5-8}$ of table~\ref{tab:excratereview} are displayed along with the one reported by Baulch \textit{et al.} \cite{Baulch1976} and a third STS reaction calculated by Sharipov \cite{Sharipov2011}. The STS reactions by Sharipov are, ordered:
\begin{align}
& \ce{CO + O2(X $^3$\Sigma^-_g) <-> TS_1 <-> CO2 + O( $^3$P)} \label{eq:Sharipov1} \\
& \ce{CO + O2(X $^3$\Sigma^-_g) <-> TS_2 <-> CO2 + O( $^1$D)}\text{ or }\ce{O($^3$P)} \label{eq:Sharipov2} \\
& \ce{CO + O2(a $^1$\Delta g) <-> CO2 + O( $^1$D)} \label{eq:Sharipov3}
\end{align}
where the TS$_1$ and TS$_2$ are two pathways in the same collision, one leads to \ce{CO2 + O( $^3$P)} and the other has a 50:50 branching towards \ce{CO2 + O( $^3$P)} and \ce{CO2 + O( $^1$D)}. The third reaction was not considered for this work as it involved the metastable \ce{O2(a)} state, the inclusion of which would make the model even more intricate. Additionally, the $T^n$ dependency on the provided rate would exceed the collisional rate violating physical consistency. But as may be seen in figure~\ref{fig:excinverse}, the third Sharipov STS rate coefficient (eq. \ref{eq:Sharipov3}) is 6 orders magnitude greater than the other rates at $1,000$ K. Furthermore, between $2,000$ and $3,000$ K this rate is 3 to 2 orders of magnitude greater than the experimental rates measured by Baulch and Thielen and the rate determined by Varga\footnote{The inversion of rates $k_1$, $k_2$ were made by other authors. $k_1$ was performed via an Arrhenius rate for the equilibrium constant \cite{Schofield1967}. Park \cite{Park1994} does not provide an inversion method for $k_2$. Rates $k_{5-8}$ inversion were performed self-consistently through the partition function in the in-house code SPARK. As such it is strange that the Varga rate is now agreeing with the experimental rates measured by Baulch and Thielen. The inversion methods should be verified and checked for consistency in all cases.}, almost reaching a rate as high as Sulzmann's measurement. The metastability of the \ce{O2(a)} state and the greater reaction rate coefficient makes an compelling case for \ce{CO2} recombination which occurs in the wakeflow of spacecraft entering Mars at a tipycal temperature of $2,000$ K. Additionally, the creation of excited \ce{O( $^1$D)} atoms will redistribute the energy to the ro-vibrational modes of \ce{CO2} as discussed in the review of Fox \& H\'{a}c \cite{Fox2018}. As such the study of the \ce{CO + O2(a $^1$\Delta g)} collision is a strong candidate for further studies. In the short term, a vibronically-specific \ce{O2} description may be included in the current model \cite{LinodaSilva2012}, henceforth extending it to include the specific treatment to the (a$^1\Delta g$) state. This will undoubtedly increase the complexity of the model but not by too much, with at least $1,500+$ reactions added in.

\subsubsection{Reduced order methods}

Despite the inescapable assumption of full separability of the vibrational modes the fact still remains that our vibrationally-specific kinetic model yields an excess of 20,000 kinetic rates. This is simply not tractable for anything else than simplified 0D and 1D geometries. Therefore, the application of reduced order models appears as a natural next step for further deployments of this kinetic scheme. Interestingly, such numerical schemes have been actively discussed in the recent years, with major strides advancing the theoretical state-of-the-art. Further, as these rely on a good quality set of kinetic data, the model in this work appears almost as tailor-made for the application of a reduced-order model as an additional layer. Two approaches are at the forefront of these reduced-order methods: Binning and Fokker--Planck methods

\begin{itemize}
\item \textbf{Binning methods}: Binning is the process in which levels and processes are bundled together thereby reducing the number of de-facto individual states to be tracked in a state-to-state simulation making it more alike a macroscopic chemistry model. Some strategies and binning algorithms are described in \cite{Liu2015,Sahai2019,SahaiThesis}. To build binned levels and processes, a state-to-state model has to be given as input. In Sahai's thesis \cite{SahaiThesis} a \ce{CO2} SSH based model with over $9,000$ levels (representing $9,000$+ ODEs to be solved) was reduced using a binning strategy to obtain a good match with state-to-state results using as few as $30$ bins. The same methods have also been applied to \ce{N2} system with over $9300$ ro-vibrational levels with similar results as \ce{CO2}. Such a reduction of problem complexity is essential to the application of state-to-state databases to spatial problems of higher order such as 2D and 3D simulations.
\item \textbf{Fokker--Planck methods}: Another method to obtain the vibrational distribution function of \ce{CO2} is to replace the master equation by a drift-diffusion Fokker-Planck equation. This approach was already being developed in the 1970s and 80s \cite{Rusanov1979,Rusanov1981} and has been recently applied to \ce{CO2} by a collaboration of authors \cite{Diomede2017,Diomede2018,Longo2019,Viegas2019}. Instead of treating the levels of \ce{CO2} as discrete, such as in the state-to-state approach, the states effectively become a continuum. This is a good approximation when levels may effectively be approximated to a continuum as is the case at the onset of \textit{vibrational chaos} for \ce{CO2}. Again, this method does not preclude the estimation of state-to-state rates for higher-levels as these are still necessary in determining coefficients necessary to solve the Fokker-Planck equation. However once more this approximation is useful in reducing computational time, as it has been reported to be up to 1000 times faster than state-to-state simulations \cite{Viegas2019}.
\end{itemize}

\subsection{Further steps beyond the state-of-the-art}

It should be apparent by now that the authors do not consider the topic for this paper to be closed by the improvements of this work, no matter how meritorious these may be. Instead one may argue that this work exhausts the more simplified venues for modeling a complex polyatomic molecule such as \ce{CO2}. Although the expected theoretical trends are found for dissociative and recombinating flows, and although quite reasonable agreement is found against experimentally available shock-tube data, it may be well the case that further progress in the accuracy and predictability of vibrationally-specific  \ce{CO2} models may only be achieved if a \textit{tabula-rasa} is made of the current approaches, and new theoretical descriptions for \ce{CO2} are brought forward. This is particularly critical for a better understanding of high-vibrational levels and dissociation dynamics in triatomic molecules such as \ce{CO2}.

Insight may be found outside of the more traditional views from the plasma chemistry community, which are to our opinion based on a mindset which excessively relies on past successes regarding the modeling of dynamic processes in atomic and diatomic molecules. Instead we may shortly review recent quantum-chemistry works regarding the structure of higher-lying levels of polyatomic molecules.

\subsubsection{Polyads}

We may start by discussing the concept of polyads, which is a useful concept when building spectroscopic Hamiltonians. There are several ways for defining the polyad but the most common is the so-called "$213$" defined thus
\begin{equation}
P_{213}=2\text{v}_1+\text{v}_2+3\text{v}_3.
\end{equation}
The polyad number $P$ may then be treated as a quantum number.

\begin{figure}
\centering
\includegraphics[width=0.7\textwidth]{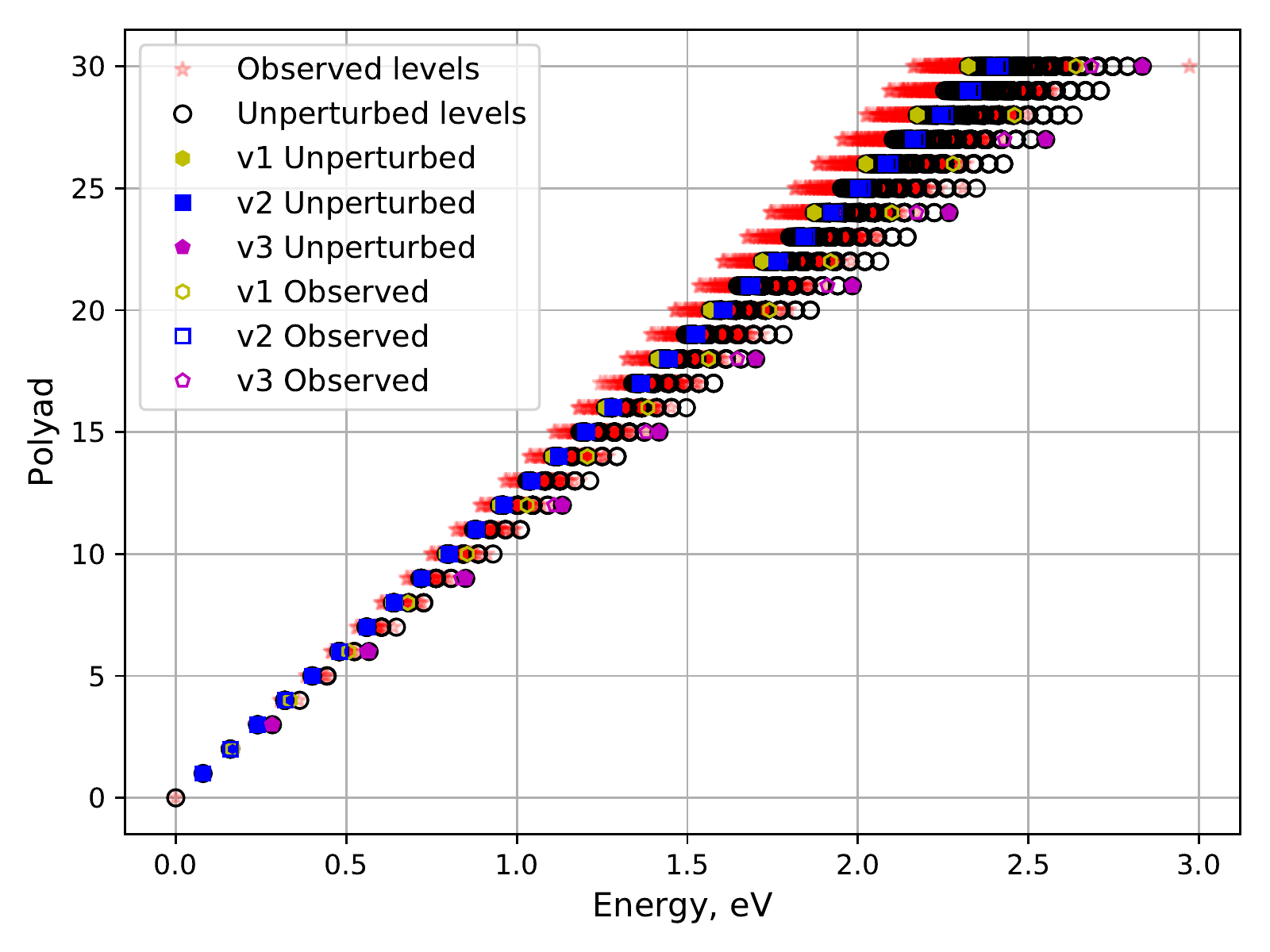}
\caption{Energy levels of \ce{CO2} in eV versus the polyad $P$ for each level. Here the polyad of a level is computed as $P=2\text{v}_1+\text{v}_2+3\text{v}_3$. This figure plots both observed levels in faded stars as well as the unperturbed levels of Ch\'{e}din's polynomial \cite{Chedin1979}. Furthermore it highlights the extreme cases of pure v$_1$, v$_2$ and v$_3$ levels as used in this work, the perturbed and unperturbed values. This figure serves as an illustration of \textit{vibrational chaos} whereas for a single high-enough energy, several polyads are crossed effectively meaning several combinations of v$_1$, v$_2$ and v$_3$ sharing the same symmetry or not have a similar energy. The energy need not be very high, at $1$ eV polyads $11$, $12$ and $13$ are crossed. The dissociation limit for the assymetric mode of the ground electronic state of \ce{CO2} is $7.42$ eV and the chaos will only increase with higher energy up to dissociation.}
\label{fig:vibchaos}
 \end{figure}

For higher poliad numbers P, we will observe a larger scattering of the energies for each individual $P+1$ levels in the polyad. This is shown in Fig. \ref{fig:vibchaos} where the energies of the different polyads are plotted against their quantum numbers, both for the perturbed (observed) levels of \ce{CO2}, and for the unperturbed levels (using Chedin's Hamiltonian\cite{Chedin1984}).
 As early as $P=6-7$, we start witnessing overlapping between the vibrational energies of adjacent polyads, up to overlapping from 4 different polyads for $P=30$. This means that the grouping of levels in "lumped" resonant levels (polyads) starts failing \footnote{for this particular $P_{213}$ polyad scheme} as soon as for about 10\% of the effective dissociation of \ce{CO2} (the $^3$B$_2$ level dissociation limit).

 In addition, the molecular dynamics for the different levels of a polyad might differ significantly, potentially affecting the transition probabilities in a collision. Kellman \cite{Kellman1986} carried out an analysis for the \ce{CO2} molecule semiclassic dynamics up to $P=8$, with an emphasis on the 2:1 Fermi resonance, using an algebraic resonance dynamics method. Trajectories for each individual quantum level were obtained in action-angle phase space, showing a varied structure. Although the strong Fermi resonance leads to the deviation of the molecule dynamics from its zero-order characteristics, as early as P=1, and up to P=8, it is also observed that different types of dynamics (so-called Type II, III, and IV) are observed for the internal quantum levels of a polyad. Then, it is not a very large leap to question to which degree collisional dynamics may differ inside a polyad, and whether it is possible to lump the polyad levels for kinetic modeling applications. This highlights how much care should be brought into defining appropriate polyad conventions, capable of achieving more or less regular internal dyamics of the molecule.

For \ce{CO2}, different polyads have been proposed and successfully applied to the interpretation of \ce{CO2} spectroscopic data. These include $P_{214}=2\text{v}_1+\text{v}_2+4\text{v}_3$, $P_{213}=2\text{v}_1+\text{v}_2+3\text{v}_3$, and $P_{212}=2\text{v}_1+\text{v}_2+2\text{v}_3$. A recent analysis by Bermudez--Monta\~{n}a et al. \cite{Bermudez-Montana2017} shows that polyad-preserving Hamiltonians based on these three notations achieve an excellent fit of the available experimental data for the main isotopologue of \ce{CO2} ($^{12}$C$^{16}$O), which goes up to 3.3 eV (26,550cm$^{-1}$). No clear distinction on the quality of the fit for each Hamiltonian was obtained, and hence no specific polyad scheme was recommended by the authors.

The limits of validity for these fitted Hamiltonians corresponds to 56\% for the crossing energy to the $^3$B$_2$ state, and 44\% of the dissociation energy for the assymetric stretch potential for the ground state. It is unlikely for these polyad-preserving Hamiltonians to faithfully reproduce highly, near-dissociative levels, where anharmonic terms increase significantly. Yet, as no valid measurements or PES models exist for these energy regions of \ce{CO2}, this statement remains somewhat speculative. It might well be possible that the inclusion of polyad-breaking interactions, using Van Vleck perturbation theory \cite{Sanchez-Castellanos2008} may be needed; or that phase-space Hamiltonians may prove more adequate in this energy region. Clearly there is a strong necessity for more theoretical studies on the near-dissociative levels of \ce{CO2}.

We note that this concept has also been discussed by Fridman \cite{Fridman2008} in a simplified way, wherein he considers a transition to a \textit{vibrational quasi-continuum} once the energy defect of the assymetric mode, compared to the symmetric and bending modes, is ``filled'' due to anharmonicity effects, leading to additional resonance effects. This transition occurs in general for

\begin{equation}
x_{as}v_{a}v_{s}\geq\Delta\omega
\end{equation}

where $x_{as}\simeq12$ cm$^{-1}$ is the assymetric-symmetric anharmonicity coefficient; $v_{a}=\text{v}_3$; $v_{s}=2\text{v}_1+\text{v}_2$; and $\Delta\omega=\Delta\omega_3\simeq300$ cm$^{-1}$.

According to this expression, we find that the energies corresponding to the critical $v_{a}$ and $v_{s}$ quantum numbers are in the $1.5$-$2.5$ eV range. This is consistent with the observed results of fig. \ref{fig:vibchaos} where different polyads start to mix energy-wise as soon as $1$ eV. Fridman does note that the onset of vibrational chaos may be delayed up to higher energies, close to the dissociation limit, as long as the assymetric vibrational temperature is higher than the symmetric/bending one ($T_{va}\gg T_{vs}$). However we have found that in heavy-impact dominated flows such as shockwaves, with $T\gg T_{v}$, this is hardly the case, where the bending mode temperature increases the fastest, owing to the low energy spacings between adjacent v$_2$ levels, with v$_1$ lagging behind, and v$_3$ even lagging further behind, with no appreciable plateau observed in the $vdf$ due to V--V--T transitions. The situation might be different for electron-impact dominated flows such as gas discharges, however this is outside the scope of our work. Still, as accidental resonances between the v$_3$ level and the other v$_1$ and v$_2$ modes may arise as early as $1$--$1.5$ eV, the possibility for resonant transitions from v$_3$ towards the other modes counteracting the so-called \textit{ladder-climbing} V--V--T processes, and therefore depleting the higher v$_3$ levels, make this a challenging proposition.

\subsubsection{Vibrational chaos}

In the last decades, polyad schemes in polyatomic molecules have been extensively discussed in relation to the concept of \textit{vibrational chaos}, which has been brought forward for improving the description of high molecular vibrations, with some success stories such as the modeling of higher vibrational motions of acetylene \cite{Kellman2007}.

The theory of \textit{vibrational chaos} starting point is the set of classical normal modes in the near-equilibrium configuration of polyatomic molecules ($ss$, $be$, and $as$ modes of \ce{CO2}). The theory focuses on how new vibrational modes are born in bifurcations, or branchings, of these original low-energy normal modes. Original modes may still persist at these higher energies, but always in an altered form. Once a set of polyads for each energy space of \ce{CO2} is agreed upon, a physically accurate kinetic model for \ce{CO2} would necessarily proceed in a layered way: The departure point for the first electronvolt of energies would be based on the classical modes of \ce{CO2}, considering the extreme states in a similar way to the model proposed in this work (this model could even include the Fermi resonance binning so dear to many authors). Then, subsequent layers of kinetic models would follow the polyad schemes more adequate for each energy stage, with kinetic rates which allow energy flow inside a polyad, and block it among different polyads \cite{Kellman2007}. This work remains to be done, starting with a better description of the near-dissociative dynamics of \ce{CO2}, a challenge that can only be tackled by physical chemists.

An appropriate example may be found in the treatment of \ce{CS2}, which is a linear triatomic molecule such as \ce{CO2}. It is particularly suited to the study of \textit{vibrational chaos} since it is molecule with heavy atoms (\ce{S}) and as such has soft vibrational modes \cite{Pique1991} (v$_1= ^{-1}$, v$_2=398$ cm$^{-1}$, v$_3=1559$ cm$^{-1}$, the \ce{CO2} modes being, respectively 1334, 667 and 2349 cm$^{-1}$). Other characteristics that make \ce{CS2} a very good candidate for studying \textit{vibrational chaos} effects is the presence of strong Fermi resonances and anharmonicities, as well as the absence of low-lying excited electronic states that may interfere in the analysis \cite{Pique1991}. The \ce{CS2} molecule has a potential well of about 36,000 cm$^{-1}$ \cite{Zhou2002}, with regularly-spaced vibrational energy levels measured up to 18,000--20,000 cm$^{-1}$ \cite{Pique1991} which means levels up to about 50\% of the dissociation energy of \ce{CS2} are measured. This is in contrast with the more "chaotic" level structure of \ce{CO2}, which has been measured up to 25,000 cm$^{-1}$, 42\% of the ground state assymetric stretch dissociation limit (see section \ref{sec:levels}). The weaker Fermi 1:2 resonance of \ce{CS2} allows showcasing a slow transition to chaotic behavior from the normal stretching and bending modes (the assymetric strech mode remaining unaffected) for increasing polyad numbers \cite{Zhou2002}.

\subsection{The need for further experimental data}

\begin{figure}
\centering
\includegraphics[width=0.49\textwidth]{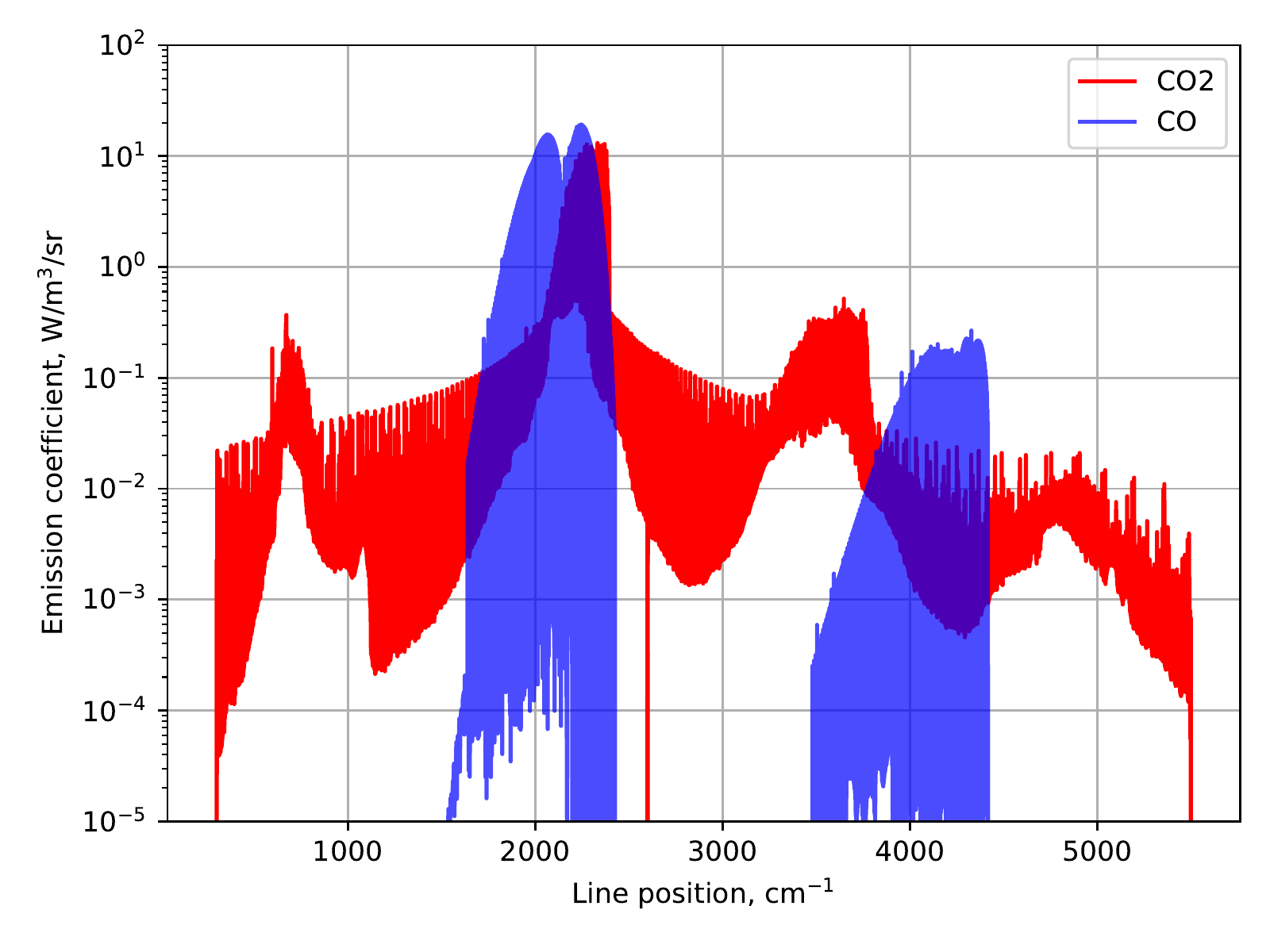}
\includegraphics[width=0.49\textwidth]{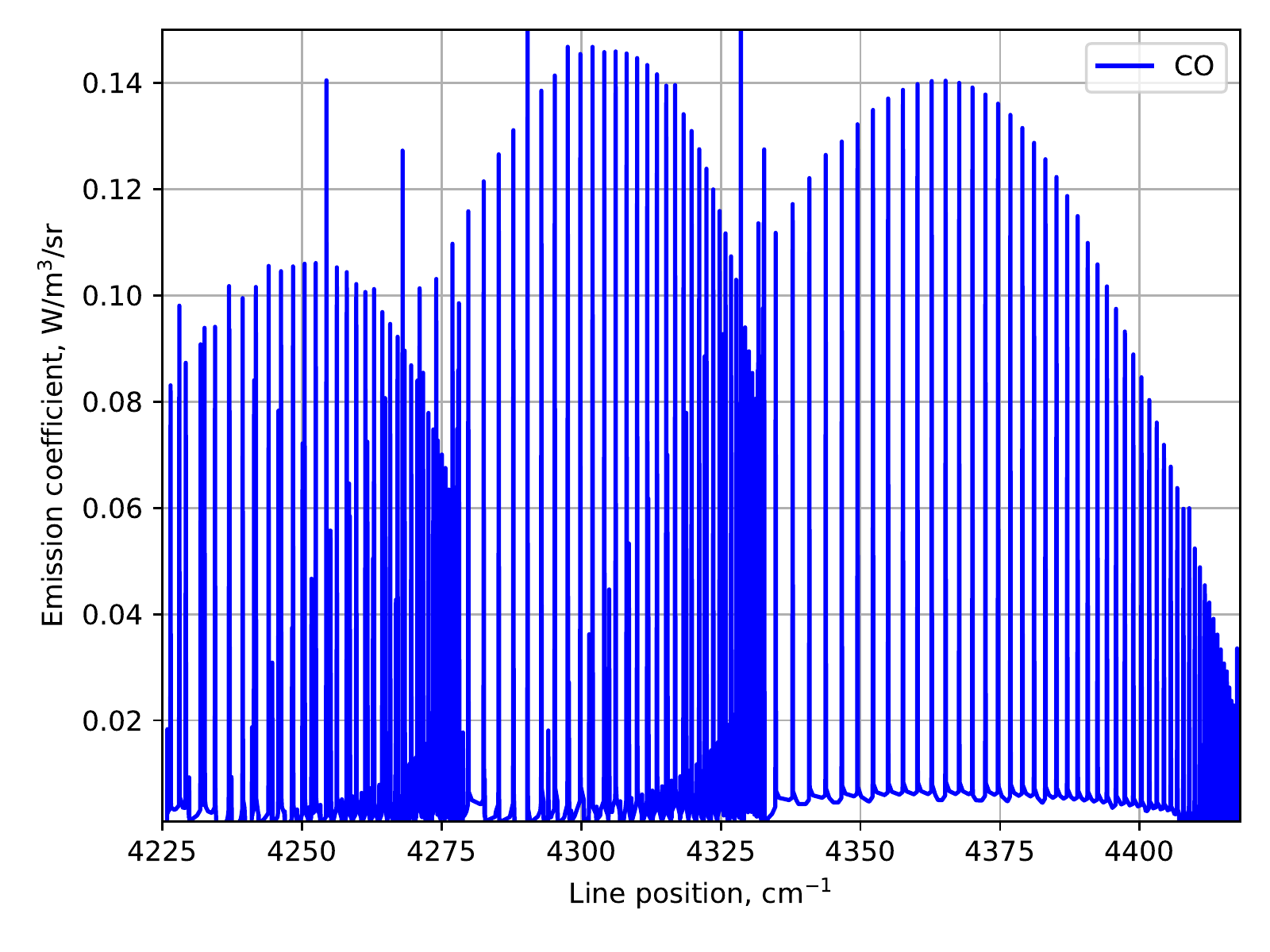}
\caption{Left: Equilibrium \ce{CO2} spectrum at 3000 K and 1 mbar, convolved with a 0.125 FWHM Gaussian Apparatus Function. The emission spectrum from \ce{CO2} is reported in red, the one from \ce{CO} is reported in blue. One may note the superposition of the fundamental band of \ce{CO} and the \ce{CO2} band in the 4.3--4.7 $\mu$m/2200--2500 cm$^{-1}$. The  other \ce{CO2} and \ce{CO} bands are well enough separated. Right: Detail for the \ce{CO} first overtone band ($\Delta$v=-2) where the rotational structure is well defined, making this a good candidate for monitoring the translational-rotational temperature in shocked flows.}
\label{fig:CO2COspectra}
\end{figure}

We conclude this work by a short discussion on the possible new experiments that would allow better calibrating our theoretical model against experimental data. Since we are chiefly concerned about heavy-impact excitation and dissociation processes in \ce{CO2}, the best-suited facilities for such analyses are undoubtedly shock-tubes. Indeed, recent shock-tube experiments have already been considered for the calibration of our model \cite{Oehlschlaeger2005,Oehlschlaeger2005a,Saxena2007,SupportMIPTReport,ReportBeck2008,Reynier2011} and there is in addition to this an extensive amount of \ce{CO2} broadband IR radiation measurements recently carried out in several test-campaigns related to Mars exploration missions in the NASA Ames EAST shock-tube \cite{Bogdanoff2009,Cruden2010,Palmer2012,Cruden2014,Brandis2015,Cruden2015,Brandis2019}. These experiments yield datasets more tailored for the study of \ce{CO2} radiation (not surprisingly since these are support campaigns for Martian planetary entry missions) rather that \ce{CO2} dissociation. To the authors knowledge, the inherently high noise/signal ratio of shock-tubes (since these are impulsive facilities, with short acquisition times) precludes the acquisition of high-resolution spectra such as the one obtained in low-pressure, steady-state plasmas (e.g. FTIR spectra). Nevertheless, even broadband radiative signals without vibrational resolution may still be fitted spectroscopically, yielding meaningful results. A possibility that could be explored would be measuring the emission bands at 4.3 and 15 $\mu$m respectively, which yield information on the relative excitation of the assymetric v$_3$ and symmetric v$_1$ modes respectively. Although the former band has a strong signal in the aforementioned NASA Ames shock-tube experiments, to our knowledge no attempts have been made yet at measuring the 15 $\mu$m bands in \ce{CO2}, and owing to their low Einstein coefficients around 10$^1$--10$^2$ (see fig. \ref{fig:posvsA}), these bands might be too diffuse to appear clearly over the background noise. On the other hand, the \ce{CO2} 4.3 $\mu$m has a certain amount of superposition with a \ce{CO} band at 4.7 $\mu$m, which might hamper the interpretation of the measured experimental spectra, if it is carried out at low resolution (see fig. \ref{fig:CO2COspectra} and discussion ahead).

In the near-UV region, probing the \ce{CO2} chemiluminescence (B$\rightarrow$X) bands yields information on the decomposition rate for \ce{CO2}. It is not possible to have absolute intensity measurements, as the transition is very diffuse and exact absorption coefficients for the transition are not known. Nevertheless one may retrieve the rate for dissociation like in the VUT-1 experiments \cite{SupportMIPTReport,ReportBeck2008,Reynier2011}, or the dissociation incubation times like in the works of Oehlschlaeger and Saxena \cite{Oehlschlaeger2005,Oehlschlaeger2005a,Saxena2007}.

Finally, other molecular radiative systems may further be considered. Further ARAS measurements (such as in Refs. \cite{Fujii1989,Burmeister1990,Eremin1993,Eremin1996}) could be carried out in conditions properly tailored from numerical test-cases, allowing the experimental determination of \ce{O}($^3$P)/\ce{O}($^1$D) ratios, as an indication of the relative ratios for direct dissociation  from the assymetric stretch mode of the ground state (X$^1\Sigma$) of \ce{CO2} versus transitions to the $^3$B$_2$ excited state and subsequent dissociation. Of particular interest would be the evolution of this ratio versus the shock speed (which induces higher translational excitation). The \ce{O2} Schumann--Runge bands can be strong emitters and absorbers in the near-UV region, providing insight on atomic oxygen recombination into \ce{O2} processes (through emission spectroscopy on the Schumann--Runge continuum), or insight on the time evolution of \ce{O2} populations (through absorption spectroscopy on the Schumann--Runge discrete bands). The probing of the \ce{CO} IR bands would also yield relevant information on the populations of one of the \ce{CO2} dissociation products. For \ce{CO} we have the fundamental band ($\Delta$v=-1) which emits around 4.7 $\mu$m, the first overtone band ($\Delta$v=-2) which emits around 2.35 $\mu$m, and the second overtone band ($\Delta$v=-3) which emits around 1.6 $\mu$m. The first band is often covered by the \ce{CO2} assymetric $v_3$ band at 4.3 $\mu$m, and the second overtone band is seldom observed due to its low intensity. This leaves us with the first overtone band, which may be observed without interference from \ce{CO2} radiation. Since this band has a regular structure, high-resolution measurements might allow fitting the rotational bands with a synthetic line-by-line code, if the signal/noise ratio is high enough. Then one would be able to determine the translational-rotational temperatures of the flow through fitting techniques. Fig. \ref{fig:CO2COspectra} presents a synthetic spectra of \ce{CO2} and \ce{CO} in equilibrium conditions at 3,000 K and 1 mbar pressure. A detail of the first overtone band of \ce{CO} at 4300 cm$^{-1}$ is also presented.

One key aspect of any experimental campaign is to have a more diversified array of shot conditions that allow reproducing slow to fast dissociation trends (with typical post-shock temperatures between 3,000 and 10,000 K) and near-equilibrium to strong nonequilibrium conditions (from 10 to 0.1 mbar pre-shock pressures). In terms of chemical compositions, the favoured approach would include the analysis of pure \ce{CO2} shocked flows versus \ce{CO2} flows highly diluted in an \ce{Ar} carrier gas. The advantage for selecting \ce{Ar} lies in its reduced mass which is quite close to \ce{CO2}. If similar aerothermodynamic conditions are achieved (similar post-shock pressure $p$ and temperature $T$) this will allow a comparison of the predictability of our kinetic model with and without reactive transitions such as \ce{CO2 + O <-> CO +O2}, since these last ones do not occur for a highly diluted \ce{CO2}--\ce{Ar} mixture. Other lighter (\ce{He}) or heavier (\ce{Kr, Xe}) dilution gases could be considered, but with a lower priority since the influence of the collisional partner of \ce{CO2} in direct dissociation reactions has already been extensively studied in the past (see ref. \cite{STELLAR} and references therein for a more detailed discussion).

\begin{figure}
\centering
\includegraphics[width=1.0\textwidth]{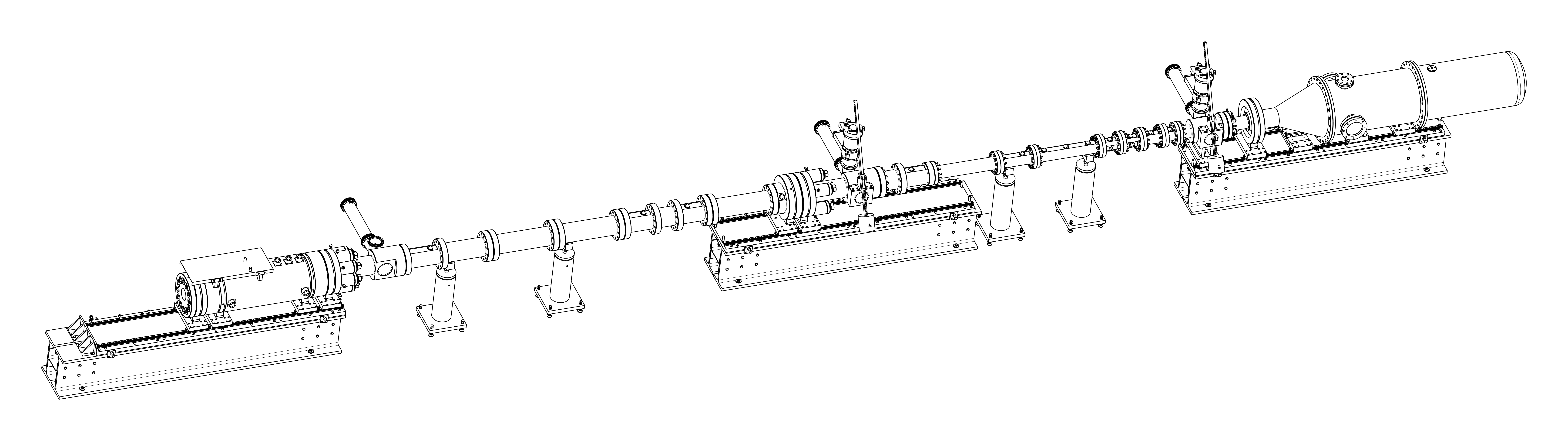}
\caption{View of the European Shock-Tube for High Enthalpy Research. ESTHER is a two-stage, combustion driven shock-tube capable of reaching shock velocities in excess of 10km/s, deploying high-speed emission and absorption diagnostics (streak-cameras+spectrometers) in the VUV, Visible, near-IR, and MWIR spectral regions}
\label{fig:ESTHER}
 \end{figure}

Our research group hosts a new generation shock-tube that has been developed in the past 10 years under funding by the European Space Agency, with the specific aim of providing Europe with a facility capable of carrying extensive investigations on nonequilibrium kinetic processes of shocked flows, for arbitrary chemical compositions. The performance mapping for the European Shock-Tube for High Enthalpy Research (ESTHER) has been recently carried out for all the Solar System planetary atmospheres \cite{DLuis2018}, including Venus and Mars \ce{CO2-N2} atmospheres. The conditions of a Venus and a Mars entry, which are of interest for planetary exploration missions nicely dovetail with the objectives of this work, as these roughly correspond to the high-temperature and low-temperature regimes of \ce{CO2} (around 10,000 and 3000 K respectively). A research programme on the kinetics of \ce{CO2} shocked flows with interest for atmospheric entry applications is currently underway under funding from the European Space Agency \cite{ESACO2Contract}, and some of the recommendations for novel experiments discussed here will be deployed in our facility (see fig. \ref{fig:ESTHER}).

\subsection{Concluding remarks}

This work aimed at furthering the state of the art on \ce{CO2} vibrationally-specific kinetics, by implementing improved theoretical approaches, including a better description of the ground and excited electronic levels manifold, improved state-specific collisional models such as the Forced Harmonic Oscillator theory --which has been extended to triatomic molecules such as \ce{CO2}-- and compiling a set of adequate chemistry rates, yielding a self-contained model capable of making predictive simulations of \ce{CO2} dissociation processes in shocked flows. However, in a certain sense, a completely adequate modeling of \ce{CO2} state-to-state processes is an endeavour still out of reach, which will require extensive theoretical and experimental work, with the application of novel approaches even further beyond the state-of-the-art of our model. A possible roadmap of theoretical improvements and new experiments has been proposed to further this goal. We hope that this work may inspire other authors to take further steps in this direction. The state-to-state rate coefficients, computed or adapted for this work, are publicly made available as part of the STELLAR database \cite{STELLAROnline}.

\section*{Acknowledgments}

The authors thank Xinchuan Huang for sharing the coefficients for the NASA--Ames--2 PES of \ce{CO2}. Additionally, the authors would like to thank the following colleagues for the many discussions and exchanges over the course of this work, Brett Cruden for providing insight on his macroscopic rates review, as well as information on the recent NASA \ce{CO2} shock-tube campaigns; Vasco Guerra and Tiago Silva for useful discussions on \ce{CO2} plasma reforming, state-to-state kinetics and the Fermi approximation; Elena Kustova and Mariia Mekhonoshina for useful discussions on \ce{CO2} state-to-state modeling; Jonathan Tennyson for discussions on vibrational chaos; Theresa Urbanietz for insight on the postprocessing of FTIR experiments of Refs. \cite{Urbanietz2018,Stewig2020}; Amal Sahai for discussions on the state-to-state binning theory; and Pedro Viegas for insight on the Fokker--Planck theory. Lastly, a heartfelt acknowledgment to Marco Panesi who hosted J. Vargas at University of Illinois at Urbana--Champaign for a year as a visiting scholar and without whom this work would not have been possible. This work has been partially supported by the Portuguese FCT, under Projects UIDB/50010/2020 and UIDP/50010/2020 and the grant PD/BD/114325/2016 (PD-F APPLAuSE) and 4000118059/16/NL/KML/fg "Standard kinetic models for \ce{CO2} dissociating flows".

\section*{Data Availability Statement}

The data that support the findings of this study are openly available in \cite{STELLAROnline}.

\bibliography{bibliography.bib}

\end{document}